\documentclass[reprint,amsmath,amssymb,aps,prb,noeprint,onecolumn]{revtex4-2}
\usepackage{physics}
\usepackage{graphicx}
\usepackage{xcolor}
\usepackage[colorlinks=true,citecolor=blue,linkcolor=blue]{hyperref}
\usepackage{ulem}
\graphicspath{{figs/}}

\newcommand{\beq}{\begin{equation}} \newcommand{\eeq}{\end{equation}}

\def\SDG{\textrm{Schr\"{o}dinger }}

\begin{document}

\title{
Analysis of switching operation in quantum conveyance 
}

\author{Yoshiaki Teranishi}
\email[]{tera@nctu.edu.tw}
\affiliation{Institute of Physics, National Yang Ming Chiao Tung University, Hsinchu, 30010, Taiwan}
\author{Satoshi Morita}
\affiliation{Graduate School of Science and Technology, Keio University, Yokohama, Kanagawa 223-8522, Japan}
\affiliation{Keio University Sustainable Quantum Artificial Intelligence Center (KSQAIC),
  Keio University, Minato-ku, Tokyo 108-8345 Japan}
\author{Seiji Miyashita}
\affiliation{JSR-UTokyo Collaboration Hub, CURIE, 
  Department of Physics, The University of Tokyo, Bunkyo-Ku, Tokyo, 113-0033, Japan}

\date{\today}

\begin{abstract}
  In this paper, we study the quantum dynamics of a particle conveyed by a moving potential well, with a focus on its survival probability. In physical systems, this process is inevitably subjected to external disturbances, such as environmental coupling, which reduce survival probability. Even when external noise is suppressed, however, intrinsic disturbances remain, leading to unwanted leakage from the trapping potential. Specifically, dynamic parameter variations induce nonadiabatic transitions.
Various types of nonadiabatic transitions arise from the time-dependence of the potential parameters. When the potential well is smoothly accelerated, the trapped particle may escape due to inertia, a phenomenon known as adiabatic tunneling. Beyond this, the switching procedures (starting and stopping the motion) exert a significant impact that depends heavily on the abruptness of the operation. Consequently, precise dynamical control of the system Hamiltonian is essential, yet it remains a highly nontrivial task.
We analytically investigate the mechanism underlying these switching effects and derive a closed-form formula to quantify them. By combining this switching contribution with the adiabatic tunneling rate, we accurately approximate the total nonadiabatic effect. Our framework reproduces survival probabilities across various acceleration protocols with high precision.
Since the switching effect is derived analytically and the adiabatic tunneling rate is determined by constant-acceleration analysis, the model is independent of specific, complex protocols. Once the fundamental parameters determined from a few numerical simulations of typical cases are established, the survival probabilities for arbitrary acceleration protocols can be predicted. This provides a practical framework for real-time quantum control without the need for full dynamical simulations.
\end{abstract}

\maketitle

%-----------------------------------------------------------

\section{Introduction}

The control and understanding of real-time dynamics in quantum systems are important issues.
They are essential not only for fundamental studies of nonequilibrium quantum phenomena, but also for a wide range of applications such as quantum computation, quantum simulation, and quantum information processing.
Precise manipulation of quantum states inevitably requires detailed knowledge of how quantum systems evolve in time under external driving fields.

A major obstacle to quantum control is decoherence, which degrades quantum coherence and limits controllability.
While dissipation caused by coupling to the environment is often regarded as the primary source of decoherence, even in the absence of such effects quantum systems can exhibit intrinsic decoherence mechanisms of purely quantum-mechanical origin.
Such mechanisms are closely related to nonadiabatic processes and quantum fluctuations, which play a crucial role in real-time quantum dynamics.

Among various quantum-mechanical mechanisms,
quantum tunneling~\cite{Razavy2013, SM2022} and Landau--Zener
transitions~\cite{Landau1932,Zener1997,Stueckelberg1932,Majorana1932} represent complementary aspects of nonadiabatic quantum dynamics.
The former captures decay induced by coupling
between energetically degenerate states across
classically forbidden regions, whereas the latter describes transitions among discrete adiabatic levels driven by temporal variations of system parameters.
Landau--Zener phenomena under ac fields~\cite{SM-QD2011,LZac1,LZac2}, as well as the behavior of metastable states under a sweeping field~\cite{Hatomura2016, Hatomura2017}, show rich and interesting phenomena.
Relaxation due to quantum fluctuations has been studied from the viewpoint of resonance states (non-Hermitian physics)~\cite{Siegert,Feshbach1958,Feshbach1962,Feshbach1967,Fano,Hazi-Taylor,Langer1937,Miller-Good,Connor1968,Dickinson1970,Mayer-Walter,Child-text,Hatano-nonhermite,moiseyev2011nonhermitian}.
Manipulating quantum systems with time-dependent fields in a desired way by means of quantum optimal control has been proposed, for example, in the control of laser-driven molecular dynamics~\cite{shi1988optimal,brif2010control}.
Real-time dynamics also plays an important role in the quantum annealing process~\cite{Kadowaki1998,Das2008,MoritaNishimori2008,Tanaka2017,DWave2011,DWave2014}.
The control of the resonance width is utilized in schemes for laser purification in molecular vibrational cooling~\cite{atabek2013proposal,leclerc2016controlling}.
Recently, protocols to overcome an energy barrier using shaped pulses have been proposed~\cite{Miyashita2023,Miyashita2024}.

For the manipulation of quantum states in real time, it is important to control nonadiabatic transitions.
When the parameters in the Hamiltonian vary slowly, a state initially in an eigenstate follows the corresponding eigenstate of the changing Hamiltonian, which is known as the adiabatic theorem.
However, when the parameters change at finite rates, nonadiabatic transitions occur, which can have undesired effects on quantum-state manipulation.

Generally, when parameters vary at finite rates, nonadiabatic transitions are induced.
In a concrete manipulation procedure, parameters are increased and decreased in various ways.
Such procedures are highly nontrivial, and their time dependence is non-uniform.

When the rate of change of parameters is much slower than the typical time scale of the system Hamiltonian, over a short period the process can be regarded as a system under constant acceleration of the parameters.
This quantum tunneling under constant acceleration is referred to as ``adiabatic tunneling.''
Besides this, for a concrete protocol of manipulation, switching procedures, i.e., ``switch-on'' and ``switch-off'' operations to start and stop the process, are necessary.
These also unavoidably give rise to nonadiabatic effects, which are described by the ``sudden approximation.''
Even if the parameter changes at the switching are continuous, the time dependence of the parameters, such as acceleration, can introduce non-analytic behavior.
It has been pointed out that such discontinuities cause nonadiabatic processes~\cite{Morita2007}.
In the present paper, we study this effect in detail and provide a concrete formula for the latter effect.
Hereafter, we refer to this effect as the ``switch disturbance.''

In the present paper, we study how nonadiabatic transitions are induced in a system of particle conveyance, where a particle is trapped in a potential well and conveyed by moving the position of the potential well.
Simple conveyance induced by an impulsive force was studied in Ref.~\cite{Conveyance0}.
Survival probabilities in the conveyance of a particle by a trapping potential under constant acceleration and under slow (gradual) acceleration were investigated in Ref.~\cite{MTM2024}, focusing on the population dynamics of adiabatic eigenstates under time-dependent acceleration.

Indeed, as a form of dynamical control in real-time quantum dynamics, quantum effects in particle conveyance using potential traps constitute a variety of important research topics.
With regard to the microscopic control of particle positions using optical potentials, techniques such as optical lattices and optical tweezers have been extensively developed, in which particles are conveyed by the motion of trapping potentials.
There have been extensive studies on particle conveyance using optical trapping potentials in ultracold gases~\cite{optical-lattice,optical-lattice2}.

The transport of charged particles and energetic neutral particles between vacuum chambers is also an important issue.
For example, in spin-dependent optical lattice potentials, where atoms are prepared in a superposition of two internal spin states, state-selective optical potentials are used to split the wave function of a single atom and transport the corresponding wave packets in two opposite directions~\cite{Spin-dependent-optlattpot}.
Fast atomic transport in a moving double-well optical lattice, whose anharmonic potential is nonseparable in the $x$--$y$ plane, has also been studied theoretically.
In such systems, configurations of acousto-optic modulators give rise to an effective moving lattice~\cite{Double-well-pot-optlattpot}.
The control and visualization of atomic motion have been extensively investigated by the Ohmori group~\cite{Ohmori2013,Ohmori2016,Ohmori2018ACR,Ohmori2018PRL,Ohmori2020,Ohmori2022,Ohmori2023PRL,Ohmori2023arxiv}.
Quantum processors using optical tweezers to transport neutral atoms are expected to realize fault-tolerant quantum computation~\cite{QuEra2022,QuEra2023,Ohmori2022}.
To form lattice structures by arranging trapped atoms, atom-by-atom assembly techniques have been realized.
In this context, it has been pointed out that particle loss induced by atomic motion constitutes an important issue~\cite{atom-by-atom-assembly1,atom-by-atom-assembly2}.
The motion of atoms trapped in Wannier--Stark ladders in an accelerating one-dimensional standing wave of light with a small oscillatory component has been studied, and particle loss at resonant points has been observed~\cite{Atomic-Wannier-Stark-Ladders}.
Particle conveyance is also an important process in ion-trap-based quantum manipulation~\cite{Cirac1995}.
The dynamics of single and multiple ions is a key ingredient in ion-trap procedures, including transport between, and separation into, spatially distinct locations in multizone linear Paul traps~\cite{ion-trap-Array1,ion-trap-Array2,Quantinuum2021,Quantinuum2023}.
Conveyance of electrons by surface acoustic waves has been discussed extensively~\cite{Tarucha2011,Byeon2021,APL119-2021}.
Coherent transport of a single electron across an array of quantum dots has also been studied~\cite{Fujita2017coherent,Mills2019shuttling,Yoneda2021coherent,Tarucha2022}.
The motion of a spin qubit through multiple quantum dots while preserving its quantum information is another important process~\cite{quantum-dots}.
Semiconductor platforms based on surface acoustic waves for such processes have been extensively investigated~\cite{SAW-platform1,SAW-platform2,SAW-platform3}.
Silicon-based platforms have also been explored, and scalable spin-qubit shuttling devices have been proposed~\cite{silicon1,silicon2,silicon3,silicon4}.
In such transport processes, nonadiabatic transitions and quantum tunneling can play a crucial role, particularly when fast or highly controlled motion is required.
For the conveyance of atomic clouds by magnetic potentials, such as gaseous Bose--Einstein condensates~\cite{BEC0,BEC1,BEC2}, ensembles of particles in magnetic traps have been transported~\cite{mag-conv,mag-conv-rev}.
In systems where Bose--Einstein condensates are adiabatically loaded into one-dimensional optical lattices, the dynamics of the condensate, such as Bloch oscillations and Landau--Zener tunneling, have been observed when the periodic potential provided by the optical lattice is accelerated~\cite{BEC1D-optical-lattices1,BEC1D-optical-lattices2,BEC1D-optical-lattices3}.
Long-distance transport of gaseous Bose--Einstein condensates has also been realized by trapping the condensate at the focus of an infrared laser and transferring the position of the laser focus through controlled acceleration, forming a condensate beam line~\cite{BEC-optTweezers}.
Although in such cases the energy structure of the trapping potential becomes complicated, 
the dominant level corresponding to the condensate ground state, together with the excited states above it, retains a similar structure.
Therefore, the concepts discussed in the present paper remain relevant and provide important insights into such transport processes.

To analyze the survival probability under conveyance, according to the above-mentioned picture,
we propose a formula in which the dominant nonadiabatic effects are decomposed into two universal contributions, namely, adiabatic tunneling during smooth evolution and switching-induced disturbances.
We analyze the mechanism of the switch disturbance in the context of the adiabatic theorem and also obtain its contribution to the survival probability explicitly.
Combining this with the contribution from adiabatic tunneling during conveyance with a smooth change of acceleration, we propose a simple formula for the survival probability.
We apply the proposed formula to various acceleration protocols and find excellent agreement with numerical simulation results, demonstrating that the survival probability is reproduced with very high accuracy.

The rate of adiabatic tunneling can be obtained by analyzing the case of constant acceleration.
The contribution from the switching disturbance is common to protocols that share the same form of acceleration near the initial switching point, and we obtain an analytical expression for this contribution.
Therefore, once the disturbance factors and the adiabatic tunneling rate under constant acceleration are known, the survival probability for a given trapping potential can be estimated without performing full dynamical simulations, regardless of the detailed form of the acceleration protocol $a(t)$.
This decomposition provides a clear and practical framework for quantifying nonadiabatic effects in real-time quantum control.

This paper is organized as follows.
In Sec.~\ref{sec_1}, we review the adiabatic theorem and nonadiabatic effects arising from changes at switching points.
In Sec.~\ref{sec_2}, the switching effect in quantum conveyance is studied.
In Sec.~\ref{sec_3}, survival probabilities in several conveyance protocols are studied.
In Sec.~\ref{sec:summary}, we summarize our findings.
In Appendix~\ref{sec_harmonic}, we provide an explicit form of the switching disturbance factor using a harmonic model.

\section{Deviation from adiabatic motion}\label{sec_1}

When the Hamiltonian changes in time, the population of each adiabatic eigenstate varies due to nonadiabatic effects. 
If the change is infinitesimally slow, the adiabatic theorem states that the populations are conserved, which is referred to as ``adiabatic motion''.
The amount of change depends on the rate of parameter variation and the energy difference between states.
For example, the Landau-Zener formula gives the amount of nonadiabatic transition at an avoided level crossing.
This is an example of a nonadiabatic change during a parameter variation process.

In practical processes, the effects of switch-on and switch-off must be taken into account.
Such effects have been studied in Landau-Zener phenomena~\cite{Morita2007} and also in conveyance problems~\cite{MTM2024}. 
The amount of nonadiabatic change depends on how smoothly parameters of the
system change at the points.

Generally, let us consider a Hamiltonian with a time dependent
part with a parameter $f(t)$ that switches from one function to
another at $t=t_0$, namely
\beq
f(t)=
\begin{cases}
f_1(t) & t\leq t_0\\
f_2(t) & t> t_0.
\end{cases}
\eeq
Assume that both functions \(f_1(t)\) and \(f_2(t)\) vary slowly enough for the adiabatic approximation
to hold within each individual interval. 
If \(f_1(t_0) \neq f_2(t_0)\), the adiabatic approximation naturally breaks down, resulting in a transition at the switching point. 
Conversely, if $f_1(t_0)=f_2(t_0)$, one might expect no transitions to occur since the adiabatic states depend solely on the instantaneous parameter $f(t)$. However, as demonstrated in
Ref.~\cite{Morita2007}, even in such a continuous case, a discontinuity in a higher-order
derivative of $f(t)$ can still induce a nonadiabatic transition.

In the case of switch-on, the parameter $f_1(t)$ is constant before
the initial time. Therefore, any time dependence in \(f_2(t)\) induces a transition. 
If the $n$-th time derivative of the adiabatic parameter exhibits a discontinuity at
the initial time, the nonadiabatic transition probability from an initial adiabatic
state \(\ket{i}\) to a final adiabatic state \(\ket{j}\) is given by 
\begin{equation}
P = \left|A_{ij}\right|^2
\label{eqn:mori1}
\end{equation}
with
\begin{equation}
A_{ij} \equiv 
\frac{\hbar^n \langle j|\hat{H}^{(n)}|i \rangle}{\Delta_{ij}^{n+1}},
\quad \hat{H}^{(n)}={d^n\hat{H}(t) \over dt^n},
\label{eqn:mori2}
\end{equation}
where
 $\Delta_{ij}$ is the energy gap between the adiabatic states $|j\rangle$ and $|i\rangle$.

Here, we focus on the case
where the time-dependent
part is given by a single operator $\hat{V}_0$ with a time-dependent coefficient $\alpha(t)$:
\begin{equation}
\hat{H}(t) = \hat{h} + \hat{V}(t),\quad \hat{V}(t) = \alpha(t)\hat{V}_0.
%\label{hamil-multi}
\end{equation}

\subsection{Basic Idea} %%%%%%%%%%%%%%%%%%%%%%%%%%%%%%
\label{sec:org849c7f6}

Let us consider the time-dependent \SDG equation
\begin{equation}
i\hbar\frac{\partial}{\partial t}\psi(t) = \hat{H}(t)\psi(t).
\label{SDG-eq-app}
\end{equation}
Here, we assume the Hamiltonian before the switching at $t=t_0$ is time-independent ($f_1(t) = \rm{const}$),
and then it starts to change in time with $f_2(t)$ after the switching.
Let a unitary matrix composed of eigenstates of $\hat{H}(t)$ be $\hat{W}_0^{-1}(t)$:
\begin{equation}
\hat{W}_0^{-1}(t)\hat{H}(t)\hat{W}_0(t) = \hat{D}_0(t).
\end{equation}
Here $D_0(t)$ is a diagonal matrix with the adiabatic eigenenergies at time $t$.

The state $\psi(t)$ is expressed in this basis as 
\begin{equation}
\phi_0(t) = \hat{W}_0^{-1}(t)\psi(t).
\end{equation}
The \SDG equation in this representation is
\begin{align}
i\hbar\frac{\partial}{\partial t}\phi_0(t) & =
\left[ \hat{D}_0(t) -i\hbar\hat{W}_0^{-1}(t)\dot{\hat{W}}_0(t)
\right]\phi_0(t) \nonumber \\
& \equiv \hat{H}_0(t)\phi_0(t).
\end{align}

In the adiabatic theorem, the second term on the right-hand side vanishes in the slow limit
(see Ref.\cite{griffiths2018introduction}).
Thus, in this limit, $\hat{H}_0(t)$ becomes diagonal.
In such a case, \(\phi_0(t)\) is given by
\beq
\phi_0(t) = e^{-\frac{i}{\hbar}\int_{0}^{t}\hat{D}_0(t^\prime)dt^\prime}\phi_0(0).
\eeq
Here, the population of each eigenstate does not change, and only the phase evolves.

If the term $\hat{W}_0^{-1}(t)\dot{\hat{W}}_0(t)$ cannot be ignored, we proceed by applying another unitary transformation $\hat{W}_1$ to diagonalize \(\hat{H}_0(t)\):
\begin{equation}
\hat{W}_1^{-1}(t)\hat{H}_0(t)\hat{W}_1(t)=\hat{D}_1(t).
\end{equation}

Similarly,
\begin{equation}
\phi_1(t) = \hat{W}_1^{-1}(t)\phi_0(t) = \hat{W}_1^{-1}\hat{W}_0^{-1}\psi(t),
\end{equation}
and the \SDG equation becomes
\begin{align}
i\hbar\frac{\partial}{\partial t}\phi_1(t) & =
\left[ \hat{D}_1(t) -i\hbar\hat{W}_1^{-1}(t)\dot{\hat{W}}_1(t)
\right]\phi_1(t) \nonumber \\
& \equiv \hat{H}_1(t)\phi_1(t).
\end{align}

Repeating this procedure \(n\) times, we obtain 
\begin{align}
\phi_n(t) &= \hat{W}_n^{-1}(t)\phi_{n-1}(t) = \hat{W}_n^{-1}\hat{W}_{n-1}^{-1}\cdots\hat{W}_0^{-1}\psi(t)
\nonumber \\
&\equiv\hat{U}_n^{-1}\psi(t),
\label{def-U}
\end{align}
with
\begin{equation}
\hat{W}_n^{-1}\hat{H}_{n-1}(t)\hat{W}_n(t)=\hat{D}_n(t).
\end{equation}

The transformed function \(\phi_n(t)\) satisfies
\begin{align}
i\hbar\frac{\partial}{\partial t}\phi_n(t) &=
\left[ \hat{D}_n(t) -i\hbar\hat{W}_n^{-1}(t)\dot{\hat{W}}_n(t)
\right]\phi_n(t) \nonumber \\
& \equiv \hat{H}_n(t)\phi_n(t).
\label{Nth-adia}
\end{align}

If the term $i\hbar\hat{W}_n^{-1}(t)\dot{\hat{W}}_n(t)$ can be neglected,
the solution is
\begin{equation}
\phi_n(t)=e^{-\frac{i}{\hbar}\int_{t_0}^t \hat{D}_n(t^{\prime})dt^{\prime}}\phi_n(t_0),
\end{equation}
and thus
\begin{equation}
\psi(t) = \hat{U}_n(t) e^{-\frac{i}{\hbar}\int_{t_0}^t \hat{D}_n(t^{\prime})dt^{\prime}}
\hat{U}_n^{-1}(t_0)\psi(t_0).
\end{equation}

This expression indicates that transitions occur due to the unitary transformations $\hat{U}_n^{-1}(t_0)$ and $\hat{U}_n(t)$ at the initial and final times, respectively, corresponding to transitions at switching points.

It should be noted that the convergence of the series of $i\hbar\hat{W}_n^{-1}(t)\dot{\hat{W}}_n(t)$ is not guaranteed. 
In such cases, the adiabatic picture of time evolution is not applicable.
Even if \(i\hbar\hat{W}_n^{-1}(t)\dot{\hat{W}}_n(t)\ne 0\), if higher-order terms are sufficiently small, this sequence of transformations provides a good description of the dynamics.

In the next section, we study the relation of our treatment to the adiabatic approximation in the nearly adiabatic limit.
In the present paper, we assume this condition is satisfied.

\subsection{Formula of nonadiabatic change in a slowly starting process}\label{sec_2b}
   
The matrix element of the term $-i\hbar\hat{W}^{-1}_0\dot{\hat{W}}_0$
between the adiabatic states $j$ and $k$ 
is given by
\begin{equation}
(\hat{Q}_0)_{jk} = -i\hbar\frac{\dot{\hat{H}}_{jk}}{(\hat{D}_0)_k-(\hat{D}_0)_j} {\rm \ for\ } j\neq k,
\end{equation}
and
\begin{equation}
(\hat{Q}_0)_{jj} = 0,
\end{equation}
with
\begin{equation}
\dot{\hat{H}}_{jk} = \left( \hat{W}^{-1}_{0}\dot{\hat{H}}\hat{W}_0\right)_{jk}.
\end{equation}
If a sudden change takes place at $t=t_0$, the time derivative $\dot{\hat{H}}$ is mathematically undefined. In this case, we can apply the sudden approximation, where the transition probability is given by $\left|\left( \hat{W}_0^{-1}(t+0)\hat{W}_0(t-0) \right)_{jk}\right|^2$. 
Next, we consider a continuous time-dependent Hamiltonian, for which the adiabatic 
approximation nearly holds, where the nonadiabatic coupling \(\hat{Q}_0\)
is small or zero. We discuss how this small nonadiabatic
coupling contributes to the dynamics. In this case, the second unitary 
transformation operator \(\hat{W}_1\) that diagonalizes the Hamiltonian \(\hat{H}_0\)
can be treated by first-order perturbation theory:
\begin{equation}
\hat{W}_1(t) = \hat{I} + \hat{\varepsilon}_1(t),
\end{equation}
where \(((\hat{\varepsilon}_1)_{jk}(t))^2 \simeq 0\).
In this case, the operator \(\hat{\varepsilon}_1\) is given by
\begin{equation}
(\hat{\varepsilon}_1)_{jk}(t) = \frac{(\hat{Q}_0)_{jk}}{(\hat{D}_1)_k-(\hat{D}_1)_j},
\end{equation}
and
\begin{equation}
(\hat{\varepsilon}_1)_{jj}(t) = 0.
\end{equation}

The diagonal matrix \(\hat{D}_1\) is equal to \(\hat{D}_0\),
because \( (\hat{Q}_0)_{jj}(t) = 0\). 
The time derivative of \(\hat{W}_1\) is given by Ref.~\cite{Messiah} as
\begin{equation}
\dot{\hat{W}}_1 = \dot{\hat{\varepsilon}}_1,
\end{equation}
with
\begin{equation}
(\dot{\hat{\varepsilon}}_1)_{jk}(t)
= \frac{(\dot{\hat{Q}}_0)_{jk} ((\hat{D}_1)_k-(\hat{D}_1)_j)
- (\hat{Q}_0)_{jk} ((\dot{\hat{D}}_1)_k-(\dot{\hat{D}}_1)_j)}
{((\hat{D}_1)_k - (\hat{D}_1)_j)^2}.
\end{equation}

This indicates that even when $\hat{Q}_0=0$ at $t=t_0$, a nonadiabatic effect exists as long as $\dot{\hat{Q}}_0$ is nonzero. 

Hereafter, we focus on the nonadiabatic effect under the condition $\hat{Q}_0=0$. 
In this case, we have 
\begin{equation}
(\dot{\hat{\varepsilon}}_1)_{jk}(t)\simeq
\frac{(\dot{\hat{Q}}_0)_{jk}}{(\hat{D}_0)_k - (\hat{D}_0)_j}.
\end{equation}
Then, as the second step, we set
\begin{equation}
\hat{Q}_1 = -i\hbar \hat{W}_1^{-1}\dot{\hat{W}}_1
= -i\hbar(\hat{I}-\hat{\varepsilon}_1)\dot{\hat{\varepsilon}}_1
\simeq -i\hbar \dot{\hat{\varepsilon}}_1,
\end{equation}
and its matrix elements are expressed as
\begin{equation}
\begin{split}
(\hat{Q}_1)_{jk} &= -i\hbar\frac{(\dot{\hat{Q}}_0)_{jk}}{(\hat{D}_0)_k-(\hat{D}_0)_j}\\
&= (-i\hbar)^2 \frac{\ddot{\hat{H}}_{jk}}{[(\hat{D}_0)_k - (\hat{D}_0)_j]^2},
\end{split}
\end{equation}
This indicates the existence of a nonadiabatic effect even when the first time derivative of the Hamiltonian vanishes. 
Furthermore, even if both the first and second time derivatives of the Hamiltonian vanish, a nonadiabatic effect can still exist.
In the same way,
\begin{equation}
(\dot{\hat{Q}}_1)_{jk}\simeq -i\hbar\frac{(\ddot{\hat{Q}}_0)_{jk}}{ (\hat{D}_0)_k - (\hat{D}_0)_j },
\end{equation}
is not necessarily zero, the contribution of the next order is given by
\begin{equation}
({\hat{Q}}_2)_{jk}\simeq (-i\hbar)^3 \frac{\hat{H}^{(3)}_{jk}}{ [(\hat{D}_0)_k - (\hat{D}_0)_j]^3 }.
\end{equation}
If all derivatives up to $(n-1)$th order vanish, then, by repeating a similar transformation, we have
\begin{equation}
({\hat{Q}}_{n-1})_{jk}\simeq (-i\hbar)^n \frac{\hat{H}^{(n)}_{jk}}{ [(\hat{D}_0)_k - (\hat{D}_0)_j]^n },
\end{equation}
and the deviation from the diagonal form is given by
\begin{equation}
\begin{split}
(\hat{\varepsilon}_n(t))_{jk} & \simeq
\frac{(\hat{Q}_{n-1})_{jk}}{(\hat{D}_0)_k-(\hat{D}_0)_j} \\
&=
(-i\hbar)^n \frac{\hat{H}^{(n)}_{jk}}{ [(\hat{D}_0)_k - (\hat{D}_0)_j]^{n+1} }.
\end{split}
\end{equation}
Then, the unitary transformation matrix defined by Eq.~(\ref{def-U})
is found to be
\begin{equation}
\hat{U}_n = \prod_{m=0}^{n}\hat{W}_m\simeq \hat{I} + \hat{\varepsilon}_n.
\end{equation}
Therefore, the transition probability from state $k$ to state $j$ at the initial time is given by
\begin{equation}
P_{k\rightarrow j} = \left|(\hat{U}_n)_{jk}\right|^2 
= \left|(\hat{\varepsilon}_n(t_0))_{jk}\right|^2.
\label{prob-multi}
\end{equation}
In the case, the Hamiltonian is separated into the time-dependent part
$\hat{h}$ and the independent part $\hat{V}$,
\begin{equation}
\hat{H}(t) = \hat{h} + \alpha(t) \hat{V}_0,
\label{hamil-multi}
\end{equation}
where $\alpha(t)$ is the adiabatic parameter and $\hat{V}_0$ is a time-independent
operator.

Noting
\beq
\alpha^{(n)}(t)={d^n\over dt^n}\alpha(t),
\eeq
the $n$-th order time derivative of the Hamiltonian $(\hat{H}^{(n)})_{jk}$ is 
\begin{equation}
(\hat{H}^{(n)})_{jk} = (\hat{V}_0)_{jk} \alpha^{(n)}(t).
\end{equation}
Assuming that $\alpha^{(m)}(t_0)=0$ for $m<n$ and $\alpha^{(n)}(t_0)\neq0$,
the transition probability due to the discontinuous change in the Hamiltonian at the initial time in Eq.~(\ref{prob-multi}) is given by
\begin{equation}
\begin{split}
P_{k\rightarrow j} &= \left|(\hat{\varepsilon}_n(t_0))_{jk}\right|^2 \\
&= 
\left|\hbar^n \frac{(\hat{V}_0)_{jk}}{ [(\hat{D}_0)_k - (\hat{D}_0)_j]^{n+1} } \alpha^{(n)}(t_0)\right|^2,
\end{split}
\label{Pkj-alpha-n}
\end{equation}
indicating that $P_{k\rightarrow j}$ is proportional to $|\alpha^{(n)}(t_0)|^2$ for all $k$ and $j$.
Thus, the total transition probability is also proportional to $|\alpha^{(n)}(t_0)|^2$.

The total transition probability is given by the sum over $j$. 
However, in cases where $(\hat{V}_0)_{jk}$ is nonzero only for a specific state $j_0$, the amount of nonadiabatic change is proportional to $P_{k\rightarrow j}$.
For example, in the case of a perturbation of the form $\alpha(t)x$ in a harmonic potential ($\hat{h}=p^2/2m+kx^2/2$), $\langle j|x|0\rangle$ is nonzero only for $j=1$.

In such cases, the amount of nonadiabatic change, when $\alpha^{(n)}(t_0)\ne 0$ and $\alpha^{(n')}(t_0) = 0$ for $n' < n$, is given by
\beq
P=\left|B_n\right|^2|\alpha^{(n)}(t_0)|^2,
\label{P-Bn-alpha}
\eeq
where
\begin{equation}
B_0 = \left| \frac{(\hat{V}_0)_{jk}}{ [(\hat{D}_0)_k - (\hat{D}_0)_j] }\right|, \quad 
B_1 = \left|\hbar \frac{(\hat{V}_0)_{jk}}{ [(\hat{D}_0)_k - (\hat{D}_0)_j]^{2} }\right|,
\end{equation}
which depend on details of the model, e.g., the form of the trapping potential and the mass of the particle $m$.
However, it should be noted that the higher-order coefficients are given by $B_0$ and $B_1$ as
\begin{equation}
B_n = B_0^{1-n}B_1^n.
\label{Bn}
\end{equation}

For example, consider nonadiabatic transitions in a process such as Landau-Zener scattering~\cite{Morita2007}.
Expressing the sweeping field from $t=0$ to $t=\tau$ as
\beq
H_z(t)=h\left[{1\over2}-f\left({t\over\tau}\right)\right],
\eeq
the Hamiltonian is given by
\beq
{\cal H}_{\rm LZ}=-H_z(t)\sigma^z-h_x\sigma^x.
\eeq
In this case, the nonadiabatic change during the process is described by the Landau-Zener formula for the transition probability at an avoided level crossing point:
\beq
P_{\rm LZ}=\exp\left(-{\pi\Delta E^2\over \hbar v}\right),
\label{P_LZ0}
\eeq
where $v$ is the rate of change of the energy difference at the avoided crossing.
In addition, the contribution from the switching points, where the external field starts at the initial time $t=0$ and stops at the final time $t=\tau$, becomes significant.
When the sweeping velocity $(v\propto\tau^{-1})$ is small, the contribution at the level crossing point becomes negligible, i.e., the state follows the ground state almost adiabatically, and the contribution from the switching points becomes dominant.

In Ref.~\cite{Morita2007}, $f_1(t)=t$, $f_2(t)=t^2(3-2t)$, $f_3(t)=t^3(10-15t+6t^2)$, and $f_4(t)=t^4(35-84t+70t^2-20t^3)$ were studied, and the excitation probabilities were presented in Fig.~3 of the paper.
Equation~(18) in that paper provides an explicit example of the relations of Eq.~(\ref{Pkj-alpha-n}) and Eq.~(\ref{Bn}).

\section{Switching effects in Quantum Conveyances}\label{sec_2}

It has also been pointed out that switching effects play important roles in the survival properties in quantum conveyance problems~\cite{MTM2024}.

We consider a quantum mechanical conveyance of a particle from the position $x = 0$ to $x = L$ over a duration $\tau$ by transporting a potential well $V_{\rm well}\left(x-x_0(t)\right)$ in which the particle is trapped.
As has been explained, in this process, the particle needs to be accelerated and
decelerated, during which the particle escapes by quantum tunneling, and disturbances at the switching points also cause
a decrease in the survival probability.
Taking into account both effects, a formula for the survival probability was proposed in our previous paper~\cite{MTM2024}.
Here, we study explicit expressions for both effects and examine how the formula works for various conveyance protocols.

\subsection{Model of conveyance}

The conveyance protocol is characterized by the time-dependent acceleration
$a(t)\,(\equiv d^2x_0(t)/dt^2)$, and we assume that the motion of the potential starts
and ends smoothly; that is, $v(0) = v(\tau) = 0$, where $v(t) \equiv dx_0(t)/dt$. 
Thus, the conveyance distance $L$ is expressed as
\begin{equation}
L = \int_0^{\tau} dt' \int_0^{t'} a(s) ds.
\label{eqn:length}
\end{equation}

For the trapping potential well, we adopt the following form,
\begin{equation}
V_\text{well}(x,t)=z\left[
\tanh^2 \left( \frac{x-x_0(t)}{w} \right) - 1
\right],
\label{potential-well}
\end{equation}
as in our previous paper~\cite{Conveyance0,MTM2024}.
The number of bound states depends on the parameters~\cite{text:Landau-Lifshitz}.
In the simulations below, we set $\hbar = m= w=z=1$. For these parameter choices,
one bound state exists.

The time-dependent Hamiltonian of the system is given by
\begin{equation}
{\cal H}(t) = \frac{p^2}{2m} + V_\text{well}(x,t),
\end{equation}
and the time evolution of the wave function is governed by the \SDG equation in the rest frame:
\begin{equation}
i \hbar \pdv{t} \ket{\Psi(t)}
={\cal H}(t) \ket{\Psi(t)}.
\label{eq:Psi-t}
\end{equation}

\subsubsection{Moving frame representation of conveyance}

If we move to a reference frame that follows the motion of $x_0(t)$, the \SDG equation is transformed as follows.
Let us consider two unitary transformations,
\begin{gather}
U_1(t) = \exp\left(-\frac{i x_0(t) p}{\hbar}\right),\\
U_2(t) = \exp\left(\frac{i m v(t) x}{\hbar}\right),
\end{gather}
which give $U_1^{\dagger}(t)\, x\, U_1(t) = x + x_0(t)$ and $U_2^{\dagger}(t)\, p\, U_2(t) = p + mv(t)$.
Successive application of these unitary transformations gives the Hamiltonian and the wave function in the moving frame as
\begin{equation}
\begin{aligned}
\mathcal{H}_\text{MF}(t) &\equiv U_2^{\dagger}(t)U_1^{\dagger}(t){\cal H}U_1(t)U_2(t)\\
&= \frac{p^2}{2m} + V_\text{well}(x, 0) + ma(t) x - \frac{mv(t)^2}{2}.
\end{aligned}
\label{eq:H-MF}
\end{equation}
\begin{equation}
\ket{\Phi(t)} \equiv U_2^{\dagger}(t) U_1^{\dagger}(t) \ket{\Psi(t)}.
\label{eq:Phi-t}
\end{equation}
This wave function satisfies the \SDG equation in the moving frame:
\begin{equation}
i\hbar \pdv{t}\ket{\Phi(t)} = \mathcal{H}_\text{MF}(t) \ket{\Phi(t)}.
\label{SDG-eq}
\end{equation}
As well known, in the case of constant acceleration, $x_0(t)=at^2/2$, ${\cal H}_{\rm MF}$ becomes time-independent, with a tilted potential $amx$.

We define the survival probability as the overlap between the wave function at time $t$ and the initial state in the moving frame,
\begin{equation}
p(t) \equiv \left|
\langle \Phi(0)|\Phi(t)\rangle
\right|^2.
\label{phi0phit}
\end{equation}
It is rewritten in the original frame as
\begin{equation}
p(t) =
\left|
\langle \Psi(0)| U_2^{\dagger}(t) U_1^{\dagger}(t) |\Psi(t) \rangle
\right|^2,
\label{defpt-move}
\end{equation}
where we use the fact that $x_0(0)=0$ and $v(0)=0$.

\subsection{Adiabatic tunneling during acceleration}
\begin{figure}[t]
$$
\begin{array}{cc}
\includegraphics[scale=0.65]{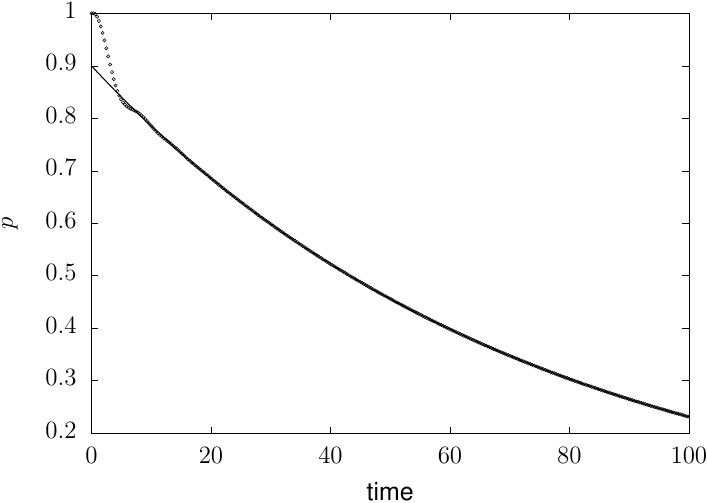} &
\includegraphics[scale=0.65]{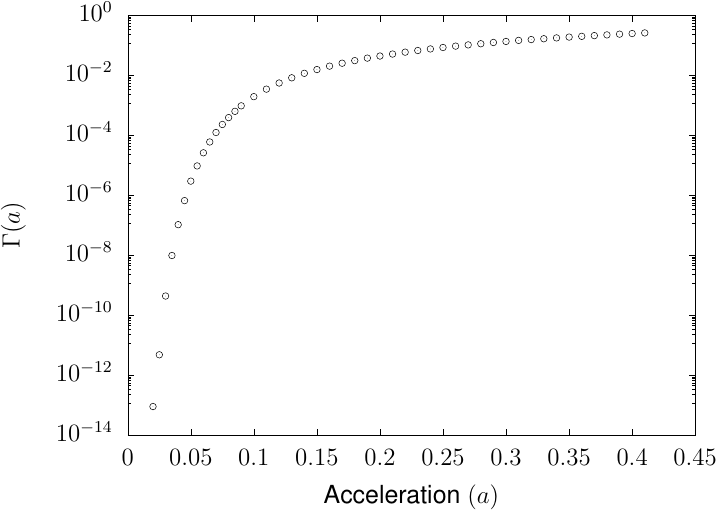}\\
({\rm a}) & ({\rm b})
\end{array}
$$ 
\caption{(a) The survival probability $p(t)$ (open circles)
for constant acceleration $a=0.145$, and an exponential fit (solid line). 
Values of parameters are $m=w=z=1$ with $\hbar=1$. 
The fitted curve is given by $p(t) = A \exp (-\Gamma t)$ with $A=0.899$ and $\Gamma = 0.0136$.
(b) Dependence of the decay rate $\Gamma(a)$ on a constant acceleration $a$.
}
\label{fig:p-t}
\end{figure}

When the trapping potential well is accelerated with a constant acceleration rate $x_0(t)=at^2/2$,
the survival probability decays exponentially at a late stage due to the tunneling effect, as depicted in Fig.~\ref{fig:p-t}(a). 
This exponential decay is not caused by a dissipative mechanism but is an intrinsic quantum mechanical process. 
The mechanism of this decay has been studied in detail from the viewpoint of the energy level structure for a metastable-type potential, and also from the viewpoint of resonance states,
in our previous paper~\cite{MTM2024}.

By fitting the exponential decay at large $t$, we obtain the decay rate $\Gamma(a)$ as a function of the constant acceleration $a$.
This decay rate corresponds to the imaginary part of the complex eigenenergy $E_{\rm res}(a)$ of the resonance state for a given acceleration $a$~\cite{Hatano-nonhermite}, due to the potential
  barrier in the acceleraation frame,
\beq
\Gamma(a) = -2 \text{Im} \left[E_{\rm res}(a)\right].
\eeq
The dependence of $\Gamma(a)$ on $a$ is plotted in Fig.~\ref{fig:p-t}(b), and is found to be a smooth function.
In the following analysis, we use interpolation from a few neighboring points to obtain $\Gamma(a)$ for a given value of $a$.

A general conveyance protocol is given by a time-dependent acceleration profile $a(t)$ spanning the duration $0\le t \le \tau$. Depending on the change in $a(t)$, the state exhibits nonadiabatic changes, which correspond to a decay of the trapping population. 
During conveyance ($0< t <\tau$), in which the acceleration rate $a(t)$
changes slowly 
compared to a typical time scale of the system, such as the oscillation period in the trapping potential determined by the energy gaps, 
the relaxation of the survival probability in a short time is approximately described by
the exponential decay with an instantaneous relaxation rate $\Gamma(a(t))$.
Using $\Gamma(a)$ obtained in Fig.~\ref{fig:p-t}(b),
the relaxation due to this mechanism over the process is given by
\beq
p_{\rm AT}(t) \equiv e^{-\int_0^{t} \Gamma(a(t')) dt'},
\label{eq:p_AT}
\eeq
for any process of $a(t)$.

This formula corresponds to an adiabatic change of the resonance state with complex eigenenergy $E_{\rm res}(a)$.
When the change in $a(t)$ is not fast, we expect that $E_{\rm res}(a)$ changes continuously.
This picture is called the ``adiabatic tunneling picture,'' originally introduced for atomic ionization phenomena induced by intense laser fields in the low-frequency limit~\cite{Keldysh1965, PPT1966, ADK1986}.
This separation of mechanisms provides a clear physical interpretation of survival decay in quantum conveyance.

\begin{figure}[t]
\centering
\includegraphics[scale=0.65]{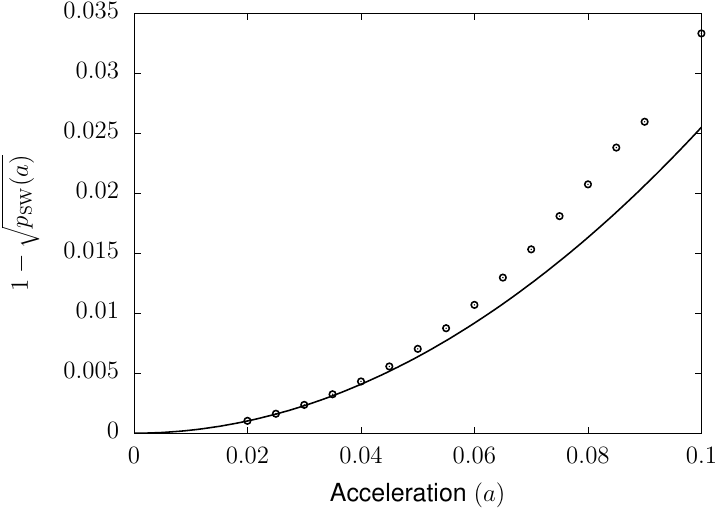}
\caption{Dependence of $1-\sqrt{p_{\rm SW}}$ on $a$ (open circles) and a fitting
$1-\sqrt{p_{\rm SW}} = 2.55 a^2$ (line). }
\label{fig:A-a}
\end{figure}

\subsection{Nonadiabatic change at switching points}

\subsubsection{Survival probability in large $\tau$ region}
\label{sec_large_tau}

When $\tau$ becomes large for a fixed $L$ in a conveyance, the magnitude of $a(t)$ decreases,
and consequently $\Gamma(a)$ also becomes small.
This leads to almost no adiabatic tunneling during the conveyance $(0<t<\tau)$, meaning
$p_\text{AT}(\tau)=e^{-\int_0^{\tau} \Gamma(a(t')) dt'}\simeq 1$.
In this limit, the survival probability is mainly determined by the switching
disturbance, which come from sudden change at the switching point.
We express this contribution to the survival probability in the form:
\beq
p(\tau) \simeq p_{\rm SW}= \left(1-d_\text{ini}[a(\cdot)]\right)
\left(1-d_\text{fin}[a(\cdot)]\right).
\label{fig:pM}
\eeq
Here, $d_\text{ini}[a(\cdot)]$ ($d_\text{fin}[a(\cdot)]$) denotes the amount of the initial (final)
drop-off depending on the function $a(t)$, which will be studied in the following section.
We use the notation $d_{\rm ini}[a(\cdot)]$ to indicate that, in general situations, the value of $d_{\rm ini}$ depends on the functional form of $a(t)$, but not on its specific magnitude.
As is indicated by Eq.~\eqref{P-Bn-alpha}, $d_{\rm ini}$ is found not to depend on the detailed form of the protocol but only on the initial power of $a(t)\simeq a_n t^n$, and it can be obtained as a function of the power $n$.

Let us consider a sudden change of the acceleration rate at $t=0$.
In the moving frame, there is no true bound state because of the inertial
potential ($ma(t)x$ in Eq.~\eqref{eq:H-MF}). Instead, there exists a group of eigenstates around
the bound state of the initial trapping potential,
which corresponds to the resonance state $|\phi_{\rm res}(a(0))\rangle$.
Thus, the initial state should be expressed as a linear combination of $|\phi_{\rm res}(a(0))\rangle$ and other states,
\begin{equation}
|\Phi(0)\rangle =
c(a(0))|\phi_{\rm res}(a(0))\rangle + |\phi_{\rm others}\rangle,
\end{equation}
with
\begin{equation}
c(a(0))=\langle\phi_{\rm res}(a(0))|\Phi(0)\rangle.
\end{equation}

The part of the wave function given by $|\phi_{\rm others}\rangle$
has higher energy, and the wave function exhibits rapid delocalization and decays
much faster than that of $|\phi_{\rm res}(a(0))\rangle$ due to rapid dephasing.
Thus, after such rapid dephasing, the wave function reduces to the resonance state as
\begin{equation}
|\Phi(t)\rangle \simeq c(a(0))|\phi_{\rm res}(a(0))\rangle,
\end{equation}
which subsequently undergoes exponential decay with the rate $\Gamma(a)$.

We expect that the system then follows adiabatic time evolution due to the time dependence of $a(t)$ unless $a(t)$ changes very rapidly, and the wave function evolves as
\begin{equation}
|\Phi(t)\rangle \simeq
c(a(0)) e^{-\int_0^t\Gamma(a(t'))/2 dt'}
e^{-i\int_0^t\varepsilon(a(t')) dt'} 
|\phi_{\rm res}(a(t))\rangle,
\end{equation}
where $\varepsilon(a)$ and $\Gamma(a)$ are defined by the complex eigenenergy as $E_{\rm res}(a) = \varepsilon(a) -i\Gamma(a)/2$.

The survival probability as defined in Eq.~(\ref{phi0phit}) at time $t$ is then
\begin{equation}
p(t) \simeq |c(a(0)) c(a(t))|^2 e^{-\int_0^{t}\Gamma(a(t')) dt'}.
\end{equation}
Therefore, $p_{\rm SW}$ is given in the form:
\begin{equation}
p_{\rm SW} = |c(a(0)) c(a(t))|^2 , 
\end{equation}
where
\begin{gather}
  d_{\rm ini}[a(\cdot)] = 1-\left|c(a(0))\right|^2,
  \label{eqn:dini}\\
  d_{\rm fin}[a(\cdot)] = 1-\left|c(a(t))\right|^2.
  \label{eqn:dfin}
\end{gather}
As we shown above, switching disturbance due to sudden change of acceleration
can be generally written by Eqs.~\eqref{eqn:dini} and~\eqref{eqn:dfin}.
Now, we estimate the factor $p_{\rm SW}$ in the case of constant acceleration $a(t)=a_0$.
Adiabatic tunneling occurs from the metastable trapped state to outside of the potential under constant acceleration, and the decay rate is $\Gamma (a_0)$, which can be evaluated for each value of $a_0$ as shown in Fig.~\ref{fig:p-t}(b).

In this case, the survival probability is given by
$p_\text{AT}(\tau)=e^{-\int_0^{\tau} \Gamma(a(t')) dt'}\simeq e^{-\Gamma(a_0)t}$ together with $p_{\rm SW}$:
\begin{equation}
p(t)\simeq p_{\rm SW}(a_0) e^{-\Gamma(a_0)t}.
\end{equation}
Here, we note that $p_{\rm SW}(a_0) = (1-d_{\rm ini})^2$, since $d_{\rm ini} = d_{\rm fin}$ in the constant acceleration case.
At small accelerations, $d_{\rm ini} = 1-\sqrt{p_{\rm SW}(a_0)}$ is proportional to $a_0^2$, which is consistent with the exact analysis for the harmonic potential shown in Appendix~\ref{sec_harmonic}.
Therefore, by fitting the dependence in Fig.~\ref{fig:A-a}
in a quadratic form of $a_0$, we estimate the coefficient as
\begin{equation}
\sqrt{p_{\rm SW}(a_0)} = |c(a_0)|^2 \simeq 1-2.55 a_0^2.
\label{Eq30}
\end{equation}
This factor will be analytically reexamined later.
Therefore, for a general time-dependent protocol $a(t)$,
the survival probability under the adiabatic tunneling picture with corrections due to initial and final disturbances is given by
\begin{equation}
p_\text{SW-AT}(t) = (1-2.55 a^2(0))(1-2.55a^2(t)) p_\text{AT}(t).
\label{eq:p-adia-corrected}
\end{equation}
The coefficient $2.55$ obtained by the fitting depends only on the mass and the shape of the trapping potential.
In the case of $a_0(0)\ne 0$ as the cos-protocol defined later in Eq.~(\ref{a-cos}),
we find $d_\text{ini/fin}[a(\cdot)] = 2.55a^2$.

Although this formula implies that $d_{\rm ini/fin}[a(\cdot)] = 0$ and $p(\tau ) = 1$, when the acceleration exhibits no discontinuity
at the initial or final time, as in the sine-protocol defined later in Eq.~(\ref{a-sin}). However, discontinuity at higher order
also contributes with a higher order dependence on the period $\tau$.
Let us now assume that the acceleration schedule $a(t)$ is given by an analytic function
in the range $0 \leq t \leq \tau$, which can be expressed through its Taylor expansion,
\beq
a(t) = \sum_n \frac{1}{n!}a^{(n)}(0)t^n.
\eeq

Since the acceleration is zero before the initial time, its derivatives approaching $t=0$ from the left are all zero.
This implies that some nonzero $a^{(n)}(0)$ exhibits a discontinuity at $t=0$, giving rise to nonadiabatic transitions.

As pointed out in~\cite{Morita2007}, when $a(t)$ begins as $a^{(n)}(0)t^n/n!$ near $t=0$,
the nonadiabatic transition probability is proportional to the square of the jump in the $n$-th derivative:
\begin{equation}
d_\text{ini}[a(\cdot)] \propto |a^{(n)}(0)|^2 .
\end{equation}

The same effect also occurs at the final time by replacing $t^n$ with $(\tau-t)^n$.

Considering the scaling with $\tau$, we have
\begin{equation}
a^{(n)}(0)\propto L\tau^{-n-2},
\label{a(n)-tau}
\end{equation}
which implies
\begin{equation}
1-p(\tau) \propto L^2\tau^{-2n-4}.
\end{equation}
This result provides a systematic classification of switching-induced nonadiabatic effects in terms of the smoothness of the acceleration protocol, clearly demonstrating that the survival probability is governed by the lowest-order discontinuity in the time derivatives of $a(t)$.

\section{Survival probability in several conveyance protocols} \label{sec_3}
Together with both contributions, 
the survival probability $p(t)$ is expected to be given by the product of
$p_\text{AT}(\tau)$ and $p_{\rm SW}$, i.e.,
\beq
p(t) \simeq p_{\rm SW-AT}(t) \equiv p_{\rm SW} p_{\rm AT}(t),
\label{p-adia}
\eeq
which provides a good approximation for the survival probability.
In this section, we investigate how the formula (\ref{p-adia}) works.

\subsection{Comparison of the Adiabatic Tunneling Approximation with direct
  numerical simulations} 

To compare the survival probability $p(t)$ with the integral form of adiabatic tunneling $p_\text{AT}(t)$,
we perform numerical simulations for the following protocols studied in our previous paper~\cite{MTM2024}.
\begin{itemize}
\item The cos-protocol:
\begin{equation}
a(t)=c_{\rm cos}\cos(\omega t), \quad
x_0(t)={c_{\rm cos}\over\omega^2}\left[1-\cos(\omega t)\right],
\label{a-cos}
\end{equation}
\item The sin-protocol:
\begin{equation}
a(t)=c_{\rm sin}\sin(2\omega t), \quad
x_0(t)=\frac{c_{\rm sin}}{2\omega}
\left[
t - \frac{\sin(2\omega t)}{2\omega}
\right],
\label{a-sin}
\end{equation}
\end{itemize}
with $\omega = \pi/\tau$.
The condition $x_0(\tau)=L$ determines the coefficients as $c_\text{cos}=\pi^2 L/2\tau^2$ and $c_\text{sin}=2\pi L/\tau^2$.

\begin{figure}[t]
$$
\begin{array}{cc}
\includegraphics[scale=0.65]{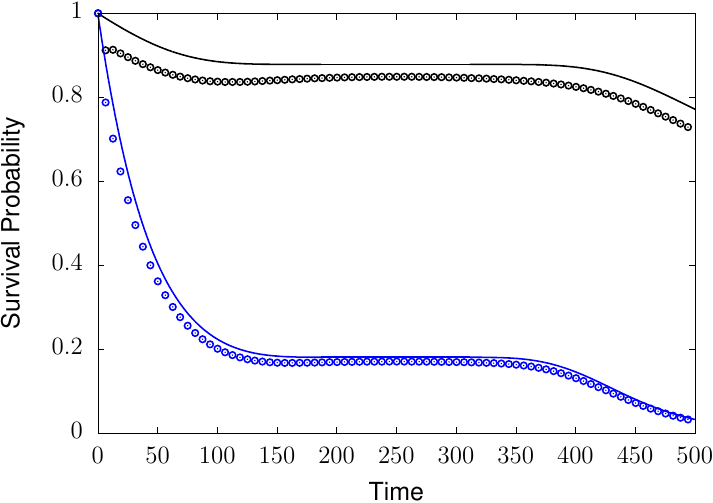} &
\includegraphics[scale=0.65]{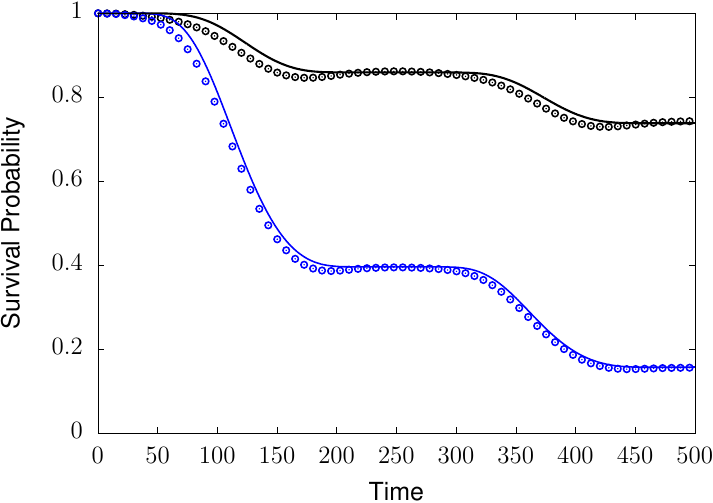} \\
({\rm a}) & ({\rm b}) \\
\end{array}
$$
\caption{Comparison of the numerically obtained $p(t)$ (open circles)
and $p_{\rm AT}(t)$ (lines) for the cases of
$\tau=500$. (a) the cos-protocol with $L=5000$ (black) and $L=8000$ (blue),
and (b) the sin-protocol with $L=8000$ (black) and $L=11000$ (blue).
}
\label{fig:prob-t-adia}
\end{figure}

In Fig.~\ref{fig:prob-t-adia}(a), the survival probability $p(t)$ for the cos-protocol with $\tau = 500$
and $L = 5000$ (black) and $L = 8000$ (blue) is plotted (lines), together with the values estimated from
$p_{\text{AT}}(t)$ in Eq.~(\ref{eq:p_AT}) (open circles).
A similar comparison for the sin-protocol is shown in Fig.~\ref{fig:prob-t-adia}(b) for $L = 8000$ (black) and $L = 11000$ (blue).

Here, we find qualitative agreement.
In the cos-protocol with $L=5000$, there is a significant deviation,
but the deviation decreases as $L$ increases.
In the sin-protocol, we observe the same tendency with smaller deviations.
The adiabatic tunneling probability $p_{\rm AT}(t)$ and the numerical data
show good agreement in both cases.
Thus, we conclude that the decay during the variation of $a(t)$
is reasonably well described by Eq.~(\ref{p-adia}).

However, deviations between $p(t)$ and $p_{\rm AT}(t)$ remain, which can be attributed to switching disturbances due to nonadiabatic changes, represented by $p_{\rm SW}(t)$, at the initial and final points.

\subsection{Comparison of the formula (\ref{p-adia}) with the direct numerical simulations} 

Now, we include $p_{\rm SW}(t)$ and compare the estimation based on the formula (\ref{p-adia}) with the results obtained by direct numerical simulations for time-dependent $a(t)$.

\begin{figure}[t]
$$
\begin{array}{cc}
\includegraphics[scale=0.65]{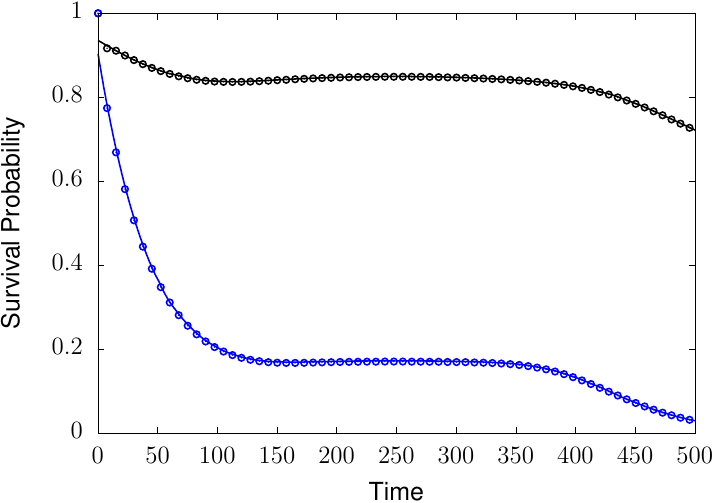} &
\includegraphics[scale=0.65]{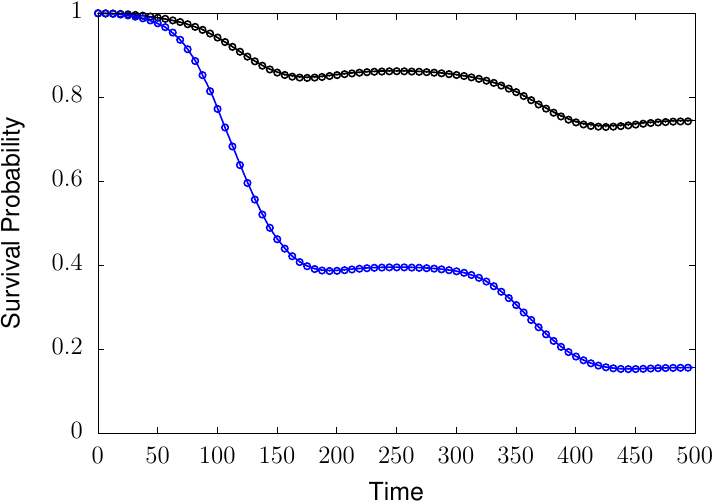}\\
({\rm a}) & ({\rm b}) \\
\end{array}
$$
\caption{Comparison of numerically obtained survival probabilities
(open circles) for $\tau=500$.
(a) cos-protocol compared with
Eq.~(\ref{eq:p-adia-corrected}) (lines)
for $L=5000$ (black) and $L=8000$ (blue). (b) sin-protocol
compared with Eq.~(\ref{eq:p-sin-corrected}) (lines)
for $L=5000$ (black) and $L=8000$ (blue).}
\label{fig:prob-t-cossin-adjust1}
\end{figure}

First, we study the case of the cos-protocol given by Eq.~(\ref{a-cos}).
Because $(1-2.55a^2(t))$ is not equal to 1 except at $t=\tau/2$, where $a(\tau/2)=0$, the adiabatic tunneling probability without the factor $p_{\rm SW}$ shows a finite deviation during the conveyance, as seen in Fig.~\ref{fig:prob-t-adia}(a).
In Fig.~\ref{fig:prob-t-cossin-adjust1}(a), we compare the numerically obtained survival probability $p(t)$
with the adiabatic tunneling probability corrected by the switching disturbance factor given by
Eq.~(\ref{eq:p-adia-corrected}).
It should be noted that $p_\text{SW-AT}(0)\neq1$ because it includes the switching disturbance at $t=0$.
Clearly, the correction factor allows us to reproduce the numerical results very well.

Next, we study the case of the sin-protocol defined by Eq.~(\ref{a-sin}). 
In the sin-protocol, the acceleration at the initial time is zero, thus $1-2.55a^2(0)=1$. 
Therefore, Eq.~(\ref{eq:p-adia-corrected}) becomes
\beq
p(t)\simeq (1-2.55a^2(t))e^{-\int_0^t\Gamma(a(t'))dt'},
\label{eq:p-sin-corrected}
\eeq
which agrees very well with the numerical results, as shown in Fig.~\ref{fig:prob-t-cossin-adjust1}(b). Within this frame $1-2.55 a^2(\tau) = 1$.
However, as shown above in Eq.~\eqref{a(n)-tau}, higher order correction remains,
which will be discussed in the following section.

Lastly, we study the case of a shifted sin-protocol, where the acceleration rate is
expressed as
\beq
a(t)=c_{\rm sin}\sin(2\omega t+\phi).
\label{eqn:shifted-sin}
\eeq
\begin{figure}[t]
\centering
\includegraphics[scale=0.65]{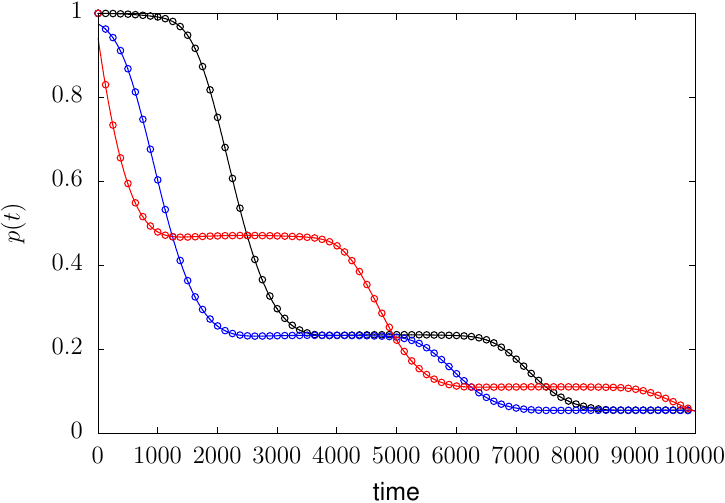}
\caption{Survival probability for the processes of
$a(t)=c_{\rm sin}\sin(2\omega t+\phi)$ (lines)
compared with $p_{\rm SW-AT}(t)$ in Eq.~(\ref{eq:p-adia-corrected}) (open circles) for phase shifts of $\phi=0$ (black),
$\phi=\pi/4$ (blue), and $\phi=\pi/2$ (red). All cases use the parameters $\tau=500$ and $c_{\rm sin}=0.91$.}
\label{fig:pfitphi-rev}
\end{figure}
In this general case, $p_\text{SW-AT}$ is still given in the form of Eq.~(\ref{eq:p-adia-corrected}).
In Fig.~\ref{fig:pfitphi-rev}, $p_\text{SW-AT}$ for $\phi=\pi/2$ and $\phi=\pi/4$ is plotted together with the numerical simulation results.
Again, we find very good agreement.

\subsection{Asymptotic Behavior in the large-$\tau$ limit}
\begin{figure}[t]
\centering
\includegraphics[scale=0.65]{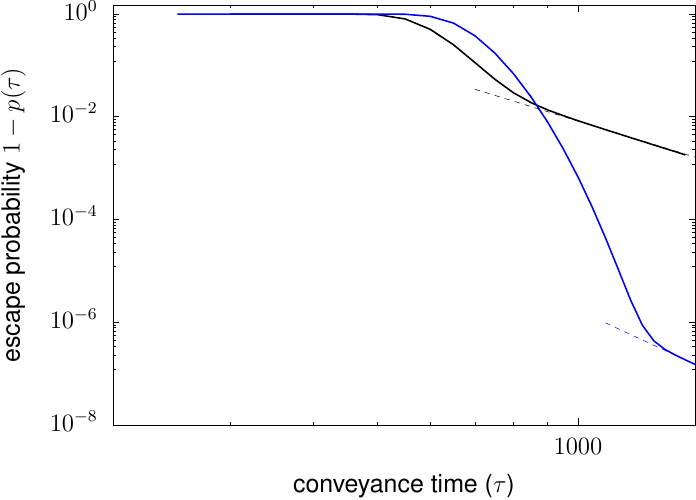}
\caption{Escape probability as a function of conveyance time $\tau$ (solid lines) 
for $L=8000$ with
asymptotic behavior in Eq.~(\ref{eqn:asym}) (dashed lines)
for the cos- (black) and sin- (blue) protocols.
}
\label{fig:cossin}
\end{figure}

As we have shown above, the escape probabilities of the cos- and sin-protocols should behave
as $1-p(\tau) \propto L^2\tau^{-4}$ and $L^2\tau^{-6}$, respectively, because of the
nonzero lowest-order terms in $a^{(0)}(0)$ and $a^{(1)}(0)$ for these protocols.
Figure~\ref{fig:cossin} shows the escape probabilities $1-p(\tau)$ for $L=8000$ as
functions of $\tau$ for the cos- and sin-protocols obtained by numerical calculation.
Clearly, both protocols exhibit the expected power-law decay for large $\tau$.

We find that the escape probabilities for large $\tau$ are well fitted by 
\beq
1-p_{\rm cos}(\tau)\simeq j_{\rm cos}\tau^{-4},\quad
1-p_{\rm sin}(\tau)\simeq j_{\rm sin}\tau^{-6},
\label{eqn:asym}
\eeq
with $j_{\rm cos}=8.24\times 10^{9}$ and $j_{\rm sin}=1.71\times 10^{12}$.
For the cos- and sin-protocols, we find $d_{\rm ini} = d_{\rm fin}$,
resulting in
$d_{\rm ini} \simeq p_{\rm escape}/2$.
The cos-protocol is defined by $a(t)=c_{\rm cos}\cos(\omega t)$ with $\omega={\pi/\tau}$, where
$a^{(0)}(0)= c_{\rm cos} = {\pi^2 L/ 2\tau^2}$. 
Using the factor $j_{\rm cos}$ estimated from the large-$\tau$ behavior,
we estimate $d_{\rm ini}$ as
\beq
d_{\rm ini}[a(\cdot)] \simeq
{2j_{\rm cos}\over \pi^4L^2} \left(a^{(0)}(0) \right)^2
\simeq 2.64 \left(a^{(0)}(0)\right)^2.
\label{eqn:d_in-cos}
\eeq
This coefficient $2.64$ agrees closely with the 2.55 in Eq.~(\ref{Eq30}),
obtained by fitting the constant-acceleration cases.
This agreement verifies the present analysis for $n=0$.

Similarly, for the case where the sin-protocol is defined by
$a(t)=c_{\rm sin}\sin(2\omega t)$ with $\omega={\pi/\tau}$,
we estimate the factor $d_{\rm ini} (= d_{\rm fin})$.
For this protocol, $a^{(1)}(0)= 2 c_{\rm sin} \omega =4\pi^2 L / \tau^3$.
Thus, with the numerically estimated value of $j_{\rm sin}=1.71\times 10^{12}$, we have
\beq
d_{\rm ini}[a(\cdot)] \simeq {j_{\rm sin}\over 32\pi^4L^2}
\left(a^{(1)}(0)\right)^2\simeq 8.61 \left(a^{(1)}(0)\right)^2,
\label{eqn:d_in-sin}
\eeq
which is proportional to $\tau^{-6}$, while Eq.~(\ref{eqn:d_in-cos}) is proportional to $\tau^{-4}$ for large $\tau$.

\begin{figure}[t]
\centering
\includegraphics[scale=0.65]{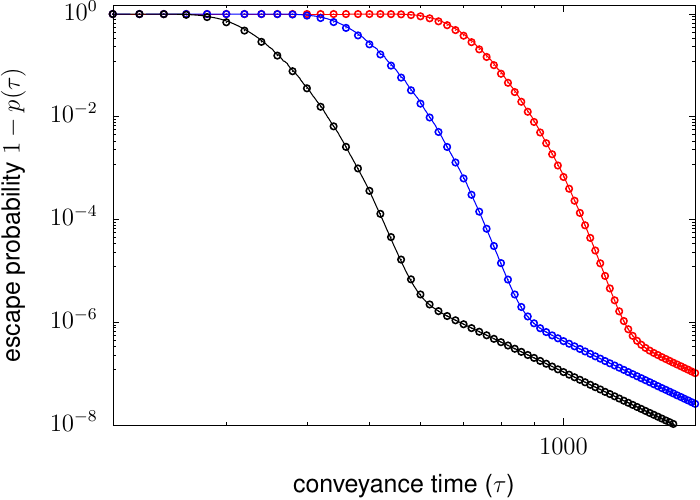}
\caption{Escape probabilities (open circles) for the sin-protocol as functions of
conveyance time $\tau$ compared with Eq.~(\ref{escape-fit-sin}) shown with lines,
for $L=2000$ (black), $L=4000$ (blue), $L=8000$ (red). 
}
\label{fig:escape-rate-sin}
\end{figure}

Taking into account this higher-order switching disturbance effect, the survival probability for the sin-protocol is refined to
\begin{equation}
p_\text{SW-AT}(\tau)=\left[1-8.61(a^{(1)}(0))^2\right]^2 p_\text{AT}(\tau).
\label{escape-fit-sin}
\end{equation}

Figure~\ref{fig:escape-rate-sin} shows numerically obtained escape probabilities
$1-p(t)$
for the sin-protocol with various $L$ compared with those for the theoretical prediction
given by Eq.~(\ref{escape-fit-sin}), exhibiting accurate agreement
not only in the large-$\tau$ limit but also in the small-$\tau$ region where adiabatic tunneling dominates.
Although Eq.~(\ref{escape-fit-sin}) is derived by fitting the numerical results for the sin-protocol with $L=8000$,
this formula can be applied to any protocol with $a^{(0)}(0)=a^{(0)}(\tau)=0$ and $a^{(1)}(0)=a^{(1)}(\tau)\neq0$.
\begin{figure}[t]
% \centering
$$
\begin{array}{cc}
\includegraphics[scale=0.65]{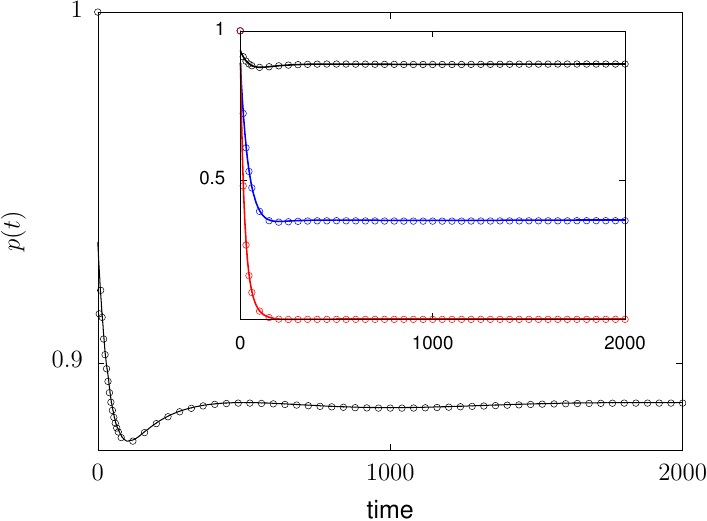} &
\includegraphics[scale=0.65]{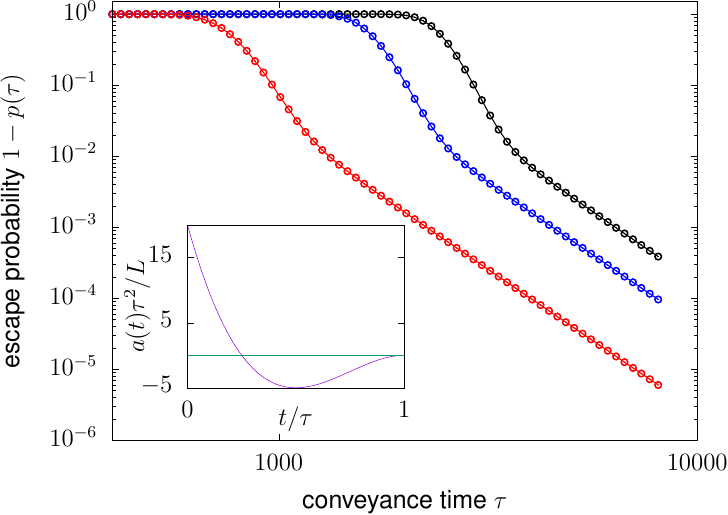} \\
({\rm a}) &({\rm b}) 
\end{array} 
$$
\caption{Comparison of the numerical results (circles)
and our analytical formula $p_{\rm SW-AT}(t)$ (lines) for the 3rd-order polynomial protocol.
(a) Instantaneous survival
probability $p(t)$ as a function of time with $\tau=2000$ for the case of 
$L=20000$. The inset includes other cases for comparison,
$L=30000$ (blue), and $L=40000$ (red).
(b) Escape probabilities as functions of conveyance time $\tau$
for the cases of
$L=5000$ (black), $L=20000$ (blue), and $L=40000$ (red), with the scaled
acceleration profile $a(t)\tau^2/L$ as a function of scaled time $t/\tau$
for the 3rd-order protocol shown in the inset.
}
\label{fig:3rd}
\end{figure}

In order to demonstrate the applicability of our theory,
we consider a conveyance protocol defined by the third-order polynomial:
\begin{equation}
a(t) = \frac{80 L}{\tau^2} \left( 1-\frac{t}{\tau}\right)^2 \left(\frac{1}{4} - \frac{t}{\tau} \right),
\label{eq:a-3rd}
\end{equation}
where the prefactor is determined to satisfy Eq.~(\ref{eqn:length}).
This protocol is shown in the inset of Fig.~\ref{fig:3rd}(b).
According to the above discussion, the conveyance probability is expected to be
well predicted by Eq.~(\ref{p-adia}). In this protocol, the acceleration
at the initial time is nonzero, indicating
that $d_{\rm ini}$ is given by Eq.~(\ref{eqn:d_in-cos}).
Therefore, the survival probability $p(t)$ is well approximated by
\begin{equation}
\begin{aligned}
p_\text{SW-AT}(t) &\simeq \left[1-2.64 (a^{(0)}(0))^2 \right]
\left[1-2.64(a^{(0)}(t))^2 \right] \\
&\quad \times p_\text{AT}(t).
\end{aligned}
\label{eq:P-3rd}
\end{equation}
Figure~\ref{fig:3rd}(a) compares this formula (lines) with the numerical results
(open circles) for $L=20000$ with $\tau=2000$, demonstrating excellent agreement.
The inset also shows good agreement in other cases with $L=30000$ and $L=40000$.
In all these cases, the survival probabilities show a sudden decrease near the initial time.
This drop originates from both the switching disturbances (initial and instantaneous) and rapid tunneling driven by the large acceleration near the initial time.
The subsequent small oscillatory behavior is attributed to the time dependence of the
instantaneous switching disturbance.
For $L=20000$ (the black line in Fig.~\ref{fig:3rd}(a)), the survival probability begins
to recover; this is due to the suppression of the instantaneous disturbance caused
by the decrease in acceleration. The following small oscillation
is also mediated by the instantaneous disturbance.
The local maximum and minimum at $t\simeq400$ and $t\simeq1000$
are attributed to the vanishing points and local extrema of the acceleration profile.
At the final time, the acceleration and its first derivative are zero,
whereas the second derivative is nonzero, giving rise to nonzero switching effect
$d_{\rm fin}$. However, this final-time switching disturbance is significantly smaller than
the initial one, making it almost invisible in Fig.~\ref{fig:3rd}(a).
Figure~\ref{fig:3rd}(b) shows the escape probabilities at the final time as functions of
conveyance time $\tau$ for various conveyance distances, which agree well with our formula Eq.~(\ref{eq:P-3rd}). A power-law
dependence is observed in the large-$\tau$ region, which is attributed to the switching
disturbance at the initial time (the first bracket in Eq.~(\ref{eq:P-3rd})).

As mentioned above, for a given form of the potential well in Eq.~(\ref{potential-well}),
the survival probability for any protocol can be obtained once we have $\Gamma(a)$
and the prefactor $p_{\rm SW}$, which can be estimated regardless of the detailed form of $a(t)$.
For large $\tau$, $p_{\rm SW}$ is solely determined by the lowest-order
coefficient $a^{(n)}$ exhibiting a discontinuity, as its contribution decays most slowly with
increasing $\tau$ according to Eq.~(\ref{a(n)-tau}).
In general, the disturbance effect
associated with a given order $n$ can be represented as
\beq
d_{\rm ini/fin}[a(\cdot)] = \left( B_n a^{(n)} \right)^2,
\label{eqn:d_ini/fin}
\eeq 
and we have found $B_0^2 = 2.64$ and $B_1^2 = 8.61$ from fitting the numerical results.
As discussed above, the coefficient $B_n$ for arbitrary order $n$ can be determined
from $B_0$ and $B_1$ as
$B_n = B_0^{1-n} B_1^n$, allowing for a
general description of higher-order effects.

\begin{figure}[t]
\centering
\includegraphics[scale=0.65]{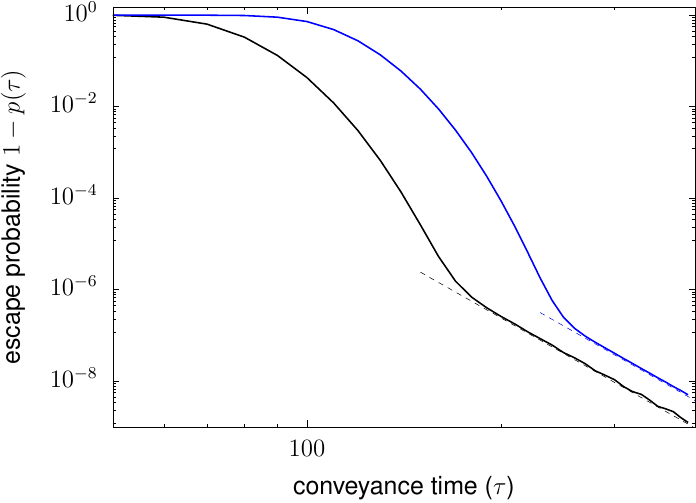}
\caption{Escape probabilities for the protocols defined by
Eq.~(\ref{eqn:5th-protocol}) with $L=125$ (black) and $L=250$ (blue)
as functions of conveyance time $\tau$. Dashed curves show the asymptotic
behavior given by Eq.~(\ref{eq:d_in-5th}).}
\label{fig:5th}
\end{figure}

To confirm the above idea, we study a conveyance protocol with higher-order terms defined by
\begin{equation}
a(t) =A_0 t^2 \left(\frac{\tau}{2} - t \right) (\tau - t)^2,
\label{eqn:5th-protocol}
\end{equation}
with $A_0 = 840 L\tau^{-7}$. This protocol exhibits discontinuities in its
second-order derivative of the acceleration at $t=0$ and $t=\tau$, with the magnitude 
of the discontinuous change being $a^{(2)}(0) = a^{(2)}(\tau) =A_0 \tau^3$.
This gives rise to the disturbance effect given by
\begin{equation}
d_{\rm ini/fin} \simeq 840^2 B_2^2 L^2 \tau^{-8},
\label{eq:d_in-5th} 
\end{equation}
with $B_2 = B_0^{-1}B_1^{2} \simeq 5.30$. The numerical escape probabilities for
$L = 125$ and $L=250$ as functions of the conveyance time $\tau$ are
shown in Fig.~\ref{fig:5th}, demonstrating excellent agreement with Eq.~(\ref{eq:d_in-5th}).

As mentioned above, in the large-$\tau$ limit, the lowest-order nonzero derivative
$\alpha^{(n)}(0)$ yields the dominant contribution.
When the trapping potential is modeled by
a harmonic function, which is a good approximation for a deep potential well,
no adiabatic tunneling occurs (i.e., $\Gamma(a)=0$). This allows us to isolate
and study the initial disturbance ($p_{\rm M}$) through exact analytical calculations
as given in Appendix~\ref{sec_harmonic}.
While the approximate form (Eq.~(\ref{eq:p-adia-corrected})) is remarkably simple,
it reproduces the actual dynamics exceptionally well.
This demonstration indicates that our approximation framework
can be utilized to predict
the survival probability for arbitrary
protocols without numerical simulations of the dynamics, once
$\Gamma(a_0)$ and $p_\text{SW}(a_0)$ for the constant acceleration case have been
determined.

\section{Summary and Discussion} \label{sec:summary}

To describe the real-time dynamics of wave functions in conveyance processes involving acceleration and deceleration, one generally needs to solve the time-dependent \SDG equation, which is computationally demanding in numerical studies.
Although disturbances at the switch-on and switch-off stages play an important role, their influence has not been thoroughly examined so far.
In the present paper we have investigated this switching effect $p_{\rm SW}$ in detail and clarified its analytical properties.
It is found that, together with the nonadiabatic change due to the tunneling effect in an accelerated system, the entire time dependence of the survival probability $p(t)$ is quantitatively well reproduced by the formula (\ref{p-adia}).

We have demonstrated the validity of the proposed formula through several examples.
First, we examined the cosine and sine protocols studied in our previous work and found excellent agreement between the survival probability obtained from numerical solutions of the time-dependent \SDG equation and $p_{\rm SW\text{-}AT}$.
We then applied the formula to the shifted sine protocol~(\ref{eqn:shifted-sin}), as well as to protocols defined by polynomial forms~(\ref{eq:a-3rd}) and~(\ref{eqn:5th-protocol}), and again confirmed its quantitative accuracy.
In particular, this factor is found to depend only on how the particle is accelerated at the switching points, for example, $a(t) \propto a_n t^n$ $(t \simeq 0)$, and is independent of the detailed form of the entire acceleration protocol $a(t)$.
We studied this effect analytically and derived an explicit formula, showing that the correction factors for $n \ge 2$ can be deduced from those for $n=0$ and $n=1$.
As long as the time dependence of $a(t)$ is slow compared with the characteristic time scale of the system, such as the vibrational period in the trapping potential, the contribution of adiabatic tunneling, denoted by $p_{\rm AT}(t)$, can be evaluated from the tunneling rate under constant acceleration, $\Gamma(a(t))$.
This rate can be determined solely from the constant-acceleration problem, independently of the detailed functional form of $a(t)$.
Therefore, once both factors are known, the survival probability for an arbitrary conveyance process can be accurately estimated by the formula (\ref{p-adia}),
$p_{\rm SW\text{-}AT} (= p_{\rm SW} p_{\rm AT})$,
without performing time-dependent simulations for each individual protocol.

Finally, we discuss the applicability of the present approach to more general situations.
Although we have focused on the specific trapping potential~(\ref{potential-well}), which was also employed in our previous work, the central idea of the present paper, namely that the survival probability $p(t)$ can be well described by the formula $p_{\rm SW\text{-}AT}$, is expected to be quite general.
For a given trapping potential, one first determines the switching factors $j_{\rm cos}$ ($n=0$) and $j_{\rm sin}$ ($n=1$), together with the tunneling rate $\Gamma(a)$ under constant acceleration.
Once these quantities are obtained, the survival probability for an arbitrary acceleration protocol $a(t)$ can be estimated in the form of $p_{\rm SW\text{-}AT}$ for that trapping potential.
It should be emphasized that the property that the switching factors for higher $n$ can be deduced from those for $n=0$ and $n=1$ holds irrespective of the detailed form of the trapping potential.
Therefore, for general trapping potentials, the present procedure is applicable: once the switching factors $d_{\rm ini/fin}$ for $n=0$ and $n=1$ and the transition rate $\Gamma(a)$ under constant acceleration are known, the survival probability is given by $p_{\rm SW\text{-}AT}(t)$.

In realistic experimental settings, the spatial dimension is three, and the situation becomes more complex.
Nevertheless, conveyance is essentially a motion along a prescribed path, and the core physics can be captured by an effective one-dimensional description.
While the present paper focuses on a single particle trapped by a potential, the conveyance of atomic clouds, such as Bose-Einstein condensates, is also an important and interesting problem.
We expect that the mechanism proposed in the present work is applicable to such systems as well.
In this paper, we have mainly considered relatively shallow trapping potentials (or light particles), where the number of bound energy levels is small.
For deep trapping potentials (or heavy particles), the number of energy levels increases, and intra-well dynamics, such as heating and excitation among bound states, become relevant.
The analysis of such regimes, including the continuous (classical) limit, will be addressed in future work.

This framework establishes a unified and predictive description of nonadiabatic effects in quantum conveyance processes.

\section*{Acknowledgments}
The present work was partially supported by JSPS KAKENHI Grant Number JP25K07146.
S. Morita is supported by the Center of Innovations for Sustainable Quantum AI
(JST Grant Number JPMJPF2221) and by JSPS KAKENHI Grant Number JP23H03818.
Y. Teranishi is supported by the National Science and Technology
Council of Taiwan (Grant Number 114-2112-M-A49-030).

\appendix

\section{Exact analysis in moving harmonic potential} \label{sec_harmonic}
In this appendix, we provide an exact analytical treatment of a particle conveyed by a moving harmonic potential. By obtaining the exact time-evolution operator through a sequence of unitary frame transformations, we rigorously verify the general formula for the switching disturbance discussed in the main text. This exactly solvable model explicitly confirms that the nonadiabatic transition probability at the switching points is determined by the lowest-order discontinuous derivative of the motion protocol.
\subsection{Problem Setting}
\label{sec:org5f2dfcc}
Here, we study the problem when the time-dependent potential \(V(x,t)\)
is given by a moving harmonic potential driven by a dipole interaction with a time-dependent
electric field \(E(t)\):
\begin{equation}
V(x,t) = \frac{m\omega^2}{2} (x-X(t))^2 + E(t)x.
\label{AppA1}
\end{equation}
This problem is equivalent to a moving harmonic potential problem without the linear potential term,
because 
\begin{equation}
  V(x,t) = \frac{m\omega^2}{2} \left(x-X(t) + \frac{E^2(t)}{m\omega^2}\right)^2
  -\frac{E^2(t)}{2m\omega^2} + X(t)E(t),
\end{equation}
and the last two terms in the right-hand side can be ignored, since it 
contributes only to the overall phase of the wavefunction. Therefore, we discuss 
the exact solution of the \SDG equation of moving potential problem, 
\begin{equation}
i\hbar\frac{\partial}{\partial t}\psi(x,t) =
\left(
\frac{\hat{p}^2}{2m} + \frac{m\omega^2}{2} (x-x_0(t))^2
\right)\psi(x,t),
\end{equation}
with
\begin{equation}
x_0(t) = X(t) - \frac{E(t)}{m\omega^2}.
\end{equation}

\subsection{Exact wavefunction for moving harmonic potential}
\label{sec:orgb40b54d}
We consider a unitary transform of the wavefunction defined by 
\begin{equation}
\phi_1(x,t) = \hat{U}[x_0(\cdot)]\psi(x,t),
\end{equation}
where $\hat{U}[x_0(\cdot)]$ is defined by 
\begin{equation}
\hat{U}[x_0(\cdot)] = e^{ -\frac{i}{\hbar}mx^{(1)}_0(t)\hat{x} } e^{ \frac{i}{\hbar}x_0(t)\hat{p} }.
\end{equation}
Here, \(x_0^{(n)}\) is the $n$-th order derivative of \(x_0(t)\) with respect to \(t\)
\beq
x_0^{(n)} \equiv \frac{d^n x_0(t)}{dt^n}. 
\eeq
The transformed wavefunction \(\phi_1(x,t)\)
satisfies the following \SDG equation:
\begin{equation}
i\hbar\frac{\partial}{\partial t}\phi_1(x,t) =
\left[
\frac{\hat{p}^2}{2m} + \frac{m\omega^2}{2} x^2 +mx_0^{(2)}x
\right]\phi_1(x,t).
\end{equation}
It should be noted that the potential terms independent of \(x\) or \(\hat{p}\) are ignored here,
since they only affect to the overall phase of wavefunction. The term \(mx_0^{(2)}x\)
is called inertial potential, which can be combined into the harmonic potential,
\begin{equation}
i\hbar\frac{\partial}{\partial t}\phi_1(x,t) =
\left[
\frac{\hat{p}^2}{2m} + \frac{m\omega^2}{2} \left(x + \frac{x_0^{(2)}(t)}{\omega^2}\right)^2 
\right]\phi_1(x,t).
\end{equation}
If \(x_0^{(2)}\) is time independent, this equation is equivalent to time-independent harmonic oscillator problem, for which exact solution is well known. 
If \(x_0^{(2)}\) is not 
constant, on the other hand, this equation is equivalent to the moving harmonic 
potential problem with the center of potential moving with \(-\frac{x_0^{(2)}(t)}{\omega^2}\).
We can apply another unitary transformation defined by 
\begin{equation}
\phi_2(x,t) = \hat{U}\left[ -\frac{ x_0^{(2)}(\cdot) }{\omega^2 }\right] \phi_1(x,t).
\label{AppA10}
\end{equation}
Then, the \SDG equation for \(\phi_2(x,t)\) is found to be
\begin{equation}
i\hbar\frac{\partial}{\partial t}\phi_2(x,t) =
\left[
\frac{\hat{p}^2}{2m} + \frac{m\omega^2}{2} \left(x - \frac{x_0^{(4)}(t)}{\omega^4}\right)^2 
\right]\phi_2(x,t).
\end{equation}
Applying similar procedures $n$ times successively, we obtain the transformed 
function \(\phi_n(x,t)\), which satisfies the \SDG equation
\begin{equation}
  \begin{aligned}
    &i\hbar\frac{\partial}{\partial t}\phi_n(x,t) \\
    & =
    \left[
      \frac{\hat{p}^2}{2m}
      + \frac{m\omega^2}{2} \left(x +\left(\frac{-1}{\omega^2}\right)^n x_0^{(2n)}(t)
    \right)^2
    \right]\phi_n(x,t).
  \end{aligned}
% i\hbar\frac{\partial}{\partial t}\phi_n(x,t) =
% \left[
% \frac{\hat{p}^2}{2m} + \frac{m\omega^2}{2} \left(x +\left (\frac{-1}{\omega^2}\right)^n x_0^{(2n)}(t)
% \right)
% \right]\phi_n(x,t).
\label{AppA12}
\end{equation}
If the motion \(x_0 (t)\) satisfies the condition:
\begin{equation}
\lim_{n\rightarrow \infty} \frac{x_0^{(2n)}}{\omega^{2n}}=0,
\end{equation}
as in cases where $x_0(t)$ is given by finite polynomials,
we can introduce the transformation of infinite successive transform,
\begin{align}
  \Phi(x,t)&=\left[ \Pi_{n=0}^{\infty}
  \hat{U}\left(\left(-\frac{1}{\omega^{2}}\right)^nx_0^{(2n)}(\cdot)\right)\right]\psi(x,t)
  \nonumber \\
  & \equiv\hat{U}_\infty[x_0(t)]\psi(x,t),
\end{align}
% $$
% \Phi(x,t)=\left[ \Pi_{n=0}^{\infty}
% \hat{U}\left(\left(-\frac{1}{\omega^{2}}\right)^nx_0^{(2n)}(\cdot)\right)\right]\psi(x,t)
% $$ \begin{equation}
% \equiv\hat{U}_\infty[x_0(t)]\psi(x,t),
% \end{equation} 
solution of which can easily be obtained.
The unitary transform with \(\hat{U}_\infty(x_0(t))\) can be obtained by
using the relation
\begin{equation}
  e^{\xi\hat{x}}e^{\eta\hat{p}} = e^{\eta\hat{p}}e^{\xi\hat{x}}e^{-i\hbar \xi\eta},
\end{equation}
and ignoring the overall phase. The result is 
\begin{equation}
\Phi(x,t) = \hat{U}_\infty[x_0(t)]\psi(x,t) = \hat{U}[\xi(t)]\psi(x,t),
\end{equation}
with
\begin{equation}
\xi(t) = \sum_{n=0}^\infty \left(\frac{-1}{\omega^2}\right)^n x_0^{(2n)}(t).
\label{xi-t-sum}
\end{equation}
For example, if \(\psi(x,t)\) is the ground state of harmonic potential:
\begin{equation}
\psi(x) = Ae^{-\frac{m\omega}{2\hbar}x^2},
\end{equation}
the transformed function $\Phi(x)$ is
\begin{equation}
\Phi(x) = Ae^{-\frac{i}{\hbar}m\xi^{(1)}x}e^{-\frac{m\omega}{2\hbar} (x-\xi(t))^2}.
\end{equation}
Thus, the overlap can be calculated by completing the square, namely
\begin{equation}
 \langle \psi|\Phi \rangle  =e^{-\frac{m\omega}{4\hbar}(\xi^2+\frac{\dot{\xi}^2}{\omega^2})}.
\end{equation}

In conclusion, the time propagation operator defined by 
\begin{equation}
\psi(x,t) = \hat{G}(t,t_0)\psi(x,t_0),
\end{equation}
is given by
\begin{equation}
\hat{G}(t,t_0) = \hat{U}^{-1}(\xi(t)) e^{-\frac{i}{\hbar}\hat{H}_0(t-t_0) }\hat{U}(\xi(t_0)).
\end{equation}
Using the relation
\begin{equation}
\hat{U}[\xi(t)]\psi(x,t) = e^{ -\frac{i}{\hbar}m\dot{\xi}(t)x }\psi(x+\xi(t),t),
\end{equation}
we find an explicit form of the transformation
\begin{equation}
\Phi(x,t) = \psi(x + \xi(t), t) e^{ -\frac{i}{\hbar} m\dot{\xi}(t)x },
\end{equation}
and the inverse transformation:
\begin{equation}
\psi(x,t) = \Phi(x-\xi(t))e^{\frac{i}{\hbar}m\dot{\xi}(t)(x-\xi(t))},
\end{equation}
which can be simplified by ignoring the global phase as 
\begin{equation}
\psi(x,t)=\Phi(x-\xi(t))e^{\frac{i}{\hbar}m\dot{\xi}(t)x}.
\end{equation}
Since \(\Phi(x,t)\) is the solution of the \SDG equation with harmonic 
potential, it is given by 
\begin{equation}
\Phi(x,t) = \sum_{j=0}^\infty c_j e^{ -\frac{i}{\hbar}\varepsilon_j t }\chi_j(x),
\end{equation}
where \(\varepsilon_k\) and \(\chi_j(x)\) are the eigenenergy and eigenstates of (non-moving) 
harmonic potential. 
Therefore, the general solution of the moving harmonic potential 
problem is given by 
\begin{equation}
\psi(x,t) = e^{\frac{i}{\hbar}m\dot{\xi}(t)x } \sum_{j=0}^\infty c_j e^{-\frac{i}{\hbar}\varepsilon_j t } \chi_j(x-\xi(t)),
\label{exact-sol}
\end{equation}
where the expansion coefficients \(c_j\)'s are determined by the initial condition:
\begin{equation}
\begin{split}
c_j &= \langle \chi_j|\hat{U}(\xi(t_0))|\psi \rangle\\
&= \int_{-\infty}^\infty e^{-\frac{i}{\hbar}m\dot{\xi}(t_0)x} \chi_j^{*}(x-\xi(t_0))\psi(x,t_0)dx\\
&= \int_{-\infty}^\infty e^{-\frac{i}{\hbar}m\dot{\xi}(t_0)x} \chi_j^{*}(x)\psi(x+\xi(t_0),t_0)dx.
\end{split}
\end{equation}

\subsection{Switch disturbance }
\label{sec:org21445b2}

Here, let us consider a motion function $x_0(t)$ switching from one function to another at $t=t_1$, namely
\begin{equation}
x_{0}(t)=\begin{cases}f_{1}(t)&t\le t_{1}\\ f_{2}(t)&t>t_{1}\end{cases}
\end{equation}
Instead of relying on the general adiabatic approximation discussed in Sec.~II, we can directly evaluate the transition probability at the switching point $t_1$ using the exact solution. The exact wavefunctions in the regions $t\le t_{1}$ and $t> t_{1}$ are respectively given by
\begin{equation}
\psi_{1}(x,t)=e^{\frac{i}{\hbar}m\dot{\xi}_{1}(t)x}\sum_{j=0}^{\infty}c_{j}^{(1)}e^{-\frac{i}{\hbar}\varepsilon_{j}t}\chi_{j}(x-\xi_{1}(t))
\end{equation}
and
\begin{equation}
\psi_{2}(x,t)=e^{\frac{i}{\hbar}m\dot{\xi}_{2}(t)x}\sum_{j=0}^{\infty}c_{j}^{(2)}e^{-\frac{i}{\hbar}\varepsilon_{j}t}\chi_{j}(x-\xi_{2}(t))
\end{equation}
where $\xi_{1}(t)$ and $\xi_{2}(t)$ are defined by applying Eq.~(A17) to $f_{1}(t)$ and $f_{2}(t)$, respectively.

Before $t_{1}$, the adiabatic parameter is constant, namely
\begin{equation}
f_{1}(t)=c
\end{equation}
and then
\begin{equation}
\xi_{1}(t)=c.
\end{equation}

Under this condition, the transition due to the sudden change of the functions at $t_{1}$ is given by
\begin{equation}
\begin{pmatrix}
c_{0}^{(2)} \\
c_{1}^{(2)} \\
c_{2}^{(2)} \\
\vdots
\end{pmatrix}
= \hat{U}(\xi_{2}(t_{1})-c)
\begin{pmatrix}
c_{0}^{(1)} \\
c_{1}^{(1)} \\
c_{2}^{(1)} \\
\vdots
\end{pmatrix}.
\end{equation}

The transition amplitude due to this sudden change in function $x_{0}(t)$ is found to be
\begin{equation}
\langle j|\hat{U}(\xi_{2}(t_{1})-c)|k\rangle = \int_{-\infty}^{\infty} dx \chi_{j}^{*}(x) \chi_{k}(x-\xi_{2}(t_{1})+c) e^{\frac{i}{\hbar}m\dot{\xi}_{2}(t_{1})x},
\end{equation}
and the survival probability amplitude is obtained when the initial state is the ground state ($k=0$) and the system remains in the ground state ($j=0$). Completing the square in the overlap integral yields
\begin{equation}
a_{\mathrm{suv}} = e^{-\frac{m\omega}{4\hbar}\left( (\xi_{2}(t_{1})-c)^{2} + \frac{\dot{\xi}_{2}^{2}(t_{1})}{\omega^{2}} \right)}.
\end{equation}
If $f_{1}=0$ and $f_{2}(t)=\frac{1}{n!}x_{0}^{(n)}(t_{1})(t-t_{1})^{n}$, we find
\begin{equation}
\xi_{2}(t_{1})=\begin{cases}\frac{(-1)^{n/2}}{\omega^{n}}x_{0}^{(n)}(t_{1}) & (n: \text{even})\\ 0 & (n: \text{odd})\end{cases}
\end{equation}
and
\begin{equation}
\dot{\xi}_{2}(t_{1})=\begin{cases}0 & (n: \text{even})\\ \frac{(-1)^{\frac{n-1}{2}}}{\omega^{n-1}}x_{0}^{(n)}(t_{1}) & (n: \text{odd})\end{cases}.
\end{equation}
Thus, we have
\begin{equation}
a_{\mathrm{suv}}=e^{-\frac{m\omega}{4\hbar}\frac{(x_{0}^{(n)}(t_{1}))^{2}}{\omega^{2n}}}.
\end{equation}
The nonadiabatic transition probability is then given by
\begin{equation}
  1-a_{\mathrm{suv}}^{2}=1-e^{-\frac{m\omega}{2\hbar}\frac{(x_{0}^{(n)}(t_{1}))^{2}}{\omega^{2n}}} \simeq \frac{m\omega}{2\hbar}\frac{(x_{0}^{(n)}(t_{1}))^{2}}{\omega^{2n}},
  \label{exact-harm}
\end{equation}
which precisely corresponds to the initial/final disturbance effect $d_{\mathrm{ini/fin}}[a(\cdot)]$.
This exact result can be compared with the general theoretical prediction discussed in Sec.~II. By applying the general formula Eq.~(3) to the present moving potential model, we can evaluate $A_{10}$ as follows. The moving potential in this case is
\begin{equation}
V(x,t)=\frac{m\omega^{2}}{2}(x-f_{2}(t))^{2}=\frac{m\omega^{2}}{2}x^{2}-m\omega^{2}f_{2}(t)x,
\end{equation}
ignoring the spatially uniform term $\frac{m\omega^{2}}{2}f_{2}(t)^{2}$ which only affects the global phase. Thus, we have the $n$-th order time derivative of the Hamiltonian
\begin{equation}
\hat{H}^{(n)}=-m\omega^{2}x_{0}^{(n)}(t_{1})\hat{x}.
\end{equation}
Using the relation
\begin{equation}
\langle j|\hat{x}|0\rangle=\delta_{j1}\sqrt{\frac{\hbar}{2m\omega}}
\end{equation}
and $\Delta_{10}=\hbar\omega$, the nonadiabatic transition probability given by Eq.~(\ref{Pkj-alpha-n}) is found to be
\begin{equation}
|A_{10}|^{2}=\left|\frac{\hbar^{n}\langle 1|\hat{H}^{(n)}|0\rangle}{\Delta_{10}^{n+1}}\right|^{2}=\frac{m\omega}{2\hbar\omega^{2n}}(x_{0}^{(n)}(t_{1}))^{2},
\end{equation}
which perfectly agrees with the current exact result derived above in Eq.~(\ref{exact-harm}).

\subsection{Example-Sinusoidal motion}
\label{sec:orgb2929c0}
Let us here take an example of moving harmonic potential problem, with the motion is given by a sinusoidal form, namely \(x_0(t) = A \cos(\Omega t)\).
Note that the driving frequency \(\Omega\) of the potential motion should be distinguished from the intrinsic frequency \(\omega\) of the harmonic trapping potential. 
In this case time derivatives of \(x_0(t)\) are found to be 
\begin{equation}
x_0^{(2n)}(t) = (-1)^n\Omega^{2n}A\cos(\Omega t)
\end{equation}
giving rise to 
\begin{equation}
\begin{split}
\xi_2(t) &=A\cos(\Omega t)\sum_{n=0}^{\infty}\left(\frac{\Omega^2}{\omega^2}\right)^n\\
&=\frac{1}{1-\Omega^2/\omega^2}A\cos(\Omega t)
=\frac{1}{1-\Omega^2/\omega^2}x_0(t).
\end{split}
\end{equation}

In the slow limit, \(\Omega/\omega \rightarrow 0\), we have \(\xi(t)\rightarrow x_0(t)\),
which corresponds to the adiabatic approximation. The correction to the adiabatic
approximation is taken into account by introducing $\Omega^2/\omega^2$ term in this
equation. Sudden changes in any orders of derivatives of the adiabatic parameter can
induce nonadiabatic effects, which can be fully captured by evaluating the sudden change in 
the effective shifted coordinate $\xi(t)$, namely $\xi_2(t_1)-\xi_1(t_1)$.

\bibliography{NonadiabaticProcess}

%Control: production of eprint (1) enabled
\begin{thebibliography}{92}%
\makeatletter
\providecommand \@ifxundefined [1]{%
 \@ifx{#1\undefined}
}%
\providecommand \@ifnum [1]{%
 \ifnum #1\expandafter \@firstoftwo
 \else \expandafter \@secondoftwo
 \fi
}%
\providecommand \@ifx [1]{%
 \ifx #1\expandafter \@firstoftwo
 \else \expandafter \@secondoftwo
 \fi
}%
\providecommand \natexlab [1]{#1}%
\providecommand \enquote  [1]{``#1''}%
\providecommand \bibnamefont  [1]{#1}%
\providecommand \bibfnamefont [1]{#1}%
\providecommand \citenamefont [1]{#1}%
\providecommand \href@noop [0]{\@secondoftwo}%
\providecommand \href [0]{\begingroup \@sanitize@url \@href}%
\providecommand \@href[1]{\@@startlink{#1}\@@href}%
\providecommand \@@href[1]{\endgroup#1\@@endlink}%
\providecommand \@sanitize@url [0]{\catcode `\\12\catcode `\$12\catcode `\&12\catcode `\#12\catcode `\^12\catcode `\_12\catcode `\%12\relax}%
\providecommand \@@startlink[1]{}%
\providecommand \@@endlink[0]{}%
\providecommand \url  [0]{\begingroup\@sanitize@url \@url }%
\providecommand \@url [1]{\endgroup\@href {#1}{\urlprefix }}%
\providecommand \urlprefix  [0]{URL }%
\providecommand \Eprint [0]{\href }%
\providecommand \doibase [0]{https://doi.org/}%
\providecommand \selectlanguage [0]{\@gobble}%
\providecommand \bibinfo  [0]{\@secondoftwo}%
\providecommand \bibfield  [0]{\@secondoftwo}%
\providecommand \translation [1]{[#1]}%
\providecommand \BibitemOpen [0]{}%
\providecommand \bibitemStop [0]{}%
\providecommand \bibitemNoStop [0]{.\EOS\space}%
\providecommand \EOS [0]{\spacefactor3000\relax}%
\providecommand \BibitemShut  [1]{\csname bibitem#1\endcsname}%
\let\auto@bib@innerbib\@empty
%</preamble>
\bibitem [{\citenamefont {Razavy}(2013)}]{Razavy2013}%
  \BibitemOpen
  \bibfield  {author} {\bibinfo {author} {\bibfnamefont {M.}~\bibnamefont {Razavy}},\ }\href {https://doi.org/10.1142/8901} {\emph {\bibinfo {title} {Quantum Theory of Tunneling}}},\ \bibinfo {edition} {2nd}\ ed.\ (\bibinfo  {publisher} {World Scientific},\ \bibinfo {address} {Singapore},\ \bibinfo {year} {2013})\BibitemShut {NoStop}%
\bibitem [{\citenamefont {Miyashita}(2022)}]{SM2022}%
  \BibitemOpen
  \bibfield  {author} {\bibinfo {author} {\bibfnamefont {S.}~\bibnamefont {Miyashita}},\ }\href {https://doi.org/10.1007/978-981-19-6668-2} {\emph {\bibinfo {title} {{Collapse of Metastability: Dynamics of First-order Phase Transition}}}},\ Fundamental Theories of Physics\ (\bibinfo  {publisher} {Springer Nature},\ \bibinfo {year} {2022})\BibitemShut {NoStop}%
\bibitem [{\citenamefont {Landau}(1932)}]{Landau1932}%
  \BibitemOpen
  \bibfield  {author} {\bibinfo {author} {\bibfnamefont {L.~D.}\ \bibnamefont {Landau}},\ }\bibfield  {title} {\bibinfo {title} {Zur {{Theorie}} der {{Energieubertragung II}}},\ }\href@noop {} {\bibfield  {journal} {\bibinfo  {journal} {Phys. Z. Sowjetunion}\ }\textbf {\bibinfo {volume} {2}},\ \bibinfo {pages} {46} (\bibinfo {year} {1932})}\BibitemShut {NoStop}%
\bibitem [{\citenamefont {Zener}(1997)}]{Zener1997}%
  \BibitemOpen
  \bibfield  {author} {\bibinfo {author} {\bibfnamefont {C.}~\bibnamefont {Zener}},\ }\bibfield  {title} {\bibinfo {title} {Non-adiabatic crossing of energy levels},\ }\href {https://doi.org/10.1098/rspa.1932.0165} {\bibfield  {journal} {\bibinfo  {journal} {Proc. R. Soc. Lond. A}\ }\textbf {\bibinfo {volume} {137}},\ \bibinfo {pages} {696} (\bibinfo {year} {1997})}\BibitemShut {NoStop}%
\bibitem [{\citenamefont {Stueckelberg}(1932)}]{Stueckelberg1932}%
  \BibitemOpen
  \bibfield  {author} {\bibinfo {author} {\bibfnamefont {E.}~\bibnamefont {Stueckelberg}},\ }\bibfield  {title} {\bibinfo {title} {Theorie der unelastischen {{St{\"o}sse}} zwischen {{Atomen}}},\ }\href {https://doi.org/10.5169/seals-110177} {\bibfield  {journal} {\bibinfo  {journal} {Helv. Phys. Acta}\ }\textbf {\bibinfo {volume} {5}},\ \bibinfo {pages} {369} (\bibinfo {year} {1932})}\BibitemShut {NoStop}%
\bibitem [{\citenamefont {Majorana}(1932)}]{Majorana1932}%
  \BibitemOpen
  \bibfield  {author} {\bibinfo {author} {\bibfnamefont {E.}~\bibnamefont {Majorana}},\ }\bibfield  {title} {\bibinfo {title} {Atomi orientati in campo magnetico variabile},\ }\href {https://doi.org/10.1007/BF02960953} {\bibfield  {journal} {\bibinfo  {journal} {Nuovo Cim.}\ }\textbf {\bibinfo {volume} {9}},\ \bibinfo {pages} {43} (\bibinfo {year} {1932})}\BibitemShut {NoStop}%
\bibitem [{\citenamefont {Miyashita}(2011)}]{SM-QD2011}%
  \BibitemOpen
  \bibfield  {author} {\bibinfo {author} {\bibfnamefont {S.}~\bibnamefont {Miyashita}},\ }\bibfield  {title} {\bibinfo {title} {{Quantum Dynamics Under Time-Dependent External Fields}},\ }\href {https://doi.org/doi:10.1166/jctn.2011.1771} {\bibfield  {journal} {\bibinfo  {journal} {J. Comput. Theor. Nanosci.}\ }\textbf {\bibinfo {volume} {8}},\ \bibinfo {pages} {919} (\bibinfo {year} {2011})}\BibitemShut {NoStop}%
\bibitem [{\citenamefont {Miyashita}\ \emph {et~al.}(1998)\citenamefont {Miyashita}, \citenamefont {Saito},\ and\ \citenamefont {De~Raedt}}]{LZac1}%
  \BibitemOpen
  \bibfield  {author} {\bibinfo {author} {\bibfnamefont {S.}~\bibnamefont {Miyashita}}, \bibinfo {author} {\bibfnamefont {K.}~\bibnamefont {Saito}},\ and\ \bibinfo {author} {\bibfnamefont {H.}~\bibnamefont {De~Raedt}},\ }\bibfield  {title} {\bibinfo {title} {{Nontrivial Response of Nanoscale Uniaxial Magnets to an Alternating Field}},\ }\href {https://doi.org/10.1103/PhysRevLett.80.1525} {\bibfield  {journal} {\bibinfo  {journal} {Phys. Rev. Lett.}\ }\textbf {\bibinfo {volume} {80}},\ \bibinfo {pages} {1525} (\bibinfo {year} {1998})}\BibitemShut {NoStop}%
\bibitem [{\citenamefont {Teranishi}\ and\ \citenamefont {Nakamura}(1998)}]{LZac2}%
  \BibitemOpen
  \bibfield  {author} {\bibinfo {author} {\bibfnamefont {Y.}~\bibnamefont {Teranishi}}\ and\ \bibinfo {author} {\bibfnamefont {H.}~\bibnamefont {Nakamura}},\ }\bibfield  {title} {\bibinfo {title} {{Control of Time-Dependent Nonadiabatic Processes by an External Field}},\ }\href {https://doi.org/10.1103/PhysRevLett.81.2032} {\bibfield  {journal} {\bibinfo  {journal} {Phys. Rev. Lett.}\ }\textbf {\bibinfo {volume} {81}},\ \bibinfo {pages} {2032} (\bibinfo {year} {1998})}\BibitemShut {NoStop}%
\bibitem [{\citenamefont {Hatomura}\ \emph {et~al.}(2016)\citenamefont {Hatomura}, \citenamefont {Barbara},\ and\ \citenamefont {Miyashita}}]{Hatomura2016}%
  \BibitemOpen
  \bibfield  {author} {\bibinfo {author} {\bibfnamefont {T.}~\bibnamefont {Hatomura}}, \bibinfo {author} {\bibfnamefont {B.}~\bibnamefont {Barbara}},\ and\ \bibinfo {author} {\bibfnamefont {S.}~\bibnamefont {Miyashita}},\ }\bibfield  {title} {\bibinfo {title} {{Quantum Stoner-Wohlfarth Model}},\ }\href {https://doi.org/10.1103/PhysRevLett.116.037203} {\bibfield  {journal} {\bibinfo  {journal} {Phys. Rev. Lett.}\ }\textbf {\bibinfo {volume} {116}},\ \bibinfo {pages} {037203} (\bibinfo {year} {2016})}\BibitemShut {NoStop}%
\bibitem [{\citenamefont {Hatomura}\ \emph {et~al.}(2017)\citenamefont {Hatomura}, \citenamefont {Mori},\ and\ \citenamefont {Miyashita}}]{Hatomura2017}%
  \BibitemOpen
  \bibfield  {author} {\bibinfo {author} {\bibfnamefont {T.}~\bibnamefont {Hatomura}}, \bibinfo {author} {\bibfnamefont {T.}~\bibnamefont {Mori}},\ and\ \bibinfo {author} {\bibfnamefont {S.}~\bibnamefont {Miyashita}},\ }\bibfield  {title} {\bibinfo {title} {Distribution of eigenstate populations and dissipative beating dynamics in uniaxial single-spin magnets},\ }\href {https://doi.org/10.1103/PhysRevB.96.134309} {\bibfield  {journal} {\bibinfo  {journal} {Phys. Rev. B}\ }\textbf {\bibinfo {volume} {96}},\ \bibinfo {pages} {134309} (\bibinfo {year} {2017})}\BibitemShut {NoStop}%
\bibitem [{\citenamefont {Siegert}(1939)}]{Siegert}%
  \BibitemOpen
  \bibfield  {author} {\bibinfo {author} {\bibfnamefont {A.~J.~F.}\ \bibnamefont {Siegert}},\ }\bibfield  {title} {\bibinfo {title} {{On the Derivation of the Dispersion Formula for Nuclear Reactions}},\ }\href {https://doi.org/10.1103/PhysRev.56.750} {\bibfield  {journal} {\bibinfo  {journal} {Phys. Rev.}\ }\textbf {\bibinfo {volume} {56}},\ \bibinfo {pages} {750} (\bibinfo {year} {1939})}\BibitemShut {NoStop}%
\bibitem [{\citenamefont {Feshbach}(1958)}]{Feshbach1958}%
  \BibitemOpen
  \bibfield  {author} {\bibinfo {author} {\bibfnamefont {H.}~\bibnamefont {Feshbach}},\ }\bibfield  {title} {\bibinfo {title} {{Unified theory of nuclear reactions}},\ }\href {https://doi.org/https://doi.org/10.1016/0003-4916(58)90007-1} {\bibfield  {journal} {\bibinfo  {journal} {Ann. Phys.}\ }\textbf {\bibinfo {volume} {5}},\ \bibinfo {pages} {357} (\bibinfo {year} {1958})}\BibitemShut {NoStop}%
\bibitem [{\citenamefont {Feshbach}(1962)}]{Feshbach1962}%
  \BibitemOpen
  \bibfield  {author} {\bibinfo {author} {\bibfnamefont {H.}~\bibnamefont {Feshbach}},\ }\bibfield  {title} {\bibinfo {title} {{A unified theory of nuclear reactions. II}},\ }\href {https://doi.org/https://doi.org/10.1016/0003-4916(62)90221-X} {\bibfield  {journal} {\bibinfo  {journal} {Ann. Phys.}\ }\textbf {\bibinfo {volume} {19}},\ \bibinfo {pages} {287} (\bibinfo {year} {1962})}\BibitemShut {NoStop}%
\bibitem [{\citenamefont {Feshbach}(1967)}]{Feshbach1967}%
  \BibitemOpen
  \bibfield  {author} {\bibinfo {author} {\bibfnamefont {H.}~\bibnamefont {Feshbach}},\ }\bibfield  {title} {\bibinfo {title} {{The unified theory of nuclear reactions: III. Overlapping resonances}},\ }\href {https://doi.org/https://doi.org/10.1016/0003-4916(67)90163-7} {\bibfield  {journal} {\bibinfo  {journal} {Ann. Phys.}\ }\textbf {\bibinfo {volume} {43}},\ \bibinfo {pages} {410} (\bibinfo {year} {1967})}\BibitemShut {NoStop}%
\bibitem [{\citenamefont {Fano}(1961)}]{Fano}%
  \BibitemOpen
  \bibfield  {author} {\bibinfo {author} {\bibfnamefont {U.}~\bibnamefont {Fano}},\ }\bibfield  {title} {\bibinfo {title} {{Effects of Configuration Interaction on Intensities and Phase Shifts}},\ }\href {https://doi.org/10.1103/PhysRev.124.1866} {\bibfield  {journal} {\bibinfo  {journal} {Phys. Rev.}\ }\textbf {\bibinfo {volume} {124}},\ \bibinfo {pages} {1866} (\bibinfo {year} {1961})}\BibitemShut {NoStop}%
\bibitem [{\citenamefont {Hazi}\ and\ \citenamefont {Taylor}(1970)}]{Hazi-Taylor}%
  \BibitemOpen
  \bibfield  {author} {\bibinfo {author} {\bibfnamefont {A.~U.}\ \bibnamefont {Hazi}}\ and\ \bibinfo {author} {\bibfnamefont {H.~S.}\ \bibnamefont {Taylor}},\ }\bibfield  {title} {\bibinfo {title} {{Stabilization Method of Calculating Resonance Energies: Model Problem}},\ }\href {https://doi.org/10.1103/PhysRevA.1.1109} {\bibfield  {journal} {\bibinfo  {journal} {Phys. Rev. A}\ }\textbf {\bibinfo {volume} {1}},\ \bibinfo {pages} {1109} (\bibinfo {year} {1970})}\BibitemShut {NoStop}%
\bibitem [{\citenamefont {Langer}(1937)}]{Langer1937}%
  \BibitemOpen
  \bibfield  {author} {\bibinfo {author} {\bibfnamefont {R.~E.}\ \bibnamefont {Langer}},\ }\bibfield  {title} {\bibinfo {title} {{On the Connection Formulas and the Solutions of the Wave Equation}},\ }\href {https://doi.org/10.1103/PhysRev.51.669} {\bibfield  {journal} {\bibinfo  {journal} {Phys. Rev.}\ }\textbf {\bibinfo {volume} {51}},\ \bibinfo {pages} {669} (\bibinfo {year} {1937})}\BibitemShut {NoStop}%
\bibitem [{\citenamefont {Miller}\ and\ \citenamefont {Good}(1953)}]{Miller-Good}%
  \BibitemOpen
  \bibfield  {author} {\bibinfo {author} {\bibfnamefont {S.~C.}\ \bibnamefont {Miller}}\ and\ \bibinfo {author} {\bibfnamefont {R.~H.}\ \bibnamefont {Good}},\ }\bibfield  {title} {\bibinfo {title} {{A WKB-Type Approximation to the Schr\"odinger Equation}},\ }\href {https://doi.org/10.1103/PhysRev.91.174} {\bibfield  {journal} {\bibinfo  {journal} {Phys. Rev.}\ }\textbf {\bibinfo {volume} {91}},\ \bibinfo {pages} {174} (\bibinfo {year} {1953})}\BibitemShut {NoStop}%
\bibitem [{\citenamefont {Connor}(1969)}]{Connor1968}%
  \BibitemOpen
  \bibfield  {author} {\bibinfo {author} {\bibfnamefont {J.}~\bibnamefont {Connor}},\ }\bibfield  {title} {\bibinfo {title} {{On the semiclassical approximation for double well potentials}},\ }\href {https://doi.org/https://doi.org/10.1016/0009-2614(69)85001-3} {\bibfield  {journal} {\bibinfo  {journal} {Chem. Phys. Lett.}\ }\textbf {\bibinfo {volume} {4}},\ \bibinfo {pages} {419} (\bibinfo {year} {1969})}\BibitemShut {NoStop}%
\bibitem [{\citenamefont {Dickinson}(1970)}]{Dickinson1970}%
  \BibitemOpen
  \bibfield  {author} {\bibinfo {author} {\bibfnamefont {A.~S.}\ \bibnamefont {Dickinson}},\ }\bibfield  {title} {\bibinfo {title} {{An approximate treatment of shape resonances in elastic scattering}},\ }\href {https://doi.org/10.1080/00268977000100511} {\bibfield  {journal} {\bibinfo  {journal} {Mol. Phys.}\ }\textbf {\bibinfo {volume} {18}},\ \bibinfo {pages} {441} (\bibinfo {year} {1970})}\BibitemShut {NoStop}%
\bibitem [{\citenamefont {Meyer}\ and\ \citenamefont {Walter}(1982)}]{Mayer-Walter}%
  \BibitemOpen
  \bibfield  {author} {\bibinfo {author} {\bibfnamefont {H.~D.}\ \bibnamefont {Meyer}}\ and\ \bibinfo {author} {\bibfnamefont {O.}~\bibnamefont {Walter}},\ }\bibfield  {title} {\bibinfo {title} {{On the calculation of S-matrix poles using the Siegert method}},\ }\href {https://doi.org/10.1088/0022-3700/15/20/013} {\bibfield  {journal} {\bibinfo  {journal} {J. Phys. B: Atom. Mol. Phys.}\ }\textbf {\bibinfo {volume} {15}},\ \bibinfo {pages} {3647} (\bibinfo {year} {1982})}\BibitemShut {NoStop}%
\bibitem [{\citenamefont {Child}(2014)}]{Child-text}%
  \BibitemOpen
  \bibfield  {author} {\bibinfo {author} {\bibfnamefont {M.~S.}\ \bibnamefont {Child}},\ }\href {https://doi.org/10.1093/acprof:oso/9780199672981.001.0001} {\emph {\bibinfo {title} {{Semiclassical Mechanics with Molecular Applications}}}}\ (\bibinfo  {publisher} {Oxford University Press},\ \bibinfo {year} {2014})\BibitemShut {NoStop}%
\bibitem [{\citenamefont {Hatano}\ \emph {et~al.}(2008)\citenamefont {Hatano}, \citenamefont {Sasada}, \citenamefont {Nakamura},\ and\ \citenamefont {Petrosky}}]{Hatano-nonhermite}%
  \BibitemOpen
  \bibfield  {author} {\bibinfo {author} {\bibfnamefont {N.}~\bibnamefont {Hatano}}, \bibinfo {author} {\bibfnamefont {K.}~\bibnamefont {Sasada}}, \bibinfo {author} {\bibfnamefont {H.}~\bibnamefont {Nakamura}},\ and\ \bibinfo {author} {\bibfnamefont {T.}~\bibnamefont {Petrosky}},\ }\bibfield  {title} {\bibinfo {title} {{Some Properties of the Resonant State in Quantum Mechanics and Its Computation}},\ }\href {https://doi.org/10.1143/PTP.119.187} {\bibfield  {journal} {\bibinfo  {journal} {Prog. Theor. Phys.}\ }\textbf {\bibinfo {volume} {119}},\ \bibinfo {pages} {187} (\bibinfo {year} {2008})}\BibitemShut {NoStop}%
\bibitem [{\citenamefont {Moiseyev}(2011)}]{moiseyev2011nonhermitian}%
  \BibitemOpen
  \bibfield  {author} {\bibinfo {author} {\bibfnamefont {N.}~\bibnamefont {Moiseyev}},\ }\href {https://doi.org/10.1017/CBO9780511976186} {\emph {\bibinfo {title} {Non-{{Hermitian Quantum Mechanics}}}}}\ (\bibinfo  {publisher} {Cambridge University Press},\ \bibinfo {address} {Cambridge},\ \bibinfo {year} {2011})\BibitemShut {NoStop}%
\bibitem [{\citenamefont {Shi}\ \emph {et~al.}(1988)\citenamefont {Shi}, \citenamefont {Woody},\ and\ \citenamefont {Rabitz}}]{shi1988optimal}%
  \BibitemOpen
  \bibfield  {author} {\bibinfo {author} {\bibfnamefont {S.}~\bibnamefont {Shi}}, \bibinfo {author} {\bibfnamefont {A.}~\bibnamefont {Woody}},\ and\ \bibinfo {author} {\bibfnamefont {H.}~\bibnamefont {Rabitz}},\ }\bibfield  {title} {\bibinfo {title} {{Optimal control of selective vibrational excitation in harmonic linear chain molecules}},\ }\href {https://doi.org/10.1063/1.454384} {\bibfield  {journal} {\bibinfo  {journal} {J. Chem. Phys.}\ }\textbf {\bibinfo {volume} {88}},\ \bibinfo {pages} {6870} (\bibinfo {year} {1988})}\BibitemShut {NoStop}%
\bibitem [{\citenamefont {Brif}\ \emph {et~al.}(2010)\citenamefont {Brif}, \citenamefont {Chakrabarti},\ and\ \citenamefont {Rabitz}}]{brif2010control}%
  \BibitemOpen
  \bibfield  {author} {\bibinfo {author} {\bibfnamefont {C.}~\bibnamefont {Brif}}, \bibinfo {author} {\bibfnamefont {R.}~\bibnamefont {Chakrabarti}},\ and\ \bibinfo {author} {\bibfnamefont {H.}~\bibnamefont {Rabitz}},\ }\bibfield  {title} {\bibinfo {title} {Control of quantum phenomena: past, present and future},\ }\href {https://doi.org/10.1088/1367-2630/12/7/075008} {\bibfield  {journal} {\bibinfo  {journal} {New J. Phys.}\ }\textbf {\bibinfo {volume} {12}},\ \bibinfo {pages} {075008} (\bibinfo {year} {2010})}\BibitemShut {NoStop}%
\bibitem [{\citenamefont {Kadowaki}\ and\ \citenamefont {Nishimori}(1998)}]{Kadowaki1998}%
  \BibitemOpen
  \bibfield  {author} {\bibinfo {author} {\bibfnamefont {T.}~\bibnamefont {Kadowaki}}\ and\ \bibinfo {author} {\bibfnamefont {H.}~\bibnamefont {Nishimori}},\ }\bibfield  {title} {\bibinfo {title} {Quantum annealing in the transverse ising model},\ }\href {https://doi.org/10.1103/PhysRevE.58.5355} {\bibfield  {journal} {\bibinfo  {journal} {Phys. Rev. E}\ }\textbf {\bibinfo {volume} {58}},\ \bibinfo {pages} {5355} (\bibinfo {year} {1998})}\BibitemShut {NoStop}%
\bibitem [{\citenamefont {Das}\ and\ \citenamefont {Chakrabarti}(2008)}]{Das2008}%
  \BibitemOpen
  \bibfield  {author} {\bibinfo {author} {\bibfnamefont {A.}~\bibnamefont {Das}}\ and\ \bibinfo {author} {\bibfnamefont {B.~K.}\ \bibnamefont {Chakrabarti}},\ }\bibfield  {title} {\bibinfo {title} {Colloquium: Quantum annealing and analog quantum computation},\ }\href {https://doi.org/10.1103/RevModPhys.80.1061} {\bibfield  {journal} {\bibinfo  {journal} {Rev. Mod. Phys.}\ }\textbf {\bibinfo {volume} {80}},\ \bibinfo {pages} {1061} (\bibinfo {year} {2008})}\BibitemShut {NoStop}%
\bibitem [{\citenamefont {Morita}\ and\ \citenamefont {Nishimori}(2008)}]{MoritaNishimori2008}%
  \BibitemOpen
  \bibfield  {author} {\bibinfo {author} {\bibfnamefont {S.}~\bibnamefont {Morita}}\ and\ \bibinfo {author} {\bibfnamefont {H.}~\bibnamefont {Nishimori}},\ }\bibfield  {title} {\bibinfo {title} {{Mathematical foundation of quantum annealing}},\ }\href {https://doi.org/10.1063/1.2995837} {\bibfield  {journal} {\bibinfo  {journal} {J. Math. Phys.}\ }\textbf {\bibinfo {volume} {49}},\ \bibinfo {pages} {125210} (\bibinfo {year} {2008})}\BibitemShut {NoStop}%
\bibitem [{\citenamefont {Tanaka}\ \emph {et~al.}(2017)\citenamefont {Tanaka}, \citenamefont {Tamura},\ and\ \citenamefont {Chakrabarti}}]{Tanaka2017}%
  \BibitemOpen
  \bibfield  {author} {\bibinfo {author} {\bibfnamefont {S.}~\bibnamefont {Tanaka}}, \bibinfo {author} {\bibfnamefont {R.}~\bibnamefont {Tamura}},\ and\ \bibinfo {author} {\bibfnamefont {B.}~\bibnamefont {Chakrabarti}},\ }\href {https://www.cambridge.org/us/universitypress/subjects/physics/condensed-matter-physics-nanoscience-and-mesoscopic-physics/quantum-spin-glasses-annealing-and-computation} {\emph {\bibinfo {title} {Quantum Spin Glasses, Annealing and Computation}}}\ (\bibinfo  {publisher} {Cambridge University Press},\ \bibinfo {year} {2017})\BibitemShut {NoStop}%
\bibitem [{\citenamefont {Johnson}\ \emph {et~al.}(2011)\citenamefont {Johnson}, \citenamefont {Amin}, \citenamefont {Gildert}, \citenamefont {Lanting}, \citenamefont {Hamze}, \citenamefont {Dickson}, \citenamefont {Harris}, \citenamefont {Berkley}, \citenamefont {Johansson}, \citenamefont {Bunyk}, \citenamefont {Chapple}, \citenamefont {Enderud}, \citenamefont {Hilton}, \citenamefont {Karimi}, \citenamefont {Ladizinsky}, \citenamefont {Ladizinsky}, \citenamefont {Oh}, \citenamefont {Perminov}, \citenamefont {Rich}, \citenamefont {Thom}, \citenamefont {Tolkacheva}, \citenamefont {Truncik}, \citenamefont {Uchaikin}, \citenamefont {Wang}, \citenamefont {Wilson},\ and\ \citenamefont {Rose}}]{DWave2011}%
  \BibitemOpen
  \bibfield  {author} {\bibinfo {author} {\bibfnamefont {M.~W.}\ \bibnamefont {Johnson}}, \bibinfo {author} {\bibfnamefont {M.~H.~S.}\ \bibnamefont {Amin}}, \bibinfo {author} {\bibfnamefont {S.}~\bibnamefont {Gildert}}, \bibinfo {author} {\bibfnamefont {T.}~\bibnamefont {Lanting}}, \bibinfo {author} {\bibfnamefont {F.}~\bibnamefont {Hamze}}, \bibinfo {author} {\bibfnamefont {N.}~\bibnamefont {Dickson}}, \bibinfo {author} {\bibfnamefont {R.}~\bibnamefont {Harris}}, \bibinfo {author} {\bibfnamefont {A.~J.}\ \bibnamefont {Berkley}}, \bibinfo {author} {\bibfnamefont {J.}~\bibnamefont {Johansson}}, \bibinfo {author} {\bibfnamefont {P.}~\bibnamefont {Bunyk}}, \bibinfo {author} {\bibfnamefont {E.~M.}\ \bibnamefont {Chapple}}, \bibinfo {author} {\bibfnamefont {C.}~\bibnamefont {Enderud}}, \bibinfo {author} {\bibfnamefont {J.~P.}\ \bibnamefont {Hilton}}, \bibinfo {author} {\bibfnamefont {K.}~\bibnamefont {Karimi}}, \bibinfo {author} {\bibfnamefont {E.}~\bibnamefont {Ladizinsky}}, \bibinfo {author} {\bibfnamefont {N.}~\bibnamefont {Ladizinsky}}, \bibinfo {author} {\bibfnamefont {T.}~\bibnamefont {Oh}}, \bibinfo {author} {\bibfnamefont {I.}~\bibnamefont {Perminov}}, \bibinfo {author} {\bibfnamefont {C.}~\bibnamefont {Rich}}, \bibinfo {author} {\bibfnamefont {M.~C.}\ \bibnamefont {Thom}}, \bibinfo {author} {\bibfnamefont {E.}~\bibnamefont {Tolkacheva}}, \bibinfo {author} {\bibfnamefont {C.~J.~S.}\ \bibnamefont {Truncik}}, \bibinfo {author} {\bibfnamefont {S.}~\bibnamefont {Uchaikin}}, \bibinfo {author} {\bibfnamefont {J.}~\bibnamefont {Wang}}, \bibinfo {author} {\bibfnamefont {B.}~\bibnamefont {Wilson}},\ and\ \bibinfo {author} {\bibfnamefont {G.}~\bibnamefont {Rose}},\ }\bibfield  {title} {\bibinfo {title} {Quantum annealing with manufactured spins},\ }\href {https://doi.org/10.1038/nature10012} {\bibfield  {journal} {\bibinfo  {journal} {Nature}\ }\textbf {\bibinfo {volume} {473}},\ \bibinfo {pages} {194} (\bibinfo {year} {2011})}\BibitemShut {NoStop}%
\bibitem [{\citenamefont {Boixo}\ \emph {et~al.}(2014)\citenamefont {Boixo}, \citenamefont {R{\o}nnow}, \citenamefont {Isakov}, \citenamefont {Wang}, \citenamefont {Wecker}, \citenamefont {Lidar}, \citenamefont {Martinis},\ and\ \citenamefont {Troyer}}]{DWave2014}%
  \BibitemOpen
  \bibfield  {author} {\bibinfo {author} {\bibfnamefont {S.}~\bibnamefont {Boixo}}, \bibinfo {author} {\bibfnamefont {T.~F.}\ \bibnamefont {R{\o}nnow}}, \bibinfo {author} {\bibfnamefont {S.~V.}\ \bibnamefont {Isakov}}, \bibinfo {author} {\bibfnamefont {Z.}~\bibnamefont {Wang}}, \bibinfo {author} {\bibfnamefont {D.}~\bibnamefont {Wecker}}, \bibinfo {author} {\bibfnamefont {D.~A.}\ \bibnamefont {Lidar}}, \bibinfo {author} {\bibfnamefont {J.~M.}\ \bibnamefont {Martinis}},\ and\ \bibinfo {author} {\bibfnamefont {M.}~\bibnamefont {Troyer}},\ }\bibfield  {title} {\bibinfo {title} {Evidence for quantum annealing with more than one hundred qubits},\ }\href {https://doi.org/10.1038/nphys2900} {\bibfield  {journal} {\bibinfo  {journal} {Nat. Phys.}\ }\textbf {\bibinfo {volume} {10}},\ \bibinfo {pages} {218} (\bibinfo {year} {2014})}\BibitemShut {NoStop}%
\bibitem [{\citenamefont {Atabek}\ \emph {et~al.}(2013)\citenamefont {Atabek}, \citenamefont {Lefebvre}, \citenamefont {Jaouadi},\ and\ \citenamefont {Desouter-Lecomte}}]{atabek2013proposal}%
  \BibitemOpen
  \bibfield  {author} {\bibinfo {author} {\bibfnamefont {O.}~\bibnamefont {Atabek}}, \bibinfo {author} {\bibfnamefont {R.}~\bibnamefont {Lefebvre}}, \bibinfo {author} {\bibfnamefont {A.}~\bibnamefont {Jaouadi}},\ and\ \bibinfo {author} {\bibfnamefont {M.}~\bibnamefont {Desouter-Lecomte}},\ }\bibfield  {title} {\bibinfo {title} {Proposal for laser purification in molecular vibrational cooling using zero-width resonances},\ }\href {https://doi.org/10.1103/PhysRevA.87.031403} {\bibfield  {journal} {\bibinfo  {journal} {Phys. Rev. A}\ }\textbf {\bibinfo {volume} {87}},\ \bibinfo {pages} {031403} (\bibinfo {year} {2013})}\BibitemShut {NoStop}%
\bibitem [{\citenamefont {Leclerc}\ \emph {et~al.}(2016)\citenamefont {Leclerc}, \citenamefont {Viennot}, \citenamefont {Jolicard}, \citenamefont {Lefebvre},\ and\ \citenamefont {Atabek}}]{leclerc2016controlling}%
  \BibitemOpen
  \bibfield  {author} {\bibinfo {author} {\bibfnamefont {A.}~\bibnamefont {Leclerc}}, \bibinfo {author} {\bibfnamefont {D.}~\bibnamefont {Viennot}}, \bibinfo {author} {\bibfnamefont {G.}~\bibnamefont {Jolicard}}, \bibinfo {author} {\bibfnamefont {R.}~\bibnamefont {Lefebvre}},\ and\ \bibinfo {author} {\bibfnamefont {O.}~\bibnamefont {Atabek}},\ }\bibfield  {title} {\bibinfo {title} {Controlling vibrational cooling with zero-width resonances: An adiabatic floquet approach},\ }\href {https://doi.org/10.1103/PhysRevA.94.043409} {\bibfield  {journal} {\bibinfo  {journal} {Phys. Rev. A}\ }\textbf {\bibinfo {volume} {94}},\ \bibinfo {pages} {043409} (\bibinfo {year} {2016})}\BibitemShut {NoStop}%
\bibitem [{\citenamefont {Miyashita}\ and\ \citenamefont {Barbara}(2023)}]{Miyashita2023}%
  \BibitemOpen
  \bibfield  {author} {\bibinfo {author} {\bibfnamefont {S.}~\bibnamefont {Miyashita}}\ and\ \bibinfo {author} {\bibfnamefont {B.}~\bibnamefont {Barbara}},\ }\bibfield  {title} {\bibinfo {title} {{How to Cross an Energy Barrier at Zero Kelvin without Tunneling Effect}},\ }\href {https://doi.org/10.1103/PhysRevLett.131.066701} {\bibfield  {journal} {\bibinfo  {journal} {Phys. Rev. Lett.}\ }\textbf {\bibinfo {volume} {131}},\ \bibinfo {pages} {066701} (\bibinfo {year} {2023})}\BibitemShut {NoStop}%
\bibitem [{\citenamefont {Miyashita}\ and\ \citenamefont {Barbara}(2024)}]{Miyashita2024}%
  \BibitemOpen
  \bibfield  {author} {\bibinfo {author} {\bibfnamefont {S.}~\bibnamefont {Miyashita}}\ and\ \bibinfo {author} {\bibfnamefont {B.}~\bibnamefont {Barbara}},\ }\bibfield  {title} {\bibinfo {title} {{Sequential resonance for giant quantum oscillations above the energy barrier and its classical counterpart}},\ }\href {https://doi.org/10.1103/PhysRevB.109.104301} {\bibfield  {journal} {\bibinfo  {journal} {Phys. Rev. B}\ }\textbf {\bibinfo {volume} {109}},\ \bibinfo {pages} {104301} (\bibinfo {year} {2024})}\BibitemShut {NoStop}%
\bibitem [{\citenamefont {Morita}(2007)}]{Morita2007}%
  \BibitemOpen
  \bibfield  {author} {\bibinfo {author} {\bibfnamefont {S.}~\bibnamefont {Morita}},\ }\bibfield  {title} {\bibinfo {title} {{Faster Annealing Schedules for Quantum Annealing}},\ }\href {https://doi.org/10.1143/JPSJ.76.104001} {\bibfield  {journal} {\bibinfo  {journal} {J. Phys. Soc. Jpn.}\ }\textbf {\bibinfo {volume} {76}},\ \bibinfo {pages} {104001} (\bibinfo {year} {2007})}\BibitemShut {NoStop}%
\bibitem [{\citenamefont {Miyashita}(2007)}]{Conveyance0}%
  \BibitemOpen
  \bibfield  {author} {\bibinfo {author} {\bibfnamefont {S.}~\bibnamefont {Miyashita}},\ }\bibfield  {title} {\bibinfo {title} {{Conveyance of Quantum Particles by a Moving Potential Well}},\ }\href {https://doi.org/10.1143/JPSJ.76.104003} {\bibfield  {journal} {\bibinfo  {journal} {J. Phys. Soc. Jpn.}\ }\textbf {\bibinfo {volume} {76}},\ \bibinfo {pages} {104003} (\bibinfo {year} {2007})}\BibitemShut {NoStop}%
\bibitem [{\citenamefont {Morita}\ \emph {et~al.}(2024)\citenamefont {Morita}, \citenamefont {Teranishi},\ and\ \citenamefont {Miyashita}}]{MTM2024}%
  \BibitemOpen
  \bibfield  {author} {\bibinfo {author} {\bibfnamefont {S.}~\bibnamefont {Morita}}, \bibinfo {author} {\bibfnamefont {Y.}~\bibnamefont {Teranishi}},\ and\ \bibinfo {author} {\bibfnamefont {S.}~\bibnamefont {Miyashita}},\ }\bibfield  {title} {\bibinfo {title} {Optimization of conveyance of quantum particles by moving potential well},\ }\href {https://doi.org/10.1103/PhysRevResearch.6.043329} {\bibfield  {journal} {\bibinfo  {journal} {Phys. Rev. Res.}\ }\textbf {\bibinfo {volume} {6}},\ \bibinfo {pages} {043329} (\bibinfo {year} {2024})}\BibitemShut {NoStop}%
\bibitem [{\citenamefont {Bloch}\ \emph {et~al.}(2008)\citenamefont {Bloch}, \citenamefont {Dalibard},\ and\ \citenamefont {Zwerger}}]{optical-lattice}%
  \BibitemOpen
  \bibfield  {author} {\bibinfo {author} {\bibfnamefont {I.}~\bibnamefont {Bloch}}, \bibinfo {author} {\bibfnamefont {J.}~\bibnamefont {Dalibard}},\ and\ \bibinfo {author} {\bibfnamefont {W.}~\bibnamefont {Zwerger}},\ }\bibfield  {title} {\bibinfo {title} {Many-body physics with ultracold gases},\ }\href {https://doi.org/10.1103/RevModPhys.80.885} {\bibfield  {journal} {\bibinfo  {journal} {Rev. Mod. Phys.}\ }\textbf {\bibinfo {volume} {80}},\ \bibinfo {pages} {885} (\bibinfo {year} {2008})}\BibitemShut {NoStop}%
\bibitem [{\citenamefont {Sch{\"a}fer}\ \emph {et~al.}(2020)\citenamefont {Sch{\"a}fer}, \citenamefont {Fukuhara}, \citenamefont {Sugawa}, \citenamefont {Takasu},\ and\ \citenamefont {Takahashi}}]{optical-lattice2}%
  \BibitemOpen
  \bibfield  {author} {\bibinfo {author} {\bibfnamefont {F.}~\bibnamefont {Sch{\"a}fer}}, \bibinfo {author} {\bibfnamefont {T.}~\bibnamefont {Fukuhara}}, \bibinfo {author} {\bibfnamefont {S.}~\bibnamefont {Sugawa}}, \bibinfo {author} {\bibfnamefont {Y.}~\bibnamefont {Takasu}},\ and\ \bibinfo {author} {\bibfnamefont {Y.}~\bibnamefont {Takahashi}},\ }\bibfield  {title} {\bibinfo {title} {Tools for quantum simulation with ultracold atoms in optical lattices},\ }\href {https://doi.org/10.1038/s42254-020-0195-3} {\bibfield  {journal} {\bibinfo  {journal} {Nat. Rev. Phys.}\ }\textbf {\bibinfo {volume} {2}},\ \bibinfo {pages} {411} (\bibinfo {year} {2020})}\BibitemShut {NoStop}%
\bibitem [{\citenamefont {Mandel}\ \emph {et~al.}(2003)\citenamefont {Mandel}, \citenamefont {Greiner}, \citenamefont {Widera}, \citenamefont {Rom}, \citenamefont {H\"ansch},\ and\ \citenamefont {Bloch}}]{Spin-dependent-optlattpot}%
  \BibitemOpen
  \bibfield  {author} {\bibinfo {author} {\bibfnamefont {O.}~\bibnamefont {Mandel}}, \bibinfo {author} {\bibfnamefont {M.}~\bibnamefont {Greiner}}, \bibinfo {author} {\bibfnamefont {A.}~\bibnamefont {Widera}}, \bibinfo {author} {\bibfnamefont {T.}~\bibnamefont {Rom}}, \bibinfo {author} {\bibfnamefont {T.~W.}\ \bibnamefont {H\"ansch}},\ and\ \bibinfo {author} {\bibfnamefont {I.}~\bibnamefont {Bloch}},\ }\bibfield  {title} {\bibinfo {title} {Coherent transport of neutral atoms in spin-dependent optical lattice potentials},\ }\href {https://doi.org/10.1103/PhysRevLett.91.010407} {\bibfield  {journal} {\bibinfo  {journal} {Phys. Rev. Lett.}\ }\textbf {\bibinfo {volume} {91}},\ \bibinfo {pages} {010407} (\bibinfo {year} {2003})}\BibitemShut {NoStop}%
\bibitem [{\citenamefont {Hauck}\ and\ \citenamefont {Stojanovi\ifmmode~\acute{c}\else \'{c}\fi{}}(2022)}]{Double-well-pot-optlattpot}%
  \BibitemOpen
  \bibfield  {author} {\bibinfo {author} {\bibfnamefont {S.~H.}\ \bibnamefont {Hauck}}\ and\ \bibinfo {author} {\bibfnamefont {V.~M.}\ \bibnamefont {Stojanovi\ifmmode~\acute{c}\else \'{c}\fi{}}},\ }\bibfield  {title} {\bibinfo {title} {Coherent atom transport via enhanced shortcuts to adiabaticity: Double-well optical lattice},\ }\href {https://doi.org/10.1103/PhysRevApplied.18.014016} {\bibfield  {journal} {\bibinfo  {journal} {Phys. Rev. Appl.}\ }\textbf {\bibinfo {volume} {18}},\ \bibinfo {pages} {014016} (\bibinfo {year} {2022})}\BibitemShut {NoStop}%
\bibitem [{\citenamefont {Katsuki}\ \emph {et~al.}(2013)\citenamefont {Katsuki}, \citenamefont {Delagnes}, \citenamefont {Hosaka}, \citenamefont {Ishioka}, \citenamefont {Chiba}, \citenamefont {Zijlstra}, \citenamefont {Garcia}, \citenamefont {Takahashi}, \citenamefont {Watanabe}, \citenamefont {Kitajima}, \citenamefont {Matsumoto}, \citenamefont {Nakamura},\ and\ \citenamefont {Ohmori}}]{Ohmori2013}%
  \BibitemOpen
  \bibfield  {author} {\bibinfo {author} {\bibfnamefont {H.}~\bibnamefont {Katsuki}}, \bibinfo {author} {\bibfnamefont {J.}~\bibnamefont {Delagnes}}, \bibinfo {author} {\bibfnamefont {K.}~\bibnamefont {Hosaka}}, \bibinfo {author} {\bibfnamefont {K.}~\bibnamefont {Ishioka}}, \bibinfo {author} {\bibfnamefont {H.}~\bibnamefont {Chiba}}, \bibinfo {author} {\bibfnamefont {E.}~\bibnamefont {Zijlstra}}, \bibinfo {author} {\bibfnamefont {M.}~\bibnamefont {Garcia}}, \bibinfo {author} {\bibfnamefont {H.}~\bibnamefont {Takahashi}}, \bibinfo {author} {\bibfnamefont {K.}~\bibnamefont {Watanabe}}, \bibinfo {author} {\bibfnamefont {M.}~\bibnamefont {Kitajima}}, \bibinfo {author} {\bibfnamefont {Y.}~\bibnamefont {Matsumoto}}, \bibinfo {author} {\bibfnamefont {K.}~\bibnamefont {Nakamura}},\ and\ \bibinfo {author} {\bibfnamefont {K.}~\bibnamefont {Ohmori}},\ }\bibfield  {title} {\bibinfo {title} {All-optical control and visualization of ultrafast two-dimensional atomic motions in a single crystal of bismuth},\ }\href {https://doi.org/10.1038/ncomms3801} {\bibfield  {journal} {\bibinfo  {journal} {Nat. Commun.}\ }\textbf {\bibinfo {volume} {4}},\ \bibinfo {pages} {2801} (\bibinfo {year} {2013})}\BibitemShut {NoStop}%
\bibitem [{\citenamefont {Takei}\ \emph {et~al.}(2016)\citenamefont {Takei}, \citenamefont {Sommer}, \citenamefont {Genes}, \citenamefont {Pupillo}, \citenamefont {Goto}, \citenamefont {Koyasu}, \citenamefont {Chiba}, \citenamefont {Weidem{\"u}ller},\ and\ \citenamefont {Ohmori}}]{Ohmori2016}%
  \BibitemOpen
  \bibfield  {author} {\bibinfo {author} {\bibfnamefont {N.}~\bibnamefont {Takei}}, \bibinfo {author} {\bibfnamefont {C.}~\bibnamefont {Sommer}}, \bibinfo {author} {\bibfnamefont {C.}~\bibnamefont {Genes}}, \bibinfo {author} {\bibfnamefont {G.}~\bibnamefont {Pupillo}}, \bibinfo {author} {\bibfnamefont {H.}~\bibnamefont {Goto}}, \bibinfo {author} {\bibfnamefont {K.}~\bibnamefont {Koyasu}}, \bibinfo {author} {\bibfnamefont {H.}~\bibnamefont {Chiba}}, \bibinfo {author} {\bibfnamefont {M.}~\bibnamefont {Weidem{\"u}ller}},\ and\ \bibinfo {author} {\bibfnamefont {K.}~\bibnamefont {Ohmori}},\ }\bibfield  {title} {\bibinfo {title} {Direct observation of ultrafast many-body electron dynamics in an ultracold {{Rydberg}} gas},\ }\href {https://doi.org/10.1038/ncomms13449} {\bibfield  {journal} {\bibinfo  {journal} {Nat. Commun.}\ }\textbf {\bibinfo {volume} {7}},\ \bibinfo {pages} {13449} (\bibinfo {year} {2016})}\BibitemShut {NoStop}%
\bibitem [{\citenamefont {Katsuki}\ \emph {et~al.}(2018)\citenamefont {Katsuki}, \citenamefont {Takei}, \citenamefont {Sommer},\ and\ \citenamefont {Ohmori}}]{Ohmori2018ACR}%
  \BibitemOpen
  \bibfield  {author} {\bibinfo {author} {\bibfnamefont {H.}~\bibnamefont {Katsuki}}, \bibinfo {author} {\bibfnamefont {N.}~\bibnamefont {Takei}}, \bibinfo {author} {\bibfnamefont {C.}~\bibnamefont {Sommer}},\ and\ \bibinfo {author} {\bibfnamefont {K.}~\bibnamefont {Ohmori}},\ }\bibfield  {title} {\bibinfo {title} {Ultrafast {{Coherent Control}} of {{Condensed Matter}} with {{Attosecond Precision}}},\ }\href {https://doi.org/10.1021/acs.accounts.7b00641} {\bibfield  {journal} {\bibinfo  {journal} {Acc. Chem. Res.}\ }\textbf {\bibinfo {volume} {51}},\ \bibinfo {pages} {1174} (\bibinfo {year} {2018})}\BibitemShut {NoStop}%
\bibitem [{\citenamefont {Liu}\ \emph {et~al.}(2018)\citenamefont {Liu}, \citenamefont {Manz}, \citenamefont {Ohmori}, \citenamefont {Sommer}, \citenamefont {Takei}, \citenamefont {Tremblay},\ and\ \citenamefont {Zhang}}]{Ohmori2018PRL}%
  \BibitemOpen
  \bibfield  {author} {\bibinfo {author} {\bibfnamefont {C.}~\bibnamefont {Liu}}, \bibinfo {author} {\bibfnamefont {J.}~\bibnamefont {Manz}}, \bibinfo {author} {\bibfnamefont {K.}~\bibnamefont {Ohmori}}, \bibinfo {author} {\bibfnamefont {C.}~\bibnamefont {Sommer}}, \bibinfo {author} {\bibfnamefont {N.}~\bibnamefont {Takei}}, \bibinfo {author} {\bibfnamefont {J.~C.}\ \bibnamefont {Tremblay}},\ and\ \bibinfo {author} {\bibfnamefont {Y.}~\bibnamefont {Zhang}},\ }\bibfield  {title} {\bibinfo {title} {{Attosecond Control of Restoration of Electronic Structure Symmetry}},\ }\href {https://doi.org/10.1103/PhysRevLett.121.173201} {\bibfield  {journal} {\bibinfo  {journal} {Phys. Rev. Lett.}\ }\textbf {\bibinfo {volume} {121}},\ \bibinfo {pages} {173201} (\bibinfo {year} {2018})}\BibitemShut {NoStop}%
\bibitem [{\citenamefont {Mizoguchi}\ \emph {et~al.}(2020)\citenamefont {Mizoguchi}, \citenamefont {Zhang}, \citenamefont {Kunimi}, \citenamefont {Tanaka}, \citenamefont {Takeda}, \citenamefont {Takei}, \citenamefont {Bharti}, \citenamefont {Koyasu}, \citenamefont {Kishimoto}, \citenamefont {Jaksch}, \citenamefont {Glaetzle}, \citenamefont {Kiffner}, \citenamefont {Masella}, \citenamefont {Pupillo}, \citenamefont {Weidem\"uller},\ and\ \citenamefont {Ohmori}}]{Ohmori2020}%
  \BibitemOpen
  \bibfield  {author} {\bibinfo {author} {\bibfnamefont {M.}~\bibnamefont {Mizoguchi}}, \bibinfo {author} {\bibfnamefont {Y.}~\bibnamefont {Zhang}}, \bibinfo {author} {\bibfnamefont {M.}~\bibnamefont {Kunimi}}, \bibinfo {author} {\bibfnamefont {A.}~\bibnamefont {Tanaka}}, \bibinfo {author} {\bibfnamefont {S.}~\bibnamefont {Takeda}}, \bibinfo {author} {\bibfnamefont {N.}~\bibnamefont {Takei}}, \bibinfo {author} {\bibfnamefont {V.}~\bibnamefont {Bharti}}, \bibinfo {author} {\bibfnamefont {K.}~\bibnamefont {Koyasu}}, \bibinfo {author} {\bibfnamefont {T.}~\bibnamefont {Kishimoto}}, \bibinfo {author} {\bibfnamefont {D.}~\bibnamefont {Jaksch}}, \bibinfo {author} {\bibfnamefont {A.}~\bibnamefont {Glaetzle}}, \bibinfo {author} {\bibfnamefont {M.}~\bibnamefont {Kiffner}}, \bibinfo {author} {\bibfnamefont {G.}~\bibnamefont {Masella}}, \bibinfo {author} {\bibfnamefont {G.}~\bibnamefont {Pupillo}}, \bibinfo {author} {\bibfnamefont {M.}~\bibnamefont {Weidem\"uller}},\ and\ \bibinfo {author} {\bibfnamefont {K.}~\bibnamefont {Ohmori}},\ }\bibfield  {title} {\bibinfo {title} {{Ultrafast Creation of Overlapping Rydberg Electrons in an Atomic BEC and Mott-Insulator Lattice}},\ }\href {https://doi.org/10.1103/PhysRevLett.124.253201} {\bibfield  {journal} {\bibinfo  {journal} {Phys. Rev. Lett.}\ }\textbf {\bibinfo {volume} {124}},\ \bibinfo {pages} {253201} (\bibinfo {year} {2020})}\BibitemShut {NoStop}%
\bibitem [{\citenamefont {Chew}\ \emph {et~al.}(2022)\citenamefont {Chew}, \citenamefont {Tomita}, \citenamefont {Mahesh}, \citenamefont {Sugawa}, \citenamefont {{de L{\'e}s{\'e}leuc}},\ and\ \citenamefont {Ohmori}}]{Ohmori2022}%
  \BibitemOpen
  \bibfield  {author} {\bibinfo {author} {\bibfnamefont {Y.}~\bibnamefont {Chew}}, \bibinfo {author} {\bibfnamefont {T.}~\bibnamefont {Tomita}}, \bibinfo {author} {\bibfnamefont {T.~P.}\ \bibnamefont {Mahesh}}, \bibinfo {author} {\bibfnamefont {S.}~\bibnamefont {Sugawa}}, \bibinfo {author} {\bibfnamefont {S.}~\bibnamefont {{de L{\'e}s{\'e}leuc}}},\ and\ \bibinfo {author} {\bibfnamefont {K.}~\bibnamefont {Ohmori}},\ }\bibfield  {title} {\bibinfo {title} {Ultrafast energy exchange between two single {{Rydberg}} atoms on a nanosecond timescale},\ }\href {https://doi.org/10.1038/s41566-022-01047-2} {\bibfield  {journal} {\bibinfo  {journal} {Nat. Photonics}\ }\textbf {\bibinfo {volume} {16}},\ \bibinfo {pages} {724} (\bibinfo {year} {2022})}\BibitemShut {NoStop}%
\bibitem [{\citenamefont {Bharti}\ \emph {et~al.}(2023)\citenamefont {Bharti}, \citenamefont {Sugawa}, \citenamefont {Mizoguchi}, \citenamefont {Kunimi}, \citenamefont {Zhang}, \citenamefont {de~L\'es\'eleuc}, \citenamefont {Tomita}, \citenamefont {Franz}, \citenamefont {Weidem\"uller},\ and\ \citenamefont {Ohmori}}]{Ohmori2023PRL}%
  \BibitemOpen
  \bibfield  {author} {\bibinfo {author} {\bibfnamefont {V.}~\bibnamefont {Bharti}}, \bibinfo {author} {\bibfnamefont {S.}~\bibnamefont {Sugawa}}, \bibinfo {author} {\bibfnamefont {M.}~\bibnamefont {Mizoguchi}}, \bibinfo {author} {\bibfnamefont {M.}~\bibnamefont {Kunimi}}, \bibinfo {author} {\bibfnamefont {Y.}~\bibnamefont {Zhang}}, \bibinfo {author} {\bibfnamefont {S.}~\bibnamefont {de~L\'es\'eleuc}}, \bibinfo {author} {\bibfnamefont {T.}~\bibnamefont {Tomita}}, \bibinfo {author} {\bibfnamefont {T.}~\bibnamefont {Franz}}, \bibinfo {author} {\bibfnamefont {M.}~\bibnamefont {Weidem\"uller}},\ and\ \bibinfo {author} {\bibfnamefont {K.}~\bibnamefont {Ohmori}},\ }\bibfield  {title} {\bibinfo {title} {Picosecond-{{Scale Ultrafast Many-Body Dynamics}} in an {{Ultracold Rydberg-Excited Atomic Mott Insulator}}},\ }\href {https://doi.org/10.1103/PhysRevLett.131.123201} {\bibfield  {journal} {\bibinfo  {journal} {Phys. Rev. Lett.}\ }\textbf {\bibinfo {volume} {131}},\ \bibinfo {pages} {123201} (\bibinfo {year} {2023})}\BibitemShut {NoStop}%
\bibitem [{\citenamefont {Bharti}\ \emph {et~al.}(2024)\citenamefont {Bharti}, \citenamefont {Sugawa}, \citenamefont {Kunimi}, \citenamefont {Chauhan}, \citenamefont {Mahesh}, \citenamefont {Mizoguchi}, \citenamefont {Matsubara}, \citenamefont {Tomita}, \citenamefont {de~L\'es\'eleuc},\ and\ \citenamefont {Ohmori}}]{Ohmori2023arxiv}%
  \BibitemOpen
  \bibfield  {author} {\bibinfo {author} {\bibfnamefont {V.}~\bibnamefont {Bharti}}, \bibinfo {author} {\bibfnamefont {S.}~\bibnamefont {Sugawa}}, \bibinfo {author} {\bibfnamefont {M.}~\bibnamefont {Kunimi}}, \bibinfo {author} {\bibfnamefont {V.~S.}\ \bibnamefont {Chauhan}}, \bibinfo {author} {\bibfnamefont {T.~P.}\ \bibnamefont {Mahesh}}, \bibinfo {author} {\bibfnamefont {M.}~\bibnamefont {Mizoguchi}}, \bibinfo {author} {\bibfnamefont {T.}~\bibnamefont {Matsubara}}, \bibinfo {author} {\bibfnamefont {T.}~\bibnamefont {Tomita}}, \bibinfo {author} {\bibfnamefont {S.}~\bibnamefont {de~L\'es\'eleuc}},\ and\ \bibinfo {author} {\bibfnamefont {K.}~\bibnamefont {Ohmori}},\ }\bibfield  {title} {\bibinfo {title} {Strong spin-motion coupling in the ultrafast dynamics of rydberg atoms},\ }\href {https://doi.org/10.1103/PhysRevLett.133.093405} {\bibfield  {journal} {\bibinfo  {journal} {Phys. Rev. Lett.}\ }\textbf {\bibinfo {volume} {133}},\ \bibinfo {pages} {093405} (\bibinfo {year} {2024})}\BibitemShut {NoStop}%
\bibitem [{\citenamefont {Bluvstein}\ \emph {et~al.}(2022)\citenamefont {Bluvstein}, \citenamefont {Levine}, \citenamefont {Semeghini}, \citenamefont {Wang}, \citenamefont {Ebadi}, \citenamefont {Kalinowski}, \citenamefont {Keesling}, \citenamefont {Maskara}, \citenamefont {Pichler}, \citenamefont {Greiner}, \citenamefont {Vuleti{\'c}},\ and\ \citenamefont {Lukin}}]{QuEra2022}%
  \BibitemOpen
  \bibfield  {author} {\bibinfo {author} {\bibfnamefont {D.}~\bibnamefont {Bluvstein}}, \bibinfo {author} {\bibfnamefont {H.}~\bibnamefont {Levine}}, \bibinfo {author} {\bibfnamefont {G.}~\bibnamefont {Semeghini}}, \bibinfo {author} {\bibfnamefont {T.~T.}\ \bibnamefont {Wang}}, \bibinfo {author} {\bibfnamefont {S.}~\bibnamefont {Ebadi}}, \bibinfo {author} {\bibfnamefont {M.}~\bibnamefont {Kalinowski}}, \bibinfo {author} {\bibfnamefont {A.}~\bibnamefont {Keesling}}, \bibinfo {author} {\bibfnamefont {N.}~\bibnamefont {Maskara}}, \bibinfo {author} {\bibfnamefont {H.}~\bibnamefont {Pichler}}, \bibinfo {author} {\bibfnamefont {M.}~\bibnamefont {Greiner}}, \bibinfo {author} {\bibfnamefont {V.}~\bibnamefont {Vuleti{\'c}}},\ and\ \bibinfo {author} {\bibfnamefont {M.~D.}\ \bibnamefont {Lukin}},\ }\bibfield  {title} {\bibinfo {title} {A quantum processor based on coherent transport of entangled atom arrays},\ }\href {https://doi.org/10.1038/s41586-022-04592-6} {\bibfield  {journal} {\bibinfo  {journal} {Nature}\ }\textbf {\bibinfo {volume} {604}},\ \bibinfo {pages} {451} (\bibinfo {year} {2022})}\BibitemShut {NoStop}%
\bibitem [{\citenamefont {Bluvstein}\ \emph {et~al.}(2023)\citenamefont {Bluvstein}, \citenamefont {Evered}, \citenamefont {Geim}, \citenamefont {Li}, \citenamefont {Zhou}, \citenamefont {Manovitz}, \citenamefont {Ebadi}, \citenamefont {Cain}, \citenamefont {Kalinowski}, \citenamefont {Hangleiter}, \citenamefont {Ataides}, \citenamefont {Maskara}, \citenamefont {Cong}, \citenamefont {Gao}, \citenamefont {Rodriguez}, \citenamefont {Karolyshyn}, \citenamefont {Semeghini}, \citenamefont {Gullans}, \citenamefont {Greiner}, \citenamefont {Vuleti{\'c}},\ and\ \citenamefont {Lukin}}]{QuEra2023}%
  \BibitemOpen
  \bibfield  {author} {\bibinfo {author} {\bibfnamefont {D.}~\bibnamefont {Bluvstein}}, \bibinfo {author} {\bibfnamefont {S.~J.}\ \bibnamefont {Evered}}, \bibinfo {author} {\bibfnamefont {A.~A.}\ \bibnamefont {Geim}}, \bibinfo {author} {\bibfnamefont {S.~H.}\ \bibnamefont {Li}}, \bibinfo {author} {\bibfnamefont {H.}~\bibnamefont {Zhou}}, \bibinfo {author} {\bibfnamefont {T.}~\bibnamefont {Manovitz}}, \bibinfo {author} {\bibfnamefont {S.}~\bibnamefont {Ebadi}}, \bibinfo {author} {\bibfnamefont {M.}~\bibnamefont {Cain}}, \bibinfo {author} {\bibfnamefont {M.}~\bibnamefont {Kalinowski}}, \bibinfo {author} {\bibfnamefont {D.}~\bibnamefont {Hangleiter}}, \bibinfo {author} {\bibfnamefont {J.~P.~B.}\ \bibnamefont {Ataides}}, \bibinfo {author} {\bibfnamefont {N.}~\bibnamefont {Maskara}}, \bibinfo {author} {\bibfnamefont {I.}~\bibnamefont {Cong}}, \bibinfo {author} {\bibfnamefont {X.}~\bibnamefont {Gao}}, \bibinfo {author} {\bibfnamefont {P.~S.}\ \bibnamefont {Rodriguez}}, \bibinfo {author} {\bibfnamefont {T.}~\bibnamefont {Karolyshyn}}, \bibinfo {author} {\bibfnamefont {G.}~\bibnamefont {Semeghini}}, \bibinfo {author} {\bibfnamefont {M.~J.}\ \bibnamefont {Gullans}}, \bibinfo {author} {\bibfnamefont {M.}~\bibnamefont {Greiner}}, \bibinfo {author} {\bibfnamefont {V.}~\bibnamefont {Vuleti{\'c}}},\ and\ \bibinfo {author} {\bibfnamefont {M.~D.}\ \bibnamefont {Lukin}},\ }\bibfield  {title} {\bibinfo {title} {Logical quantum processor based on reconfigurable atom arrays},\ }\href {https://doi.org/10.1038/s41586-023-06927-3} {\bibfield  {journal} {\bibinfo  {journal} {Nature}\ }\textbf {\bibinfo {volume} {626}},\ \bibinfo {pages} {58} (\bibinfo {year} {2023})}\BibitemShut {NoStop}%
\bibitem [{\citenamefont {Endres}\ \emph {et~al.}(2016)\citenamefont {Endres}, \citenamefont {Bernien}, \citenamefont {Keesling}, \citenamefont {Levine}, \citenamefont {Anschuetz}, \citenamefont {Krajenbrink}, \citenamefont {Senko}, \citenamefont {Vuletic}, \citenamefont {Greiner},\ and\ \citenamefont {Lukin}}]{atom-by-atom-assembly1}%
  \BibitemOpen
  \bibfield  {author} {\bibinfo {author} {\bibfnamefont {M.}~\bibnamefont {Endres}}, \bibinfo {author} {\bibfnamefont {H.}~\bibnamefont {Bernien}}, \bibinfo {author} {\bibfnamefont {A.}~\bibnamefont {Keesling}}, \bibinfo {author} {\bibfnamefont {H.}~\bibnamefont {Levine}}, \bibinfo {author} {\bibfnamefont {E.~R.}\ \bibnamefont {Anschuetz}}, \bibinfo {author} {\bibfnamefont {A.}~\bibnamefont {Krajenbrink}}, \bibinfo {author} {\bibfnamefont {C.}~\bibnamefont {Senko}}, \bibinfo {author} {\bibfnamefont {V.}~\bibnamefont {Vuletic}}, \bibinfo {author} {\bibfnamefont {M.}~\bibnamefont {Greiner}},\ and\ \bibinfo {author} {\bibfnamefont {M.~D.}\ \bibnamefont {Lukin}},\ }\bibfield  {title} {\bibinfo {title} {Atom-by-atom assembly of defect-free one-dimensional cold atom arrays},\ }\href {https://doi.org/10.1126/science.aah3752} {\bibfield  {journal} {\bibinfo  {journal} {Science}\ }\textbf {\bibinfo {volume} {354}},\ \bibinfo {pages} {1024} (\bibinfo {year} {2016})}\BibitemShut {NoStop}%
\bibitem [{\citenamefont {Schymik}\ \emph {et~al.}(2020)\citenamefont {Schymik}, \citenamefont {Lienhard}, \citenamefont {Barredo}, \citenamefont {Scholl}, \citenamefont {Williams}, \citenamefont {Browaeys},\ and\ \citenamefont {Lahaye}}]{atom-by-atom-assembly2}%
  \BibitemOpen
  \bibfield  {author} {\bibinfo {author} {\bibfnamefont {K.-N.}\ \bibnamefont {Schymik}}, \bibinfo {author} {\bibfnamefont {V.}~\bibnamefont {Lienhard}}, \bibinfo {author} {\bibfnamefont {D.}~\bibnamefont {Barredo}}, \bibinfo {author} {\bibfnamefont {P.}~\bibnamefont {Scholl}}, \bibinfo {author} {\bibfnamefont {H.}~\bibnamefont {Williams}}, \bibinfo {author} {\bibfnamefont {A.}~\bibnamefont {Browaeys}},\ and\ \bibinfo {author} {\bibfnamefont {T.}~\bibnamefont {Lahaye}},\ }\bibfield  {title} {\bibinfo {title} {Enhanced atom-by-atom assembly of arbitrary tweezer arrays},\ }\href {https://doi.org/10.1103/PhysRevA.102.063107} {\bibfield  {journal} {\bibinfo  {journal} {Phys. Rev. A}\ }\textbf {\bibinfo {volume} {102}},\ \bibinfo {pages} {063107} (\bibinfo {year} {2020})}\BibitemShut {NoStop}%
\bibitem [{\citenamefont {Wilkinson}\ \emph {et~al.}(1996)\citenamefont {Wilkinson}, \citenamefont {Bharucha}, \citenamefont {Madison}, \citenamefont {Niu},\ and\ \citenamefont {Raizen}}]{Atomic-Wannier-Stark-Ladders}%
  \BibitemOpen
  \bibfield  {author} {\bibinfo {author} {\bibfnamefont {S.~R.}\ \bibnamefont {Wilkinson}}, \bibinfo {author} {\bibfnamefont {C.~F.}\ \bibnamefont {Bharucha}}, \bibinfo {author} {\bibfnamefont {K.~W.}\ \bibnamefont {Madison}}, \bibinfo {author} {\bibfnamefont {Q.}~\bibnamefont {Niu}},\ and\ \bibinfo {author} {\bibfnamefont {M.~G.}\ \bibnamefont {Raizen}},\ }\bibfield  {title} {\bibinfo {title} {Observation of atomic wannier-stark ladders in an accelerating optical potential},\ }\href {https://doi.org/10.1103/PhysRevLett.76.4512} {\bibfield  {journal} {\bibinfo  {journal} {Phys. Rev. Lett.}\ }\textbf {\bibinfo {volume} {76}},\ \bibinfo {pages} {4512} (\bibinfo {year} {1996})}\BibitemShut {NoStop}%
\bibitem [{\citenamefont {Cirac}\ and\ \citenamefont {Zoller}(1995)}]{Cirac1995}%
  \BibitemOpen
  \bibfield  {author} {\bibinfo {author} {\bibfnamefont {J.~I.}\ \bibnamefont {Cirac}}\ and\ \bibinfo {author} {\bibfnamefont {P.}~\bibnamefont {Zoller}},\ }\bibfield  {title} {\bibinfo {title} {Quantum computations with cold trapped ions},\ }\href {https://doi.org/10.1103/PhysRevLett.74.4091} {\bibfield  {journal} {\bibinfo  {journal} {Phys. Rev. Lett.}\ }\textbf {\bibinfo {volume} {74}},\ \bibinfo {pages} {4091} (\bibinfo {year} {1995})}\BibitemShut {NoStop}%
\bibitem [{\citenamefont {Bowler}\ \emph {et~al.}(2012)\citenamefont {Bowler}, \citenamefont {Gaebler}, \citenamefont {Lin}, \citenamefont {Tan}, \citenamefont {Hanneke}, \citenamefont {Jost}, \citenamefont {Home}, \citenamefont {Leibfried},\ and\ \citenamefont {Wineland}}]{ion-trap-Array1}%
  \BibitemOpen
  \bibfield  {author} {\bibinfo {author} {\bibfnamefont {R.}~\bibnamefont {Bowler}}, \bibinfo {author} {\bibfnamefont {J.}~\bibnamefont {Gaebler}}, \bibinfo {author} {\bibfnamefont {Y.}~\bibnamefont {Lin}}, \bibinfo {author} {\bibfnamefont {T.~R.}\ \bibnamefont {Tan}}, \bibinfo {author} {\bibfnamefont {D.}~\bibnamefont {Hanneke}}, \bibinfo {author} {\bibfnamefont {J.~D.}\ \bibnamefont {Jost}}, \bibinfo {author} {\bibfnamefont {J.~P.}\ \bibnamefont {Home}}, \bibinfo {author} {\bibfnamefont {D.}~\bibnamefont {Leibfried}},\ and\ \bibinfo {author} {\bibfnamefont {D.~J.}\ \bibnamefont {Wineland}},\ }\bibfield  {title} {\bibinfo {title} {Coherent diabatic ion transport and separation in a multizone trap array},\ }\href {https://doi.org/10.1103/PhysRevLett.109.080502} {\bibfield  {journal} {\bibinfo  {journal} {Phys. Rev. Lett.}\ }\textbf {\bibinfo {volume} {109}},\ \bibinfo {pages} {080502} (\bibinfo {year} {2012})}\BibitemShut {NoStop}%
\bibitem [{\citenamefont {Kielpinski}\ \emph {et~al.}(2002)\citenamefont {Kielpinski}, \citenamefont {Monroe},\ and\ \citenamefont {Wineland}}]{ion-trap-Array2}%
  \BibitemOpen
  \bibfield  {author} {\bibinfo {author} {\bibfnamefont {D.}~\bibnamefont {Kielpinski}}, \bibinfo {author} {\bibfnamefont {C.}~\bibnamefont {Monroe}},\ and\ \bibinfo {author} {\bibfnamefont {D.~J.}\ \bibnamefont {Wineland}},\ }\bibfield  {title} {\bibinfo {title} {Architecture for a large-scale ion-trap quantum computer},\ }\href {https://doi.org/10.1038/nature00784} {\bibfield  {journal} {\bibinfo  {journal} {Nature}\ }\textbf {\bibinfo {volume} {417}},\ \bibinfo {pages} {709} (\bibinfo {year} {2002})}\BibitemShut {NoStop}%
\bibitem [{\citenamefont {Pino}\ \emph {et~al.}(2021)\citenamefont {Pino}, \citenamefont {Dreiling}, \citenamefont {Figgatt}, \citenamefont {Gaebler}, \citenamefont {Moses}, \citenamefont {Allman}, \citenamefont {Baldwin}, \citenamefont {{Foss-Feig}}, \citenamefont {Hayes}, \citenamefont {Mayer}, \citenamefont {{Ryan-Anderson}},\ and\ \citenamefont {Neyenhuis}}]{Quantinuum2021}%
  \BibitemOpen
  \bibfield  {author} {\bibinfo {author} {\bibfnamefont {J.~M.}\ \bibnamefont {Pino}}, \bibinfo {author} {\bibfnamefont {J.~M.}\ \bibnamefont {Dreiling}}, \bibinfo {author} {\bibfnamefont {C.}~\bibnamefont {Figgatt}}, \bibinfo {author} {\bibfnamefont {J.~P.}\ \bibnamefont {Gaebler}}, \bibinfo {author} {\bibfnamefont {S.~A.}\ \bibnamefont {Moses}}, \bibinfo {author} {\bibfnamefont {M.~S.}\ \bibnamefont {Allman}}, \bibinfo {author} {\bibfnamefont {C.~H.}\ \bibnamefont {Baldwin}}, \bibinfo {author} {\bibfnamefont {M.}~\bibnamefont {{Foss-Feig}}}, \bibinfo {author} {\bibfnamefont {D.}~\bibnamefont {Hayes}}, \bibinfo {author} {\bibfnamefont {K.}~\bibnamefont {Mayer}}, \bibinfo {author} {\bibfnamefont {C.}~\bibnamefont {{Ryan-Anderson}}},\ and\ \bibinfo {author} {\bibfnamefont {B.}~\bibnamefont {Neyenhuis}},\ }\bibfield  {title} {\bibinfo {title} {Demonstration of the trapped-ion quantum {{CCD}} computer architecture},\ }\href {https://doi.org/10.1038/s41586-021-03318-4} {\bibfield  {journal} {\bibinfo  {journal} {Nature}\ }\textbf {\bibinfo {volume} {592}},\ \bibinfo {pages} {209} (\bibinfo {year} {2021})}\BibitemShut {NoStop}%
\bibitem [{\citenamefont {Moses}\ \emph {et~al.}(2023)\citenamefont {Moses}, \citenamefont {Baldwin}, \citenamefont {Allman}, \citenamefont {Ancona}, \citenamefont {Ascarrunz}, \citenamefont {Barnes}, \citenamefont {Bartolotta}, \citenamefont {Bjork}, \citenamefont {Blanchard}, \citenamefont {Bohn}, \citenamefont {Bohnet}, \citenamefont {Brown}, \citenamefont {Burdick}, \citenamefont {Burton}, \citenamefont {Campbell}, \citenamefont {Campora}, \citenamefont {Carron}, \citenamefont {Chambers}, \citenamefont {Chan}, \citenamefont {Chen}, \citenamefont {Chernoguzov}, \citenamefont {Chertkov}, \citenamefont {Colina}, \citenamefont {Curtis}, \citenamefont {Daniel}, \citenamefont {DeCross}, \citenamefont {Deen}, \citenamefont {Delaney}, \citenamefont {Dreiling}, \citenamefont {Ertsgaard}, \citenamefont {Esposito}, \citenamefont {Estey}, \citenamefont {Fabrikant}, \citenamefont {Figgatt}, \citenamefont {Foltz}, \citenamefont {Foss-Feig}, \citenamefont {Francois}, \citenamefont {Gaebler}, \citenamefont {Gatterman}, \citenamefont {Gilbreth}, \citenamefont {Giles}, \citenamefont {Glynn}, \citenamefont {Hall}, \citenamefont {Hankin}, \citenamefont {Hansen}, \citenamefont {Hayes}, \citenamefont {Higashi}, \citenamefont {Hoffman}, \citenamefont {Horning}, \citenamefont {Hout}, \citenamefont {Jacobs}, \citenamefont {Johansen}, \citenamefont {Jones}, \citenamefont {Karcz}, \citenamefont {Klein}, \citenamefont {Lauria}, \citenamefont {Lee}, \citenamefont {Liefer}, \citenamefont {Lu}, \citenamefont {Lucchetti}, \citenamefont {Lytle}, \citenamefont {Malm}, \citenamefont {Matheny}, \citenamefont {Mathewson}, \citenamefont {Mayer}, \citenamefont {Miller}, \citenamefont {Mills}, \citenamefont {Neyenhuis}, \citenamefont {Nugent}, \citenamefont {Olson}, \citenamefont {Parks}, \citenamefont {Price}, \citenamefont {Price}, \citenamefont {Pugh}, \citenamefont {Ransford}, \citenamefont {Reed}, \citenamefont {Roman}, \citenamefont {Rowe}, \citenamefont {Ryan-Anderson}, \citenamefont {Sanders}, \citenamefont {Sedlacek}, \citenamefont {Shevchuk}, \citenamefont {Siegfried}, \citenamefont {Skripka}, \citenamefont {Spaun}, \citenamefont {Sprenkle}, \citenamefont {Stutz}, \citenamefont {Swallows}, \citenamefont {Tobey}, \citenamefont {Tran}, \citenamefont {Tran}, \citenamefont {Vogt}, \citenamefont {Volin}, \citenamefont {Walker}, \citenamefont {Zolot},\ and\ \citenamefont {Pino}}]{Quantinuum2023}%
  \BibitemOpen
  \bibfield  {author} {\bibinfo {author} {\bibfnamefont {S.~A.}\ \bibnamefont {Moses}}, \bibinfo {author} {\bibfnamefont {C.~H.}\ \bibnamefont {Baldwin}}, \bibinfo {author} {\bibfnamefont {M.~S.}\ \bibnamefont {Allman}}, \bibinfo {author} {\bibfnamefont {R.}~\bibnamefont {Ancona}}, \bibinfo {author} {\bibfnamefont {L.}~\bibnamefont {Ascarrunz}}, \bibinfo {author} {\bibfnamefont {C.}~\bibnamefont {Barnes}}, \bibinfo {author} {\bibfnamefont {J.}~\bibnamefont {Bartolotta}}, \bibinfo {author} {\bibfnamefont {B.}~\bibnamefont {Bjork}}, \bibinfo {author} {\bibfnamefont {P.}~\bibnamefont {Blanchard}}, \bibinfo {author} {\bibfnamefont {M.}~\bibnamefont {Bohn}}, \bibinfo {author} {\bibfnamefont {J.~G.}\ \bibnamefont {Bohnet}}, \bibinfo {author} {\bibfnamefont {N.~C.}\ \bibnamefont {Brown}}, \bibinfo {author} {\bibfnamefont {N.~Q.}\ \bibnamefont {Burdick}}, \bibinfo {author} {\bibfnamefont {W.~C.}\ \bibnamefont {Burton}}, \bibinfo {author} {\bibfnamefont {S.~L.}\ \bibnamefont {Campbell}}, \bibinfo {author} {\bibfnamefont {J.~P.}\ \bibnamefont {Campora}}, \bibinfo {author} {\bibfnamefont {C.}~\bibnamefont {Carron}}, \bibinfo {author} {\bibfnamefont {J.}~\bibnamefont {Chambers}}, \bibinfo {author} {\bibfnamefont {J.~W.}\ \bibnamefont {Chan}}, \bibinfo {author} {\bibfnamefont {Y.~H.}\ \bibnamefont {Chen}}, \bibinfo {author} {\bibfnamefont {A.}~\bibnamefont {Chernoguzov}}, \bibinfo {author} {\bibfnamefont {E.}~\bibnamefont {Chertkov}}, \bibinfo {author} {\bibfnamefont {J.}~\bibnamefont {Colina}}, \bibinfo {author} {\bibfnamefont {J.~P.}\ \bibnamefont {Curtis}}, \bibinfo {author} {\bibfnamefont {R.}~\bibnamefont {Daniel}}, \bibinfo {author} {\bibfnamefont {M.}~\bibnamefont {DeCross}}, \bibinfo {author} {\bibfnamefont {D.}~\bibnamefont {Deen}}, \bibinfo {author} {\bibfnamefont {C.}~\bibnamefont {Delaney}}, \bibinfo {author} {\bibfnamefont {J.~M.}\ \bibnamefont {Dreiling}}, \bibinfo {author} {\bibfnamefont {C.~T.}\ \bibnamefont {Ertsgaard}}, \bibinfo {author} {\bibfnamefont {J.}~\bibnamefont {Esposito}}, \bibinfo {author} {\bibfnamefont {B.}~\bibnamefont {Estey}}, \bibinfo {author} {\bibfnamefont {M.}~\bibnamefont {Fabrikant}}, \bibinfo {author} {\bibfnamefont {C.}~\bibnamefont {Figgatt}}, \bibinfo {author} {\bibfnamefont {C.}~\bibnamefont {Foltz}}, \bibinfo {author} {\bibfnamefont {M.}~\bibnamefont {Foss-Feig}}, \bibinfo {author} {\bibfnamefont {D.}~\bibnamefont {Francois}}, \bibinfo {author} {\bibfnamefont {J.~P.}\ \bibnamefont {Gaebler}}, \bibinfo {author} {\bibfnamefont {T.~M.}\ \bibnamefont {Gatterman}}, \bibinfo {author} {\bibfnamefont {C.~N.}\ \bibnamefont {Gilbreth}}, \bibinfo {author} {\bibfnamefont {J.}~\bibnamefont {Giles}}, \bibinfo {author} {\bibfnamefont {E.}~\bibnamefont {Glynn}}, \bibinfo {author} {\bibfnamefont {A.}~\bibnamefont {Hall}}, \bibinfo {author} {\bibfnamefont {A.~M.}\ \bibnamefont {Hankin}}, \bibinfo {author} {\bibfnamefont {A.}~\bibnamefont {Hansen}}, \bibinfo {author} {\bibfnamefont {D.}~\bibnamefont {Hayes}}, \bibinfo {author} {\bibfnamefont {B.}~\bibnamefont {Higashi}}, \bibinfo {author} {\bibfnamefont {I.~M.}\ \bibnamefont {Hoffman}}, \bibinfo {author} {\bibfnamefont {B.}~\bibnamefont {Horning}}, \bibinfo {author} {\bibfnamefont {J.~J.}\ \bibnamefont {Hout}}, \bibinfo {author} {\bibfnamefont {R.}~\bibnamefont {Jacobs}}, \bibinfo {author} {\bibfnamefont {J.}~\bibnamefont {Johansen}}, \bibinfo {author} {\bibfnamefont {L.}~\bibnamefont {Jones}}, \bibinfo {author} {\bibfnamefont {J.}~\bibnamefont {Karcz}}, \bibinfo {author} {\bibfnamefont {T.}~\bibnamefont {Klein}}, \bibinfo {author} {\bibfnamefont {P.}~\bibnamefont {Lauria}}, \bibinfo {author} {\bibfnamefont {P.}~\bibnamefont {Lee}}, \bibinfo {author} {\bibfnamefont {D.}~\bibnamefont {Liefer}}, \bibinfo {author} {\bibfnamefont {S.~T.}\ \bibnamefont {Lu}}, \bibinfo {author} {\bibfnamefont {D.}~\bibnamefont {Lucchetti}}, \bibinfo {author} {\bibfnamefont {C.}~\bibnamefont {Lytle}}, \bibinfo {author} {\bibfnamefont {A.}~\bibnamefont {Malm}}, \bibinfo {author} {\bibfnamefont {M.}~\bibnamefont {Matheny}}, \bibinfo {author} {\bibfnamefont {B.}~\bibnamefont {Mathewson}}, \bibinfo {author} {\bibfnamefont {K.}~\bibnamefont {Mayer}}, \bibinfo {author} {\bibfnamefont {D.~B.}\ \bibnamefont {Miller}}, \bibinfo {author} {\bibfnamefont {M.}~\bibnamefont {Mills}}, \bibinfo {author} {\bibfnamefont {B.}~\bibnamefont {Neyenhuis}}, \bibinfo {author} {\bibfnamefont {L.}~\bibnamefont {Nugent}}, \bibinfo {author} {\bibfnamefont {S.}~\bibnamefont {Olson}}, \bibinfo {author} {\bibfnamefont {J.}~\bibnamefont {Parks}}, \bibinfo {author} {\bibfnamefont {G.~N.}\ \bibnamefont {Price}}, \bibinfo {author} {\bibfnamefont {Z.}~\bibnamefont {Price}}, \bibinfo {author} {\bibfnamefont {M.}~\bibnamefont {Pugh}}, \bibinfo {author} {\bibfnamefont {A.}~\bibnamefont {Ransford}}, \bibinfo {author} {\bibfnamefont {A.~P.}\ \bibnamefont {Reed}}, \bibinfo {author} {\bibfnamefont {C.}~\bibnamefont {Roman}}, \bibinfo {author} {\bibfnamefont {M.}~\bibnamefont {Rowe}}, \bibinfo {author} {\bibfnamefont {C.}~\bibnamefont {Ryan-Anderson}}, \bibinfo {author} {\bibfnamefont {S.}~\bibnamefont {Sanders}}, \bibinfo {author} {\bibfnamefont {J.}~\bibnamefont {Sedlacek}}, \bibinfo {author} {\bibfnamefont {P.}~\bibnamefont {Shevchuk}}, \bibinfo {author} {\bibfnamefont {P.}~\bibnamefont {Siegfried}}, \bibinfo {author} {\bibfnamefont {T.}~\bibnamefont {Skripka}}, \bibinfo {author} {\bibfnamefont {B.}~\bibnamefont {Spaun}}, \bibinfo {author} {\bibfnamefont {R.~T.}\ \bibnamefont {Sprenkle}}, \bibinfo {author} {\bibfnamefont {R.~P.}\ \bibnamefont {Stutz}}, \bibinfo {author} {\bibfnamefont {M.}~\bibnamefont {Swallows}}, \bibinfo {author} {\bibfnamefont {R.~I.}\ \bibnamefont {Tobey}}, \bibinfo {author} {\bibfnamefont {A.}~\bibnamefont {Tran}}, \bibinfo {author} {\bibfnamefont {T.}~\bibnamefont {Tran}}, \bibinfo {author} {\bibfnamefont {E.}~\bibnamefont {Vogt}}, \bibinfo {author} {\bibfnamefont {C.}~\bibnamefont {Volin}}, \bibinfo {author} {\bibfnamefont {J.}~\bibnamefont {Walker}}, \bibinfo {author} {\bibfnamefont {A.~M.}\ \bibnamefont {Zolot}},\ and\ \bibinfo {author} {\bibfnamefont {J.~M.}\ \bibnamefont {Pino}},\ }\bibfield  {title} {\bibinfo {title} {{A Race-Track Trapped-Ion Quantum Processor}},\ }\href {https://doi.org/10.1103/PhysRevX.13.041052} {\bibfield  {journal} {\bibinfo  {journal} {Phys. Rev. X}\ }\textbf {\bibinfo {volume} {13}},\ \bibinfo {pages} {041052} (\bibinfo {year} {2023})}\BibitemShut {NoStop}%
\bibitem [{\citenamefont {Hermelin}\ \emph {et~al.}(2011{\natexlab{a}})\citenamefont {Hermelin}, \citenamefont {Takada}, \citenamefont {Yamamoto}, \citenamefont {Tarucha}, \citenamefont {Wieck}, \citenamefont {Saminadayar}, \citenamefont {B{\"a}uerle},\ and\ \citenamefont {Meunier}}]{Tarucha2011}%
  \BibitemOpen
  \bibfield  {author} {\bibinfo {author} {\bibfnamefont {S.}~\bibnamefont {Hermelin}}, \bibinfo {author} {\bibfnamefont {S.}~\bibnamefont {Takada}}, \bibinfo {author} {\bibfnamefont {M.}~\bibnamefont {Yamamoto}}, \bibinfo {author} {\bibfnamefont {S.}~\bibnamefont {Tarucha}}, \bibinfo {author} {\bibfnamefont {A.~D.}\ \bibnamefont {Wieck}}, \bibinfo {author} {\bibfnamefont {L.}~\bibnamefont {Saminadayar}}, \bibinfo {author} {\bibfnamefont {C.}~\bibnamefont {B{\"a}uerle}},\ and\ \bibinfo {author} {\bibfnamefont {T.}~\bibnamefont {Meunier}},\ }\bibfield  {title} {\bibinfo {title} {{Electrons Surfing on a Sound Wave as a Platform for Quantum Optics with Flying Electrons}},\ }\href {https://doi.org/10.1038/nature10416} {\bibfield  {journal} {\bibinfo  {journal} {Nature}\ }\textbf {\bibinfo {volume} {477}},\ \bibinfo {pages} {435} (\bibinfo {year} {2011}{\natexlab{a}})}\BibitemShut {NoStop}%
\bibitem [{\citenamefont {Byeon}\ \emph {et~al.}(2021)\citenamefont {Byeon}, \citenamefont {Nasyedkin}, \citenamefont {Lane}, \citenamefont {Beysengulov}, \citenamefont {Zhang}, \citenamefont {Loloee},\ and\ \citenamefont {Pollanen}}]{Byeon2021}%
  \BibitemOpen
  \bibfield  {author} {\bibinfo {author} {\bibfnamefont {H.}~\bibnamefont {Byeon}}, \bibinfo {author} {\bibfnamefont {K.}~\bibnamefont {Nasyedkin}}, \bibinfo {author} {\bibfnamefont {J.~R.}\ \bibnamefont {Lane}}, \bibinfo {author} {\bibfnamefont {N.~R.}\ \bibnamefont {Beysengulov}}, \bibinfo {author} {\bibfnamefont {L.}~\bibnamefont {Zhang}}, \bibinfo {author} {\bibfnamefont {R.}~\bibnamefont {Loloee}},\ and\ \bibinfo {author} {\bibfnamefont {J.}~\bibnamefont {Pollanen}},\ }\bibfield  {title} {\bibinfo {title} {Piezoacoustics for precision control of electrons floating on helium},\ }\href {https://doi.org/10.1038/s41467-021-24452-7} {\bibfield  {journal} {\bibinfo  {journal} {Nat. Commun.}\ }\textbf {\bibinfo {volume} {12}},\ \bibinfo {pages} {4150} (\bibinfo {year} {2021})}\BibitemShut {NoStop}%
\bibitem [{\citenamefont {Edlbauer}\ \emph {et~al.}(2021)\citenamefont {Edlbauer}, \citenamefont {Wang}, \citenamefont {Ota}, \citenamefont {Richard}, \citenamefont {Jadot}, \citenamefont {Mortemousque}, \citenamefont {Okazaki}, \citenamefont {Nakamura}, \citenamefont {Kodera}, \citenamefont {Kaneko}, \citenamefont {Ludwig}, \citenamefont {Wieck}, \citenamefont {Urdampilleta}, \citenamefont {Meunier}, \citenamefont {Bäuerle},\ and\ \citenamefont {Takada}}]{APL119-2021}%
  \BibitemOpen
  \bibfield  {author} {\bibinfo {author} {\bibfnamefont {H.}~\bibnamefont {Edlbauer}}, \bibinfo {author} {\bibfnamefont {J.}~\bibnamefont {Wang}}, \bibinfo {author} {\bibfnamefont {S.}~\bibnamefont {Ota}}, \bibinfo {author} {\bibfnamefont {A.}~\bibnamefont {Richard}}, \bibinfo {author} {\bibfnamefont {B.}~\bibnamefont {Jadot}}, \bibinfo {author} {\bibfnamefont {P.-A.}\ \bibnamefont {Mortemousque}}, \bibinfo {author} {\bibfnamefont {Y.}~\bibnamefont {Okazaki}}, \bibinfo {author} {\bibfnamefont {S.}~\bibnamefont {Nakamura}}, \bibinfo {author} {\bibfnamefont {T.}~\bibnamefont {Kodera}}, \bibinfo {author} {\bibfnamefont {N.-H.}\ \bibnamefont {Kaneko}}, \bibinfo {author} {\bibfnamefont {A.}~\bibnamefont {Ludwig}}, \bibinfo {author} {\bibfnamefont {A.~D.}\ \bibnamefont {Wieck}}, \bibinfo {author} {\bibfnamefont {M.}~\bibnamefont {Urdampilleta}}, \bibinfo {author} {\bibfnamefont {T.}~\bibnamefont {Meunier}}, \bibinfo {author} {\bibfnamefont {C.}~\bibnamefont {Bäuerle}},\ and\ \bibinfo {author} {\bibfnamefont {S.}~\bibnamefont {Takada}},\ }\bibfield  {title} {\bibinfo {title} {{In-flight distribution of an electron within a surface acoustic wave}},\ }\href {https://doi.org/10.1063/5.0062491} {\bibfield  {journal} {\bibinfo  {journal} {Appl. Phys. Lett.}\ }\textbf {\bibinfo {volume} {119}},\ \bibinfo {pages} {114004} (\bibinfo {year} {2021})}\BibitemShut {NoStop}%
\bibitem [{\citenamefont {Fujita}\ \emph {et~al.}(2017)\citenamefont {Fujita}, \citenamefont {Baart}, \citenamefont {Reichl}, \citenamefont {Wegscheider},\ and\ \citenamefont {Vandersypen}}]{Fujita2017coherent}%
  \BibitemOpen
  \bibfield  {author} {\bibinfo {author} {\bibfnamefont {T.}~\bibnamefont {Fujita}}, \bibinfo {author} {\bibfnamefont {T.~A.}\ \bibnamefont {Baart}}, \bibinfo {author} {\bibfnamefont {C.}~\bibnamefont {Reichl}}, \bibinfo {author} {\bibfnamefont {W.}~\bibnamefont {Wegscheider}},\ and\ \bibinfo {author} {\bibfnamefont {L.~M.~K.}\ \bibnamefont {Vandersypen}},\ }\bibfield  {title} {\bibinfo {title} {Coherent shuttle of electron-spin states},\ }\href {https://doi.org/10.1038/s41534-017-0024-4} {\bibfield  {journal} {\bibinfo  {journal} {npj Quantum Information}\ }\textbf {\bibinfo {volume} {3}},\ \bibinfo {pages} {22} (\bibinfo {year} {2017})}\BibitemShut {NoStop}%
\bibitem [{\citenamefont {Mills}\ \emph {et~al.}(2019)\citenamefont {Mills}, \citenamefont {Zajac}, \citenamefont {Gullans}, \citenamefont {Schupp}, \citenamefont {Hazard},\ and\ \citenamefont {Petta}}]{Mills2019shuttling}%
  \BibitemOpen
  \bibfield  {author} {\bibinfo {author} {\bibfnamefont {A.~R.}\ \bibnamefont {Mills}}, \bibinfo {author} {\bibfnamefont {D.~M.}\ \bibnamefont {Zajac}}, \bibinfo {author} {\bibfnamefont {M.~J.}\ \bibnamefont {Gullans}}, \bibinfo {author} {\bibfnamefont {F.~J.}\ \bibnamefont {Schupp}}, \bibinfo {author} {\bibfnamefont {T.~M.}\ \bibnamefont {Hazard}},\ and\ \bibinfo {author} {\bibfnamefont {J.~R.}\ \bibnamefont {Petta}},\ }\bibfield  {title} {\bibinfo {title} {Shuttling a single charge across a one-dimensional array of silicon quantum dots},\ }\href {https://doi.org/10.1038/s41467-019-08970-z} {\bibfield  {journal} {\bibinfo  {journal} {Nat. Commun.}\ }\textbf {\bibinfo {volume} {10}},\ \bibinfo {pages} {1063} (\bibinfo {year} {2019})}\BibitemShut {NoStop}%
\bibitem [{\citenamefont {Yoneda}\ \emph {et~al.}(2021)\citenamefont {Yoneda}, \citenamefont {Huang}, \citenamefont {Feng}, \citenamefont {Yang}, \citenamefont {Chan}, \citenamefont {Tanttu}, \citenamefont {Gilbert}, \citenamefont {Leon}, \citenamefont {Hudson}, \citenamefont {Itoh}, \citenamefont {Morello}, \citenamefont {Bartlett}, \citenamefont {Laucht}, \citenamefont {Saraiva},\ and\ \citenamefont {Dzurak}}]{Yoneda2021coherent}%
  \BibitemOpen
  \bibfield  {author} {\bibinfo {author} {\bibfnamefont {J.}~\bibnamefont {Yoneda}}, \bibinfo {author} {\bibfnamefont {W.}~\bibnamefont {Huang}}, \bibinfo {author} {\bibfnamefont {M.}~\bibnamefont {Feng}}, \bibinfo {author} {\bibfnamefont {C.~H.}\ \bibnamefont {Yang}}, \bibinfo {author} {\bibfnamefont {K.~W.}\ \bibnamefont {Chan}}, \bibinfo {author} {\bibfnamefont {T.}~\bibnamefont {Tanttu}}, \bibinfo {author} {\bibfnamefont {W.}~\bibnamefont {Gilbert}}, \bibinfo {author} {\bibfnamefont {R.~C.~C.}\ \bibnamefont {Leon}}, \bibinfo {author} {\bibfnamefont {F.~E.}\ \bibnamefont {Hudson}}, \bibinfo {author} {\bibfnamefont {K.~M.}\ \bibnamefont {Itoh}}, \bibinfo {author} {\bibfnamefont {A.}~\bibnamefont {Morello}}, \bibinfo {author} {\bibfnamefont {S.~D.}\ \bibnamefont {Bartlett}}, \bibinfo {author} {\bibfnamefont {A.}~\bibnamefont {Laucht}}, \bibinfo {author} {\bibfnamefont {A.}~\bibnamefont {Saraiva}},\ and\ \bibinfo {author} {\bibfnamefont {A.~S.}\ \bibnamefont {Dzurak}},\ }\bibfield  {title} {\bibinfo {title} {Coherent spin qubit transport in silicon},\ }\href {https://doi.org/10.1038/s41467-021-24371-7} {\bibfield  {journal} {\bibinfo  {journal} {Nat. Commun.}\ }\textbf {\bibinfo {volume} {12}},\ \bibinfo {pages} {4114} (\bibinfo {year} {2021})}\BibitemShut {NoStop}%
\bibitem [{\citenamefont {Noiri}\ \emph {et~al.}(2022)\citenamefont {Noiri}, \citenamefont {Takeda}, \citenamefont {Nakajima}, \citenamefont {Kobayashi}, \citenamefont {Sammak}, \citenamefont {Scappucci},\ and\ \citenamefont {Tarucha}}]{Tarucha2022}%
  \BibitemOpen
  \bibfield  {author} {\bibinfo {author} {\bibfnamefont {A.}~\bibnamefont {Noiri}}, \bibinfo {author} {\bibfnamefont {K.}~\bibnamefont {Takeda}}, \bibinfo {author} {\bibfnamefont {T.}~\bibnamefont {Nakajima}}, \bibinfo {author} {\bibfnamefont {T.}~\bibnamefont {Kobayashi}}, \bibinfo {author} {\bibfnamefont {A.}~\bibnamefont {Sammak}}, \bibinfo {author} {\bibfnamefont {G.}~\bibnamefont {Scappucci}},\ and\ \bibinfo {author} {\bibfnamefont {S.}~\bibnamefont {Tarucha}},\ }\bibfield  {title} {\bibinfo {title} {A shuttling-based two-qubit logic gate for linking distant silicon quantum processors},\ }\href {https://doi.org/10.1038/s41467-022-33453-z} {\bibfield  {journal} {\bibinfo  {journal} {Nat. Commun.}\ }\textbf {\bibinfo {volume} {13}},\ \bibinfo {pages} {5740} (\bibinfo {year} {2022})}\BibitemShut {NoStop}%
\bibitem [{\citenamefont {Freedman}\ \emph {et~al.}(2024)\citenamefont {Freedman}, \citenamefont {Storey}, \citenamefont {Daniel~Dominguez}, \citenamefont {Magri}, \citenamefont {Otterstrom},\ and\ \citenamefont {Eichenfield}}]{quantum-dots}%
  \BibitemOpen
  \bibfield  {author} {\bibinfo {author} {\bibfnamefont {J.~M.}\ \bibnamefont {Freedman}}, \bibinfo {author} {\bibfnamefont {M.~J.}\ \bibnamefont {Storey}}, \bibinfo {author} {\bibfnamefont {A.}~\bibnamefont {Daniel~Dominguez}, \bibfnamefont {Daniel~Leenheer}}, \bibinfo {author} {\bibfnamefont {S.}~\bibnamefont {Magri}}, \bibinfo {author} {\bibfnamefont {N.~T.}\ \bibnamefont {Otterstrom}},\ and\ \bibinfo {author} {\bibfnamefont {M.}~\bibnamefont {Eichenfield}},\ }\bibfield  {title} {\bibinfo {title} {Coherent spin qubit shuttling through germanium quantum dots},\ }\href {https://doi.org/https://doi.org/10.1038/s41467-024-49358-y} {\bibfield  {journal} {\bibinfo  {journal} {Nat. Commun.}\ }\textbf {\bibinfo {volume} {15}},\ \bibinfo {pages} {5716} (\bibinfo {year} {2024})}\BibitemShut {NoStop}%
\bibitem [{\citenamefont {Hermelin}\ \emph {et~al.}(2011{\natexlab{b}})\citenamefont {Hermelin}, \citenamefont {Takada}, \citenamefont {Yamamoto}, \citenamefont {Tarucha}, \citenamefont {Wieck}, \citenamefont {Saminadayar}, \citenamefont {B\"{a}uerle},\ and\ \citenamefont {Meunier}}]{SAW-platform1}%
  \BibitemOpen
  \bibfield  {author} {\bibinfo {author} {\bibfnamefont {S.}~\bibnamefont {Hermelin}}, \bibinfo {author} {\bibfnamefont {S.}~\bibnamefont {Takada}}, \bibinfo {author} {\bibfnamefont {M.}~\bibnamefont {Yamamoto}}, \bibinfo {author} {\bibfnamefont {S.}~\bibnamefont {Tarucha}}, \bibinfo {author} {\bibfnamefont {A.~D.}\ \bibnamefont {Wieck}}, \bibinfo {author} {\bibfnamefont {L.}~\bibnamefont {Saminadayar}}, \bibinfo {author} {\bibfnamefont {C.}~\bibnamefont {B\"{a}uerle}},\ and\ \bibinfo {author} {\bibfnamefont {T.}~\bibnamefont {Meunier}},\ }\bibfield  {title} {\bibinfo {title} {Electrons surfing on a sound wave as a platform for quantum optics with flying electrons},\ }\href {https://doi.org/doi:10.1038/nature10416} {\bibfield  {journal} {\bibinfo  {journal} {Nature}\ }\textbf {\bibinfo {volume} {477}},\ \bibinfo {pages} {435} (\bibinfo {year} {2011}{\natexlab{b}})}\BibitemShut {NoStop}%
\bibitem [{\citenamefont {McNeil}\ \emph {et~al.}(2011)\citenamefont {McNeil}, \citenamefont {M.}, \citenamefont {Ford}, \citenamefont {Barnes}, \citenamefont {~}, \citenamefont {Jones}, \citenamefont {Farrer},\ and\ \citenamefont {Ritchie}}]{SAW-platform2}%
  \BibitemOpen
  \bibfield  {author} {\bibinfo {author} {\bibfnamefont {R.~P.~G.}\ \bibnamefont {McNeil}}, \bibinfo {author} {\bibfnamefont {K.}~\bibnamefont {M.}}, \bibinfo {author} {\bibfnamefont {C.~J.~B.}\ \bibnamefont {Ford}}, \bibinfo {author} {\bibfnamefont {C.~H.~W.}\ \bibnamefont {Barnes}}, \bibinfo {author} {\bibfnamefont {D.}~\bibnamefont {~}, \bibfnamefont {Anderson}}, \bibinfo {author} {\bibfnamefont {G.~A.~C.}\ \bibnamefont {Jones}}, \bibinfo {author} {\bibfnamefont {I.}~\bibnamefont {Farrer}},\ and\ \bibinfo {author} {\bibfnamefont {D.~A.}\ \bibnamefont {Ritchie}},\ }\bibfield  {title} {\bibinfo {title} {On-demand single-electron transfer between distant quantum dots},\ }\href {https://doi.org/doi:10.1038/nature10444} {\bibfield  {journal} {\bibinfo  {journal} {Nature}\ }\textbf {\bibinfo {volume} {477}},\ \bibinfo {pages} {439} (\bibinfo {year} {2011})}\BibitemShut {NoStop}%
\bibitem [{\citenamefont {Takada}\ \emph {et~al.}(2019)\citenamefont {Takada}, \citenamefont {Edlbauer}, \citenamefont {Lepage}, \citenamefont {Wang}, \citenamefont {Mortemousque}, \citenamefont {Georgiou}, \citenamefont {Barnes}, \citenamefont {Ford}, \citenamefont {Yuan}, \citenamefont {Santos}, \citenamefont {Waintal}, \citenamefont {Ludwig}, \citenamefont {Wieck}, \citenamefont {Urdampilleta}, \citenamefont {Tristan},\ and\ \citenamefont {B\"{a}uerle}}]{SAW-platform3}%
  \BibitemOpen
  \bibfield  {author} {\bibinfo {author} {\bibfnamefont {S.}~\bibnamefont {Takada}}, \bibinfo {author} {\bibfnamefont {H.}~\bibnamefont {Edlbauer}}, \bibinfo {author} {\bibfnamefont {H.~V.}\ \bibnamefont {Lepage}}, \bibinfo {author} {\bibfnamefont {J.}~\bibnamefont {Wang}}, \bibinfo {author} {\bibfnamefont {P.-A.}\ \bibnamefont {Mortemousque}}, \bibinfo {author} {\bibfnamefont {G.}~\bibnamefont {Georgiou}}, \bibinfo {author} {\bibfnamefont {C.~H.~W.}\ \bibnamefont {Barnes}}, \bibinfo {author} {\bibfnamefont {C.~J.~B.}\ \bibnamefont {Ford}}, \bibinfo {author} {\bibfnamefont {M.}~\bibnamefont {Yuan}}, \bibinfo {author} {\bibfnamefont {P.~V.}\ \bibnamefont {Santos}}, \bibinfo {author} {\bibfnamefont {X.}~\bibnamefont {Waintal}}, \bibinfo {author} {\bibfnamefont {A.}~\bibnamefont {Ludwig}}, \bibinfo {author} {\bibfnamefont {A.~D.}\ \bibnamefont {Wieck}}, \bibinfo {author} {\bibfnamefont {M.}~\bibnamefont {Urdampilleta}}, \bibinfo {author} {\bibfnamefont {M.}~\bibnamefont {Tristan}},\ and\ \bibinfo {author} {\bibfnamefont {C.}~\bibnamefont {B\"{a}uerle}},\ }\bibfield  {title} {\bibinfo {title} {Sound-driven single-electron transfer in a circuit of coupled quantum rails},\ }\href {https://doi.org/https://doi.org/10.1038/s41467-019-12514-w} {\bibfield  {journal} {\bibinfo  {journal} {Nat. Commun.}\ }\textbf {\bibinfo {volume} {10}},\ \bibinfo {pages} {4557} (\bibinfo {year} {2019})}\BibitemShut {NoStop}%
\bibitem [{\citenamefont {Seidler}\ \emph {et~al.}(2022)\citenamefont {Seidler}, \citenamefont {Struck}, \citenamefont {Xue}, \citenamefont {Focke}, \citenamefont {Trellenkamp}, \citenamefont {Bluhm},\ and\ \citenamefont {Schreiber}}]{silicon1}%
  \BibitemOpen
  \bibfield  {author} {\bibinfo {author} {\bibfnamefont {I.}~\bibnamefont {Seidler}}, \bibinfo {author} {\bibfnamefont {T.}~\bibnamefont {Struck}}, \bibinfo {author} {\bibfnamefont {R.}~\bibnamefont {Xue}}, \bibinfo {author} {\bibfnamefont {N.}~\bibnamefont {Focke}}, \bibinfo {author} {\bibfnamefont {S.}~\bibnamefont {Trellenkamp}}, \bibinfo {author} {\bibfnamefont {H.}~\bibnamefont {Bluhm}},\ and\ \bibinfo {author} {\bibfnamefont {L.~R.}\ \bibnamefont {Schreiber}},\ }\bibfield  {title} {\bibinfo {title} {Conveyor-mode single-electron shuttling in {{Si/SiGe}} for a scalable quantum computing architecture},\ }\href {https://doi.org/https://doi.org/10.1038/s41534-022-00615-2} {\bibfield  {journal} {\bibinfo  {journal} {npj Quantum Inf.}\ }\textbf {\bibinfo {volume} {8}},\ \bibinfo {pages} {100} (\bibinfo {year} {2022})}\BibitemShut {NoStop}%
\bibitem [{\citenamefont {Struck}\ \emph {et~al.}(2004)\citenamefont {Struck}, \citenamefont {Volmer}, \citenamefont {Visser}, \citenamefont {Offermann}, \citenamefont {Xue}, \citenamefont {Tu}, \citenamefont {Trellenkamp}, \citenamefont {Cywi\ifmmode~\acute{n}\else \'{n}\fi{}ski}, \citenamefont {Bluhm},\ and\ \citenamefont {Schreiber}}]{silicon2}%
  \BibitemOpen
  \bibfield  {author} {\bibinfo {author} {\bibfnamefont {T.}~\bibnamefont {Struck}}, \bibinfo {author} {\bibfnamefont {M.}~\bibnamefont {Volmer}}, \bibinfo {author} {\bibfnamefont {L.}~\bibnamefont {Visser}}, \bibinfo {author} {\bibfnamefont {T.}~\bibnamefont {Offermann}}, \bibinfo {author} {\bibfnamefont {R.}~\bibnamefont {Xue}}, \bibinfo {author} {\bibfnamefont {J.-S.}\ \bibnamefont {Tu}}, \bibinfo {author} {\bibfnamefont {S.}~\bibnamefont {Trellenkamp}}, \bibinfo {author} {\bibfnamefont {L.}~\bibnamefont {Cywi\ifmmode~\acute{n}\else \'{n}\fi{}ski}}, \bibinfo {author} {\bibfnamefont {H.}~\bibnamefont {Bluhm}},\ and\ \bibinfo {author} {\bibfnamefont {L.~R.}\ \bibnamefont {Schreiber}},\ }\bibfield  {title} {\bibinfo {title} {Spin-epr-pair separation by conveyor-mode single electron shuttling in {{Si/SiGe}}},\ }\href {https://doi.org/https://doi.org/10.1038/s41467-024-45583-7} {\bibfield  {journal} {\bibinfo  {journal} {Nat. Commun.}\ }\textbf {\bibinfo {volume} {15}},\ \bibinfo {pages} {1325} (\bibinfo {year} {2004})}\BibitemShut {NoStop}%
\bibitem [{\citenamefont {K\"{u}nne}\ \emph {et~al.}(2024)\citenamefont {K\"{u}nne}, \citenamefont {Willmes}, \citenamefont {Oberl\"{a}nder}, \citenamefont {Gorjaew}, \citenamefont {Teske}, \citenamefont {Bhardwaj}, \citenamefont {Beer}, \citenamefont {Kammerloher}, \citenamefont {Otten}, \citenamefont {Seidler}, \citenamefont {Xue}, \citenamefont {Schreiber},\ and\ \citenamefont {Bluhm}}]{silicon3}%
  \BibitemOpen
  \bibfield  {author} {\bibinfo {author} {\bibfnamefont {M.}~\bibnamefont {K\"{u}nne}}, \bibinfo {author} {\bibfnamefont {A.}~\bibnamefont {Willmes}}, \bibinfo {author} {\bibfnamefont {M.}~\bibnamefont {Oberl\"{a}nder}}, \bibinfo {author} {\bibfnamefont {C.}~\bibnamefont {Gorjaew}}, \bibinfo {author} {\bibfnamefont {J.~D.}\ \bibnamefont {Teske}}, \bibinfo {author} {\bibfnamefont {H.}~\bibnamefont {Bhardwaj}}, \bibinfo {author} {\bibfnamefont {M.}~\bibnamefont {Beer}}, \bibinfo {author} {\bibfnamefont {E.}~\bibnamefont {Kammerloher}}, \bibinfo {author} {\bibfnamefont {R.}~\bibnamefont {Otten}}, \bibinfo {author} {\bibfnamefont {I.}~\bibnamefont {Seidler}}, \bibinfo {author} {\bibfnamefont {R.}~\bibnamefont {Xue}}, \bibinfo {author} {\bibfnamefont {L.~R.}\ \bibnamefont {Schreiber}},\ and\ \bibinfo {author} {\bibfnamefont {H.}~\bibnamefont {Bluhm}},\ }\bibfield  {title} {\bibinfo {title} {The spinbus architecture for scaling spin qubits with electron shuttling},\ }\href {https://doi.org/https://doi.org/10.1038/s41467-024-49182-4} {\bibfield  {journal} {\bibinfo  {journal} {Nat. Commun.}\ }\textbf {\bibinfo {volume} {15}},\ \bibinfo {pages} {4977} (\bibinfo {year} {2024})}\BibitemShut {NoStop}%
\bibitem [{\citenamefont {Langrock}\ \emph {et~al.}(2023)\citenamefont {Langrock}, \citenamefont {Krzywda}, \citenamefont {Focke}, \citenamefont {Seidler}, \citenamefont {Schreiber},\ and\ \citenamefont {Cywi\ifmmode~\acute{n}\else \'{n}\fi{}ski}}]{silicon4}%
  \BibitemOpen
  \bibfield  {author} {\bibinfo {author} {\bibfnamefont {V.}~\bibnamefont {Langrock}}, \bibinfo {author} {\bibfnamefont {J.~A.}\ \bibnamefont {Krzywda}}, \bibinfo {author} {\bibfnamefont {N.}~\bibnamefont {Focke}}, \bibinfo {author} {\bibfnamefont {I.}~\bibnamefont {Seidler}}, \bibinfo {author} {\bibfnamefont {L.~R.}\ \bibnamefont {Schreiber}},\ and\ \bibinfo {author} {\bibfnamefont {L.}~\bibnamefont {Cywi\ifmmode~\acute{n}\else \'{n}\fi{}ski}},\ }\bibfield  {title} {\bibinfo {title} {Blueprint of a scalable spin qubit shuttle device for coherent mid-range qubit transfer in disordered {{Si/SiGe/SiO}}$_{2}$},\ }\href {https://doi.org/10.1103/PRXQuantum.4.020305} {\bibfield  {journal} {\bibinfo  {journal} {PRX Quantum}\ }\textbf {\bibinfo {volume} {4}},\ \bibinfo {pages} {020305} (\bibinfo {year} {2023})}\BibitemShut {NoStop}%
\bibitem [{\citenamefont {Ketterle}\ \emph {et~al.}(1999)\citenamefont {Ketterle}, \citenamefont {Durfee},\ and\ \citenamefont {{Stamper-Kurn}}}]{BEC0}%
  \BibitemOpen
  \bibfield  {author} {\bibinfo {author} {\bibfnamefont {W.}~\bibnamefont {Ketterle}}, \bibinfo {author} {\bibfnamefont {D.~S.}\ \bibnamefont {Durfee}},\ and\ \bibinfo {author} {\bibfnamefont {D.~M.}\ \bibnamefont {{Stamper-Kurn}}},\ }\bibinfo {title} {Making, probing and understanding {{Bose-Einstein}} condensates},\ in\ \href {https://doi.org/10.3254/978-1-61499-225-7-67} {\emph {\bibinfo {booktitle} {Bose-{{Einstein Condensation}} in {{Atomic Gases}}}}},\ \bibinfo {editor} {edited by\ \bibinfo {editor} {\bibfnamefont {M.}~\bibnamefont {Inguscio}}, \bibinfo {editor} {\bibfnamefont {S.}~\bibnamefont {Stringari}},\ and\ \bibinfo {editor} {\bibfnamefont {C.~E.}\ \bibnamefont {Wieman}}}\ (\bibinfo  {publisher} {IOS Press},\ \bibinfo {address} {Amsterdam},\ \bibinfo {year} {1999})\ pp.\ \bibinfo {pages} {67--176}\BibitemShut {NoStop}%
\bibitem [{\citenamefont {Ketterle}(2000)}]{BEC1}%
  \BibitemOpen
  \bibfield  {author} {\bibinfo {author} {\bibfnamefont {W.}~\bibnamefont {Ketterle}},\ }\bibfield  {title} {\bibinfo {title} {Bose-einstein condensation in dilute atomic gases: atomic physics meets condensed matter physics},\ }\href {https://doi.org/https://doi.org/10.1016/S0921-4526(99)01413-1} {\bibfield  {journal} {\bibinfo  {journal} {Physica B: Cond. Matt.}\ }\textbf {\bibinfo {volume} {280}},\ \bibinfo {pages} {11} (\bibinfo {year} {2000})}\BibitemShut {NoStop}%
\bibitem [{\citenamefont {Anglin}\ and\ \citenamefont {Ketterle}(2002)}]{BEC2}%
  \BibitemOpen
  \bibfield  {author} {\bibinfo {author} {\bibfnamefont {J.~R.}\ \bibnamefont {Anglin}}\ and\ \bibinfo {author} {\bibfnamefont {W.}~\bibnamefont {Ketterle}},\ }\bibfield  {title} {\bibinfo {title} {Bose--{{Einstein}} condensation of atomic gases},\ }\href {https://doi.org/10.1038/416211a} {\bibfield  {journal} {\bibinfo  {journal} {Nature}\ }\textbf {\bibinfo {volume} {416}},\ \bibinfo {pages} {211} (\bibinfo {year} {2002})}\BibitemShut {NoStop}%
\bibitem [{\citenamefont {H\"ansel}\ \emph {et~al.}(2001)\citenamefont {H\"ansel}, \citenamefont {Reichel}, \citenamefont {Hommelhoff},\ and\ \citenamefont {H\"ansch}}]{mag-conv}%
  \BibitemOpen
  \bibfield  {author} {\bibinfo {author} {\bibfnamefont {W.}~\bibnamefont {H\"ansel}}, \bibinfo {author} {\bibfnamefont {J.}~\bibnamefont {Reichel}}, \bibinfo {author} {\bibfnamefont {P.}~\bibnamefont {Hommelhoff}},\ and\ \bibinfo {author} {\bibfnamefont {T.~W.}\ \bibnamefont {H\"ansch}},\ }\bibfield  {title} {\bibinfo {title} {Magnetic conveyor belt for transporting and merging trapped atom clouds},\ }\href {https://doi.org/10.1103/PhysRevLett.86.608} {\bibfield  {journal} {\bibinfo  {journal} {Phys. Rev. Lett.}\ }\textbf {\bibinfo {volume} {86}},\ \bibinfo {pages} {608} (\bibinfo {year} {2001})}\BibitemShut {NoStop}%
\bibitem [{\citenamefont {Fort\'agh}\ and\ \citenamefont {Zimmermann}(2007)}]{mag-conv-rev}%
  \BibitemOpen
  \bibfield  {author} {\bibinfo {author} {\bibfnamefont {J.}~\bibnamefont {Fort\'agh}}\ and\ \bibinfo {author} {\bibfnamefont {C.}~\bibnamefont {Zimmermann}},\ }\bibfield  {title} {\bibinfo {title} {Magnetic microtraps for ultracold atoms},\ }\href {https://doi.org/10.1103/RevModPhys.79.235} {\bibfield  {journal} {\bibinfo  {journal} {Rev. Mod. Phys.}\ }\textbf {\bibinfo {volume} {79}},\ \bibinfo {pages} {235} (\bibinfo {year} {2007})}\BibitemShut {NoStop}%
\bibitem [{\citenamefont {Cristiani}\ \emph {et~al.}(2002)\citenamefont {Cristiani}, \citenamefont {Morsch}, \citenamefont {M\"uller}, \citenamefont {Ciampini},\ and\ \citenamefont {Arimondo}}]{BEC1D-optical-lattices1}%
  \BibitemOpen
  \bibfield  {author} {\bibinfo {author} {\bibfnamefont {M.}~\bibnamefont {Cristiani}}, \bibinfo {author} {\bibfnamefont {O.}~\bibnamefont {Morsch}}, \bibinfo {author} {\bibfnamefont {J.~H.}\ \bibnamefont {M\"uller}}, \bibinfo {author} {\bibfnamefont {D.}~\bibnamefont {Ciampini}},\ and\ \bibinfo {author} {\bibfnamefont {E.}~\bibnamefont {Arimondo}},\ }\bibfield  {title} {\bibinfo {title} {Experimental properties of {{Bose-Einstein}} condensates in one-dimensional optical lattices: {{Bloch}} oscillations, {{Landau-Zener}} tunneling, and mean-field effects},\ }\href {https://doi.org/10.1103/PhysRevA.65.063612} {\bibfield  {journal} {\bibinfo  {journal} {Phys. Rev. A}\ }\textbf {\bibinfo {volume} {65}},\ \bibinfo {pages} {063612} (\bibinfo {year} {2002})}\BibitemShut {NoStop}%
\bibitem [{\citenamefont {Choi}\ and\ \citenamefont {Niu}(1999)}]{BEC1D-optical-lattices2}%
  \BibitemOpen
  \bibfield  {author} {\bibinfo {author} {\bibfnamefont {D.-I.}\ \bibnamefont {Choi}}\ and\ \bibinfo {author} {\bibfnamefont {Q.}~\bibnamefont {Niu}},\ }\bibfield  {title} {\bibinfo {title} {{{Bose-Einstein}} condensates in an optical lattice},\ }\href {https://doi.org/10.1103/PhysRevLett.82.2022} {\bibfield  {journal} {\bibinfo  {journal} {Phys. Rev. Lett.}\ }\textbf {\bibinfo {volume} {82}},\ \bibinfo {pages} {2022} (\bibinfo {year} {1999})}\BibitemShut {NoStop}%
\bibitem [{\citenamefont {Wu}\ and\ \citenamefont {Niu}(2003)}]{BEC1D-optical-lattices3}%
  \BibitemOpen
  \bibfield  {author} {\bibinfo {author} {\bibfnamefont {B.}~\bibnamefont {Wu}}\ and\ \bibinfo {author} {\bibfnamefont {Q.}~\bibnamefont {Niu}},\ }\bibfield  {title} {\bibinfo {title} {Superfluidity of {{Bose-Einstein}} condensate in an optical lattice: {{Landau-Zener}} tunnelling and dynamical instability},\ }\href {https://doi.org/10.1088/1367-2630/5/1/104} {\bibfield  {journal} {\bibinfo  {journal} {New J. Phys.}\ }\textbf {\bibinfo {volume} {5}},\ \bibinfo {pages} {104} (\bibinfo {year} {2003})}\BibitemShut {NoStop}%
\bibitem [{\citenamefont {Gustavson}\ \emph {et~al.}(2001)\citenamefont {Gustavson}, \citenamefont {Chikkatur}, \citenamefont {Leanhardt}, \citenamefont {G\"orlitz}, \citenamefont {Gupta}, \citenamefont {Pritchard},\ and\ \citenamefont {Ketterle}}]{BEC-optTweezers}%
  \BibitemOpen
  \bibfield  {author} {\bibinfo {author} {\bibfnamefont {T.~L.}\ \bibnamefont {Gustavson}}, \bibinfo {author} {\bibfnamefont {A.~P.}\ \bibnamefont {Chikkatur}}, \bibinfo {author} {\bibfnamefont {A.~E.}\ \bibnamefont {Leanhardt}}, \bibinfo {author} {\bibfnamefont {A.}~\bibnamefont {G\"orlitz}}, \bibinfo {author} {\bibfnamefont {S.}~\bibnamefont {Gupta}}, \bibinfo {author} {\bibfnamefont {D.~E.}\ \bibnamefont {Pritchard}},\ and\ \bibinfo {author} {\bibfnamefont {W.}~\bibnamefont {Ketterle}},\ }\bibfield  {title} {\bibinfo {title} {Transport of bose-einstein condensates with optical tweezers},\ }\href {https://doi.org/10.1103/PhysRevLett.88.020401} {\bibfield  {journal} {\bibinfo  {journal} {Phys. Rev. Lett.}\ }\textbf {\bibinfo {volume} {88}},\ \bibinfo {pages} {020401} (\bibinfo {year} {2001})}\BibitemShut {NoStop}%
\bibitem [{\citenamefont {Griffiths}\ and\ \citenamefont {Schroeter}(2018)}]{griffiths2018introduction}%
  \BibitemOpen
  \bibfield  {author} {\bibinfo {author} {\bibfnamefont {D.~J.}\ \bibnamefont {Griffiths}}\ and\ \bibinfo {author} {\bibfnamefont {D.~F.}\ \bibnamefont {Schroeter}},\ }\href@noop {} {\emph {\bibinfo {title} {Introduction to Quantum Mechanics}}},\ \bibinfo {edition} {3rd}\ ed.\ (\bibinfo  {publisher} {Cambridge University Press},\ \bibinfo {year} {2018})\BibitemShut {NoStop}%
\bibitem [{\citenamefont {Messiah}(1976)}]{Messiah}%
  \BibitemOpen
  \bibfield  {author} {\bibinfo {author} {\bibfnamefont {A.}~\bibnamefont {Messiah}},\ }\bibfield  {title} {\bibinfo {title} {Quantum mechanics},\ }\href@noop {} {\bibfield  {journal} {\bibinfo  {journal} {Wiley, New York}\ } (\bibinfo {year} {1976})}\BibitemShut {NoStop}%
\bibitem [{\citenamefont {Landau}\ and\ \citenamefont {Lifshitz}(1965)}]{text:Landau-Lifshitz}%
  \BibitemOpen
  \bibfield  {author} {\bibinfo {author} {\bibfnamefont {L.~D.}\ \bibnamefont {Landau}}\ and\ \bibinfo {author} {\bibfnamefont {E.~M.}\ \bibnamefont {Lifshitz}},\ }\href@noop {} {\emph {\bibinfo {title} {{Quantum Mechanics: Non-Relativistic Theory}}}},\ Course of Theoretical Physics\ (\bibinfo  {publisher} {Pergamon Press},\ \bibinfo {year} {1965})\BibitemShut {NoStop}%
\bibitem [{\citenamefont {Keldysh}(1965)}]{Keldysh1965}%
  \BibitemOpen
  \bibfield  {author} {\bibinfo {author} {\bibfnamefont {L.~V.}\ \bibnamefont {Keldysh}},\ }\bibfield  {title} {\bibinfo {title} {{Ionization in the field of a strong electromagnetic wave}},\ }\href {http://www.jetp.ras.ru/cgi-bin/e/index/e/20/5/p1307?a=list} {\bibfield  {journal} {\bibinfo  {journal} {Sov. Phys. JETP.}\ }\textbf {\bibinfo {volume} {20}},\ \bibinfo {pages} {1307} (\bibinfo {year} {1965})}\BibitemShut {NoStop}%
\bibitem [{\citenamefont {{Perelomov}}\ \emph {et~al.}(1966)\citenamefont {{Perelomov}}, \citenamefont {{Popov}},\ and\ \citenamefont {{Terent'ev}}}]{PPT1966}%
  \BibitemOpen
  \bibfield  {author} {\bibinfo {author} {\bibfnamefont {A.~M.}\ \bibnamefont {{Perelomov}}}, \bibinfo {author} {\bibfnamefont {V.~S.}\ \bibnamefont {{Popov}}},\ and\ \bibinfo {author} {\bibfnamefont {M.~V.}\ \bibnamefont {{Terent'ev}}},\ }\bibfield  {title} {\bibinfo {title} {{Ionization of Atoms in an Alternating Electric Field}},\ }\href {http://www.jetp.ras.ru/cgi-bin/e/index/e/23/5/p924?a=list} {\bibfield  {journal} {\bibinfo  {journal} {Sov. Phys. JETP.}\ }\textbf {\bibinfo {volume} {23}},\ \bibinfo {pages} {924} (\bibinfo {year} {1966})}\BibitemShut {NoStop}%
\bibitem [{\citenamefont {{Ammosov}}\ \emph {et~al.}(1986)\citenamefont {{Ammosov}}, \citenamefont {{Delone}},\ and\ \citenamefont {{Krainov}}}]{ADK1986}%
  \BibitemOpen
  \bibfield  {author} {\bibinfo {author} {\bibfnamefont {M.~V.}\ \bibnamefont {{Ammosov}}}, \bibinfo {author} {\bibfnamefont {N.~B.}\ \bibnamefont {{Delone}}},\ and\ \bibinfo {author} {\bibfnamefont {V.~P.}\ \bibnamefont {{Krainov}}},\ }\bibfield  {title} {\bibinfo {title} {{Tunnel ionization of complex atoms and of atomic ions in an alternating electromagnetic field}},\ }\href {http://www.jetp.ras.ru/cgi-bin/e/index/e/64/6/p1191?a=list} {\bibfield  {journal} {\bibinfo  {journal} {Sov. Phys. JETP.}\ }\textbf {\bibinfo {volume} {64}},\ \bibinfo {pages} {1191} (\bibinfo {year} {1986})}\BibitemShut {NoStop}%
\end{thebibliography}%
\end{document}